\documentclass[twocolumn,aps,showpacs,amssymb,floatfix]{revtex4}

\usepackage{graphicx}
\usepackage{bm}
\usepackage{amsmath}
\usepackage{amssymb}
\usepackage{amsfonts}
\usepackage{float}
\usepackage{hyperref}
\usepackage{dsfont}  
\usepackage{slashed}  
\usepackage{booktabs}
\usepackage{multirow}
\usepackage{subfigure}
\usepackage[sort&compress]{natbib}
\usepackage{placeins}
\usepackage{wasysym}

\newcommand{\be}{\begin{equation}}  
\newcommand{\ee}{\end{equation}}  
\newcommand{\beq}{\begin{eqnarray}}  
\newcommand{\eeq}{\end{eqnarray}}

\newcommand{\Dlr}{\buildrel \leftrightarrow \over D\raise-1pt\hbox{}}

\begin{document}

\title{\hspace*{14cm}  DESY 15-123\\[3ex]
Nucleon   and pion structure with lattice QCD simulations at physical value of the pion mass}
\date{\today}

\author{
  A.~Abdel-Rehim$^{1}$,
  C.~Alexandrou$^{1,2}$,
  M.~Constantinou$^{1,2}$,
  P.~Dimopoulos$^{3,4}$,
  R.~Frezzotti$^{4,5}$, 
  K.~Hadjiyiannakou$^{1,2}$,
  K.~Jansen$^{6}$,
  Ch.~Kallidonis$^{1}$,
  B.~Kostrzewa$^{6,7}$,
  G.~Koutsou$^{1}$,
  M.~Mangin-Brinet$^{8}$,
  M.~Oehm$^{9}$,
  G.~C.~Rossi$^{4,5}$,
  C.~Urbach$^{9}$,
  U. Wenger$^{10}$}

\affiliation{
  $^1$Computation-based Science and Technology Research Center,
  The Cyprus Institute,
  20 Kavafi Str.,
  Nicosia 2121,
  Cyprus \\
  $^2$Department of Physics,
  University of Cyprus,
  P.O. Box 20537,
  1678 Nicosia,
  Cyprus\\ 
$^3$ Centro Fermi - Museo Storico della Fisica e Centro Studi e Ricerche Enrico Fermi Compendio del Viminale, Piazza del Viminiale 1, I-00184, Rome, Italy\\
$^4$Dipartimento di Fisica, Universit\'a di Roma “Tor Vergata” Via della Ricerca Scientifica 1, I-00133 Rome, Italy\\
$^5$ INFN, Sezione di “Tor Vergata”
Via della Ricerca Scientifica 1, I-00133 Rome, Italy\\
  $^6$NIC, DESY,
  Platanenallee 6,
  D-15738 Zeuthen,
  Germany\\
$^7$ Institut f\"ur Physik, Humboldt-Universit\"at zu Berlin, Newtonstr. 15,
12489 Berlin, Germany\\
$^8$ Theory Group, Lab. de Physique Subatomique et de Cosmologie, 38026 Grenoble, France\\
  $^9$HISKP (Theory), Bonn University,
  Nussallee 14-16,
  Bonn,
  Germany\\
$^{10}$ Albert Einstein Center for Fundamental Physics, University of Bern, CH-3012 Bern, Switzerland
}

\begin{abstract}
We present results on the nucleon scalar, axial and tensor charges as
well as on the momentum fraction, and  the  helicity  and transversity
moments. The pion momentum fraction is also presented. 
The computation of these key observables is carried out
  using
lattice QCD simulations at a physical value of the pion mass. The
evaluation is based on gauge configurations generated with two
degenerate sea quarks of twisted mass fermions with a clover term. We
investigate excited states contributions with the nucleon quantum numbers  by
analyzing three sink-source time separations. We find that, for the
scalar charge, excited states contribute significantly and to a less degree to the nucleon
momentum fraction and helicity moment.  Our result for the nucleon axial charge agrees
with the experimental value. Furthermore, we predict a value of 1.027(62) in the
$\overline{\text{MS}}$ scheme at $2$ GeV for the 
isovector  nucleon tensor charge directly at the physical point. The pion momentum fraction is
found to be   $\langle x\rangle_{u-d}^{\pi^\pm}=0.214(15)(^{+12}_{-9})$ in the $\overline{\rm MS}$ at 2~GeV. 
\end{abstract}  


\maketitle 

\begin{figure}
  \begin{center}

  \end{center}
\end{figure}

\newcommand{\Op}{\mathcal{O}}
\newcommand{\C}{\mathcal{C}}
\newcommand{\eins}{\mathds{1}}

\bibliographystyle{apsrev}

\section{Introduction}
The nucleon axial-vector coupling or nucleon axial charge $g_A$ is
experimentally a well known quantity determined from the $\beta$-decay
of the neutron. It is a key parameter for understanding the chiral
structure of the nucleon and a quantity that has been studied
extensively in chiral effective
theories~\cite{Bernard:1995dp,Hemmert:2003cb}. A description of baryon
properties in chiral effective theory requires as an input $g_A$ and
thus its value at the chiral limit and its dependence on the pion mass
constitute important information that lattice QCD can provide.  Its
importance for phenomenology as well as the fact that it is rather
straightforward to compute in lattice QCD have made it one of the
most studied quantities  within different fermion
discretization schemes~\cite{Alexandrou:2010cm, Constantinou:2014tga,
  Alexandrou:2013cda, Alexandrou:2014yha, Syritsyn:2014saa, Bali:2013gya, Horsley:2013ayv}. In lattice QCD, $g_A$ is determined
directly from the zero momentum transfer nucleon matrix element of the
axial-vector current without requiring any extrapolation from finite
momentum transfer calculations as, for example, is required for the
anomalous magnetic moment of the nucleon. In addition, being an
isovector quantity, it does not receive any contributions from the
coupling of the current to closed quark loops and thus one only needs
to compute the connected contribution with well established lattice
QCD techniques.  Therefore, $g_A$ has come to be regarded as a prime
benchmark quantity for the computation of lattice QCD matrix
elements. Postdiction of the value of $g_A$ within lattice QCD is,
therefore, regarded as an essential step before the reliable
prediction of other couplings and form factors for which   the same
formalism is used.

Unlike $g_A$, the nucleon scalar and tensor charges are not well
known. Limits on the value of the scalar and tensor coupling constants
arise from $0^+ \rightarrow 0^+$ nuclear decays and the radiative pion
decay $\pi\rightarrow e\nu \gamma$, respectively. They have become the
focus of planned experiments to search for physics beyond the familiar
weak interactions of the Standard Model sought in the decay of
ultra-cold neutrons~\cite{Bhattacharya:2011qm}. The computation of the
tensor charge is particularly timely since new experiments using
polarized $^3$He/Proton at Jefferson lab aim at increasing the
experimental accuracy of its measurement by an order of
magnitude~\cite{Gao:2010av}. In addition, experiments at LHC are
expected to increase the limits to contributions arising from tensor
and scalar interactions by an order of magnitude making these
observables interesting probes of new physics originating at the TeV
scale.  Computing the scalar charge will also provide input for dark
matter searches.  Experiments, which aim at a direct detection of dark
matter, are based on measuring the recoil energy of a nucleon hit by a
dark matter candidate. In many supersymmetric
scenarios~\cite{Ellis:2010kf} and in some Kaluza-Klein extensions of
the standard model~\cite{Servant:2002hb,Bertone:2010ww} the dark
matter nucleon interaction is mediated through a Higgs boson. In such
a case the theoretical expression of the spin independent scattering
amplitude at zero momentum transfer involves the quark content of the
nucleon or the nucleon sigma-term, which is closely related
to the scalar charge. In fact, this contributes the
largest uncertainty on the nucleon dark matter cross section.
Therefore, computing the scalar $g_S$ and tensor $g_T$ charges of the
nucleon within lattice QCD will provide useful input for the ongoing
experimental searches for beyond the standard model physics.

Another experimental frontier that provides information on the quark
and gluon structure of a hadron, is the measurement of parton
distribution functions (PDFs) in a variety of high energy processes
such as deep-inelastic lepton scattering and Drell-Yan in
hadron-hadron collisions. PDFs give, to leading twist, the probability
of finding a specific parton in the hadron carrying certain momentum
and spin, in the infinite momentum frame. Their universal nature
relies on factorization theorems that allow differential
cross-sections to be written in terms of a convolution of certain
process-dependent coefficients that encode the hard perturbative
physics and process-independent PDFs that describe the soft,
non-perturbative physics at a factorization energy scale
$\mu$~\cite{Collins:1985ue,Brock:1993sz}.  Because these PDFs are
light-cone correlation functions it is not straightforward to
calculate them directly in Euclidean space. Instead, one calculates
Mellin moments of the PDFs expressed in terms of hadron matrix
elements of local operators, which through the operator product
expansion are related to the original light-cone correlation
functions. Mellin moments are measured or extracted from
phenomenological analyses in deep-inelastic scattering experiments and
thus they can be directly compared to lattice results when converted
to the same energy scale $\mu$.

In this work, we consider the three first moments that one can
construct, namely the first moment of the spin-independent (or
unpolarized) $q=q_\downarrow+q_\uparrow$, helicity (or polarized)
$\Delta q=q_\downarrow-q_\uparrow,$ and transversity $\delta
q=q_\top+q_\perp$ distributions, which are define as follows:

\begin{eqnarray}
   \langle x\rangle_q         &=& \int_{0}^1 x \left [q(x)+\bar{q}(x)\right] dx \\
   \langle x\rangle_{\Delta q} &=& \int_{0}^1 x \left [\Delta q(x)-\Delta \bar{q}(x)\right] dx \\
   \langle x\rangle_{\delta q} &=& \int_{0}^1 x \left[\delta q(x)+\delta \bar{q}(x)\right] dx \>,
\end{eqnarray}
where $q_{\downarrow}$ and $q_{\uparrow}$ correspond respectively, to
quarks with helicity aligned and anti-aligned with that of a
longitudinally polarized target, and $q_\top$ and $q_{\perp}$
correspond to quarks with spin aligned and anti-aligned with that of a
transversely polarized target. These moments, at leading twist, can
be extracted from the hadron matrix elements of one-derivative vector,
axial-vector and tensor operators at zero momentum transfer. Thus, they
constitute the next level of observables in terms of complexity that
can be computed in lattice QCD after the coupling constants that do
not involve derivative operators. The unpolarized and polarized
moments $\langle x\rangle_q$ and $\langle x\rangle_{\Delta q}$ of the
nucleon are measured experimentally and thus lattice QCD provides a
postdiction, while a computation of the nucleon transversity $\langle
x\rangle_{\delta q}$ provides a prediction.
It is worth mentioning a new approach proposed recently for measuring
directly the PDFs within lattice QCD~\cite{Ji:2013dva,Xiong:2013bka},
which is currently under
investigation~\cite{Lin:2014zya,Alexandrou:2014pna}. 

In this paper, we extend the analysis  of meson masses, the muon 
anomalous magnetic moment $g-2$  and the meson decay constants
considered in Ref.~\cite{Abdel-Rehim:2015pwa}, to the nucleon matrix elements for the
three first Mellin moments while for the pion we compute the momentum
fraction. While the present paper builds on the methodology developed
in
Refs.~\cite{Alexandrou:2010hf,Alexandrou:2011nr,Alexandrou:2011db,Dinter:2011sg,Alexandrou:2013joa},
this work presents the first evaluation of these six quantities
directly at the physical value of the pion mass.  This is a
substantial step forward since it avoids chiral extrapolations, which
are often difficult and can lead to rather large systematic
uncertainties.

The paper is organized as follows: In section II we define the nucleon
and pion matrix elements, in section III we explain the lattice
methodology, in section IV we give the simulation details and in
section V our results. Section VI summarizes our findings and gives
our conclusions.

\section{Matrix elements}
\subsection{Nucleon}

We are interested in extracting the forward nucleon matrix elements $
\langle N(p)|\Op|N(p)\rangle$, with $p$ the nucleon initial and final
momentum. We consider the complete set of local and one-derivative
operators, yielding a non-zero result. The local scalar, axial-vector,
and tensor operators are:

\be
\Op_{S^a} = \bar{q}\frac{\tau^a}{2} q \>, \quad
\Op^\mu_{A^a} = \bar{q}\gamma_5\gamma^\mu\frac{\tau^a}{2} q\>,\quad
\Op^{\mu\nu}_{T^a} =\bar{q}\sigma^{\mu\nu}\frac{\tau^a}{2}q\>.
\label{eq:localops}
\ee
We do not consider the vector operator $\overline{\psi}(x) \gamma_mu \psi(x)$
since this yields the renormalization constant $Z_V$, which we calculate separately using our RI-MOM setup, as explained in section~V.
If one instead uses  the lattice conserved 
Noether current, which we typically do in in our computation of the nucleon electromagnetic form factors,
we  
 the forward matrix element trivially yields
the electric charge.

The one-derivative vector, axial-vector, and
tensor operators are given by

\beq
\Op_{V^a}^{\mu\nu} &=& \bar{q}\gamma^{\{\mu} \Dlr^{\nu\}} \frac{\tau^a}{2}q,\nonumber\\
\Op_{A^a}^{\mu\nu} &=& \bar{q}\gamma^{\{\mu} \Dlr^{\nu\}} \gamma_5 \frac{\tau^a}{2}q,\nonumber\\
\Op_{T^a}^{\mu\nu\rho} &=& \bar{q}\sigma^{[\mu\{\nu]} \Dlr^{\rho\}}\frac{\tau^a}{2} q,
\label{eq:derivops}
\eeq
where $\bar{q}=(\bar{u},\bar{d})$,
\[
\stackrel{\leftrightarrow}{D}_\mu = \frac{1}{2}\left(\stackrel{\rightarrow}{D}_\mu -
\stackrel{\leftarrow}{D}_\mu\right)\,,\quad D_\mu=\frac{1}{2}
\left(\bigtriangledown_\mu+\bigtriangledown_\mu^*\right)
\]
and $\bigtriangledown_\mu$ ($\bigtriangledown_\mu^*$) is the usual
forward (backward) derivative on the lattice. The curly (square)
brackets represent a symmetrization (anti-symmetrization) over pairs
of indices, with the symmetrization accompanied by subtraction of
the trace. 

In what follows all expressions will be given in Euclidean time.  
For example, the one-derivative vector current that will  be used
in the computations of both nucleon and pion momentum fractions,  in Euclidean time and setting $\mu=\nu=4$, is given by:
\beq
\Op_{V^a}^{44} &=& \bar{q}\frac{3}{4}[\gamma^4 \Dlr^4 -
  \frac{1}{3}\sum_{k=1}^{3}\gamma^k\Dlr^k]\frac{\tau^a}{2}q.
\label{eq:deriv44}
\eeq
 In this
work, we consider the isovector quantities obtained from
Eqs.~(\ref{eq:localops}) and (\ref{eq:derivops}) by using the Pauli
matrix $\tau^3$.
We also consider the isoscalar combination obtained
by replacing $\tau^a$ by unity. The individual
up- and down-quark combinations can be extracted form the isovector and isoscalar quantities which are equivalent to replacing
$\tau^a$ with the projectors onto the up- or down- quarks. 
The isoscalar combination and the up- and
down-quark contributions receive disconnected contributions. Our high
statistics study using an $N_f=2+1+1$ ensemble of twisted mass
fermions with pion mass of 373~MeV has shown that the disconnected
contributions for the tensor isoscalar charge and the isoscalar first
moments are very small compared to the
connected~\cite{Abdel-Rehim:2013wlz,Alexandrou:2013wca}. In the same
study, the disconnected contributions to the isoscalar axial and scalar
charge were found to be about (7- 10)\% of the connected.  In this
work, we will only compute the connected contributions.  The
disconnected contributions will need at least an order of magnitude
more statistics and will be presented in a follow-up publication.

For zero momentum transfer,  the nucleon matrix
elements of the local operators in Eq.~(\ref{eq:localops}) can be
decomposed in the following form factors:
\begin{widetext}
\beq \langle N(p,s^\prime)| \Op_{S}| N(p,s)\rangle &=& \bar
u_N(p,s^\prime)\Bigl[\frac{1}{2}G_S(0)\Bigr]u_N(p,s),\label{eq:OpS}\\
 \langle N(p,s')|\mathcal{O}^\mu_{A}|N(p,s)\rangle &=& i \bar{u}_N(p,s') 
 \Bigl[\frac{1}{2}G_A(0)\gamma^\mu\gamma_5\Bigr]
 u_N(p,s),\label{eq:OpA}\\ 
\langle N(p,s')|\mathcal{O}^{\mu\nu}_{T}|N(p,s)\rangle &=& \bar{u}_N(p,s')
\Bigl[\frac{1}{2}
A_{T10}(0)\sigma^{\mu\nu}
\Bigr]
u_N(p,s).\label{eq:OpT}
\eeq
\end{widetext}
Thus, the scalar matrix element at zero momentum transfer yields the
form factor $G_S(0)\equiv g_S$, the local axial-vector $G_A(0)\equiv
g_A$ and the local tensor matrix element yields $A_{T10}(0)\equiv
g_T$. In all these quantities the operators are either the isovector
or isoscalar combinations, or individual up- or down-quark
contributions.  At
non-zero momentum, additional form-factors arise in the decomposition
of Eqs.~(\ref{eq:OpA}) and (\ref{eq:OpT}). Namely, the induced
pseudo-scalar $G_p(Q^2)$ appears as the second form factor in the
decomposition of the matrix element of the axial-vector and the form
factors $B_{T10}(Q^2)$ and $\tilde{A}_{T10}(Q^2)$ appear in the
decomposition of the nucleon matrix element of the tensor
operator, where  $Q^2$ is the momentum transfer
square in Euclidean time. These cannot be extracted at zero momentum transfer and
will not be considered in this work. 

The corresponding decomposition for the one-derivative operators in
Eq.~(\ref{eq:derivops}) is given by:
\begin{widetext}
  \beq
  \langle N(p,s^\prime)| \Op_{V}^{\mu\nu}|N(p,s)\rangle &=&
  \bar{u}_N(p,s^\prime)\Bigl[
    \frac{1}{2} A_{20}(0) \gamma^{\{\mu}p^{\nu\}}
    \Bigr] u_N(p,s), \\
  \langle N(p,s^\prime)| \Op_{A}^{\mu\nu}|N(p,s)\rangle &=& i
  \bar u_N(p,s^\prime)\Bigl[
    \frac{1}{2} \tilde{A}_{20}(0)\gamma^{\{\mu}p^{\nu\}}\gamma^5
    \Bigr] u_N(p,s),\\
  \langle N(p,s^\prime)| \Op_{T}^{\mu\nu\rho}|N(p,s)\rangle &=& i
  \bar u_N(p,s^\prime)\Bigl[ 
    \frac{1}{2}A_{T20}(0)\sigma^{[\mu\{\nu]}p^{\rho\}}
    \Bigr]
  u_N(p,s).
  \eeq
\end{widetext}

The momentum fraction, helicity moment, and the transversity moment
are obtained from the above forward matrix elements by $\langle
x\rangle_{q}=A_{20}^{q}(0)$, $\langle x\rangle_{\Delta
  q}=\tilde{A}^{q}_{20}(0)$, and $\langle x\rangle_{\delta
  q}=A^{q}_{T20}(0)$ respectively.  Here we use the generic symbol $q$
to denote the quark combination, where $q = u+d$ will denote the
isoscalar combination, $q=u-d$ will denote the isovector combination
and $q=u$ or $q=d$ denotes the individual up- and down-quark
contributions. For instance, the isovector helicity moment will be
denoted as $\langle x\rangle_{\Delta u-\Delta d} = \tilde{A}^{u-d}_{20}(0)$.  
For uniformity in our notation we will
also write $g_A^{u-d}$ for the nucleon axial charge, despite the fact that the
measured axial charge is understood to be an isovector quantity.

\subsection{Pion}

The isovector momentum fraction of the pion $\langle x\rangle^{\pi^\pm}_{u-d}$, can
be extracted from the corresponding pion matrix element of the
one-derivative vector operator. Specifically we use the following
operator, sometimes also denoted as $\mathcal{O}_{v2b}$,
\begin{equation}
  \mathcal{O}^{44}(x)=\frac{2}{3}\Op^{44}_{V^3}(x)
\end{equation}
where $\Op^{44}_{V^3}(x)$ is given in
Eq.~\ref{eq:deriv44}.  As in the case of the
nucleon, no external momentum is needed in our calculation, which is
advantageous since an external momentum increases the noise to signal
ratio.

\section{Lattice Methodology}
\subsection{Correlation functions}

In order to compute hadron matrix elements we need to calculate the
appropriate three-point function. We first present the setup for the
nucleon matrix elements for the special case ${\bf q}={\bf 0}$.  The
three-point function is then given by
\begin{widetext}
  \beq 
  G^{\mu_1,...,\mu_n}_{\rm 3pt}(\Gamma^\nu, {\bf p},  t_s, t_{\rm ins}) &=& 
  \sum_{{\bf x}_s,{\bf x}_{\rm ins}}\> e^{-i({\bf x}_s-{\bf x}_0)\cdot{\bf p}}\>
  \Gamma^\nu_{\beta\alpha}\langle J_\alpha({\bf x}_s,t_s)\Op^{\mu_1,...,\mu_n}_{\Gamma}
 ({\bf x}_{\rm ins}, t_{\rm ins})\bar{J}_\beta({\bf x}_0,t_0)\rangle 
  \label{eq:correlators}
  \eeq
\end{widetext}
  where $x_0$, $x_{\rm ins}$ and $x_s$ are
the source, insertion and sink coordinates respectively. In order to
cancel unknown overlaps of the interpolating field with the nucleon
state as well as the time evolution in Euclidean time we construct
ratios of the three-point function with the two-point function, which
is given by \be G_{\rm 2pt}({\bf 0}, t_s) = \sum_{{\bf
    x}_s}\Gamma^4_{\beta\alpha}\langle J_\alpha({\bf
  x}_s,t_s)\bar{J}_\beta({\bf x}_0,t_0)\rangle.
\label{eq:two-point} 
\ee
The projection matrices are 
\be
\Gamma^4 = \frac{1}{4}(\eins+\gamma_4),\,\,\,\Gamma^k = \Gamma^4i\gamma_5\gamma_k.
\label{eq:proj}
\ee
We use the proton interpolating operators: 
\be
J_{\alpha}(x) = \epsilon^{abc}u_{\alpha}^a(x)[u^{\top b}(x) C\gamma_5
  d^c(x)] 
\ee 
with $a$, $b$ and $c$ denoting color components. We employ Gaussian
smeared quark fields~\cite{Alexandrou:1992ti,Gusken:1989ad} to increase
the overlap with the proton state and decrease overlap with excited
states. The smeared interpolating fields are given by 
\beq q_{\rm
  smear}^a(t,{\bf x}) &=& \sum_{{\bf y}} F^{ab}({\bf x},{\bf
y};U(t))\ q^b(t,{\bf y})\,,\\ F &=& (\eins + {a_G} H)^{N_G} \,,
\nonumber\\ H({\bf x},{\bf y}; U(t)) &=& \sum_{i=1}^3[U_i(x)
  \delta_{x,y-\hat\imath} + U_i^\dagger(x-\hat\imath)
  \delta_{x,y+\hat\imath}]\,. \nonumber 
\eeq 
We apply APE-smearing to the gauge fields $U_\mu$ entering the hopping
matrix $H$. The parameters for the Gaussian smearing $a_G$ and $N_G$
are optimized using the nucleon ground state~\cite{Alexandrou:2008tn}
such as to give a root mean square radius of about 0.5~fm. We use
$(N_G,a_G)=(50,4)$ and $(N_{\rm APE},a_{\rm APE})=(50,0.5)$.

\begin{figure}
\includegraphics[width=\linewidth]{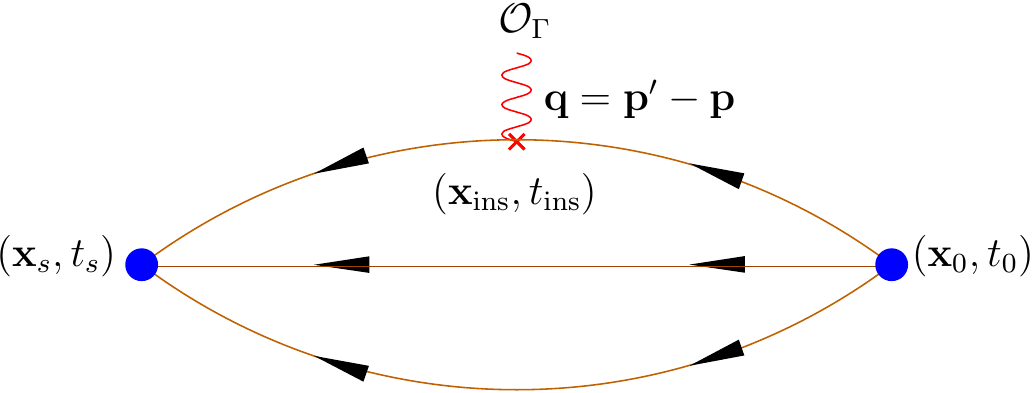}\\\vspace*{0.8cm}
\includegraphics[width=\linewidth]{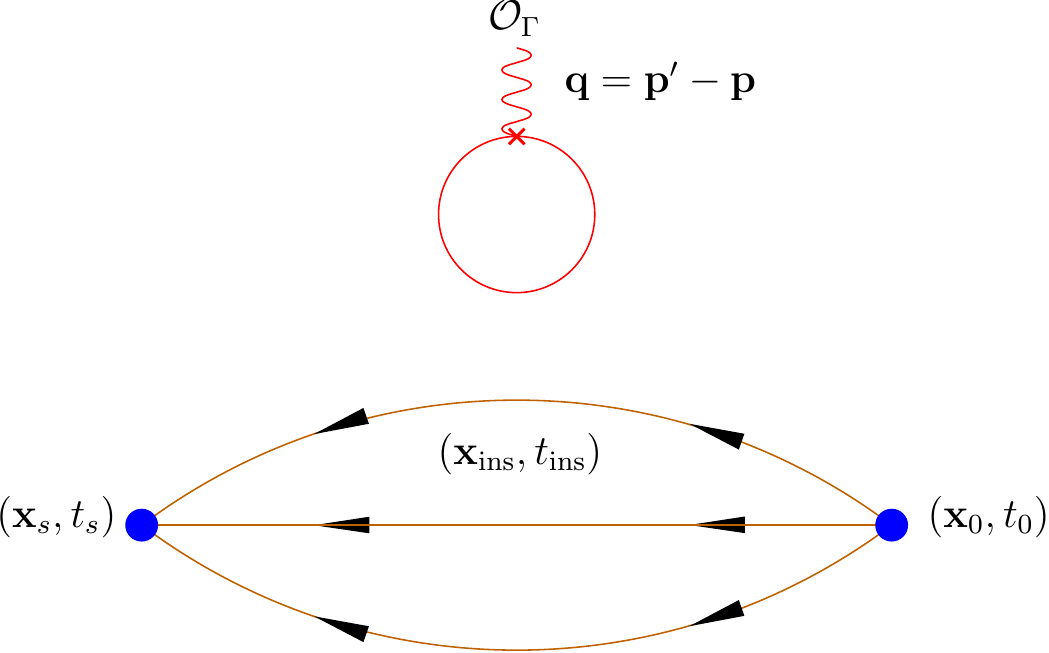}
\caption{Connected (upper) and disconnected (lower) contributions to
  nucleon three-point functions.}
\label{fig:diagram}
\end{figure}

For the case of isovector quantities, the so-called disconnected
contributions arising from the coupling of the operators to a sea
quark shown schematically in the lower panel of
Fig.~\ref{fig:diagram}, are zero in the isospin limit up to lattice
cut-off effects. Since we work with an automatic ${\cal O}(a)$-improved
action we expect cut-off effects to be small for our action and
lattice spacing.  Thus, these correlators can be calculated by
evaluating the connected diagram of Fig.~\ref{fig:diagram} (upper
panel) for which we employ sequential inversions through the
sink~\cite{Dolgov:2002zm}. For the case of isoscalar quantities, the
disconnected diagrams do not vanish and need to be computed.  The
calculation of disconnected contributions needs special techniques and
at least an order of magnitude more statistics than the connected
ones~\cite{Alexandrou:2013wca}. In a high-statistics analysis using
$N_f=2+1+1$ twisted mass fermions at pion mass of 373~MeV we found
that they contribute about 7\% to the isoscalar axial charge while
they are negligible for the tensor charge, $\langle x \rangle_q$ and
$\langle x \rangle_{\Delta q}$~\cite{Abdel-Rehim:2013wlz}. In this
work we restrict ourselves to the calculation of the connected
contributions.

For evaluation of the connected contribution we take the nucleon
creation operator at a position $x_0=({\bf x}_0,t_0) $, referred to as
the source position, which is randomized over gauge-configurations to
reduce autocorrelations, and the annihilation operator at a later time
$t_s$, the sink position $x_s$. The current couples to a quark at an
intermediate time $t_{\rm ins}$, the insertion time.  In this
calculation, we fix the sink and source to momentum ${\bf p} {=} {\bf
  0}$ and thus the current carries zero momentum. We employ the
sequential inversion method through the sink requiring one set of
sequential inversions per choice of the sink time-slice $t_s$ and sink
projector. Thus, within this approach, at a fixed sink-source time
separation we obtain results for all insertion times as well as for
any operator $\Op^{\{\mu_1\cdots\mu_n\}}_\Gamma$.  We perform separate
inversions for the four projection matrices $\Gamma^4$ and $\Gamma^k$
given in Eq.~(\ref{eq:proj}).

Using the two- and three-point functions of
Eqs.~(\ref{eq:correlators}) and (\ref{eq:two-point}) we form the ratio
\be 
R(\Gamma^\lambda,t_s,t_{\rm ins})=
\frac{G_{\rm 3pt}(\Gamma^\lambda,{\bf 0},t_s,t_{\rm ins}) }{G_{\rm 2pt}({\bf 0}, t_s)}
\label{eq:ratio}
\ee
where all time separations are given relative to the source time
$t_0$. For large time separations, the two-point function in the
denominator cancels unknown overlaps of the nucleon interpolating
operators with the nucleon spinors as well as the Euclidean time
evolution, such that the desired matrix element is isolated. However,
in order to identify when the large time limit sets in, one has to
carefully study the time dependence of the ratio since excited states
contamination can affect the value of the ratio. For arbitrary times
the contributions from excited states in the ratio are
\beq
& &R(\Gamma^\lambda,t_s,t_{\rm ins}) \propto  \nonumber\\
& &\hspace*{-0.6cm} \frac{\sum_{n',n}\langle
  J|n'\rangle\langle n|\bar{J}\rangle \langle n'| \mathcal{O}_\Gamma | n \rangle
  e^{-E_{n'}(t_s-t_{\rm ins})} e^{-E_{n}(t_{\rm ins}-t_0)}}{\sum_{n}|\langle J|n\rangle|^2e^{-E_n(t_s-t_0)}} \nonumber\\
{}
\label{eq:ratiofull}
\eeq
where $|n\rangle$ is the $n^{\rm th}$ eigenstate of the QCD
Hamiltonian with the quantum numbers of the nucleon and $E_n$ is the
energy of the state in the rest frame of the nucleon. Denoting with
$|0\rangle=|N\rangle$ the nucleon ground state, with
$|1\rangle=|N'\rangle$ and $|2\rangle=|N''\rangle$ the first and
second excited states, and with $\Delta = E_{N'}-m_N$ and $\Delta' =
E_{N''}-E_{N'}$ their respective energy gaps, the ratio in
Eq.~(\ref{eq:ratiofull}) yields
\beq & & R(\Gamma^\lambda,t_s,t_{\rm
  ins})\propto\nonumber\\
& & \hspace*{-0.5cm} \frac{\mathcal{M}+\mathcal{R}e^{-\Delta(t_s-t_{\rm
      ins})}+\mathcal{R}^\dagger \>e^{-\Delta (t_{\rm ins}-t_0)}+
  O(e^{-\Delta'(t_{\rm ins}-t_0)})}{1+|\mathcal{C}|^2e^{-\Delta (t_s-t_0)}+O(e^{-\Delta'(t_s-t_0)})} \nonumber \\
{}
\label{eq:ratioexc}
\eeq
where $\mathcal{M}=\langle N | \mathcal{O}_\Gamma|N\rangle$ is the
desired matrix element, $\mathcal{C}=\frac{\langle
  J|N'\rangle}{\langle J|N\rangle}$, and
$\mathcal{R}=\mathcal{C}\langle N'|\mathcal{O}_\Gamma|N\rangle$.  If
the exponential terms are small compared to ${\cal M}$ and to unity,
then we have what is referred to as ground-state dominance and the
ratio yields the desired ground state matrix element.

The computation of the pion momentum fraction is carried out along the
same lines as for the nucleon. To extract the desired matrix element
$\langle\pi({\bf 0})|\mathcal{O}^{44}|\pi({\bf 0})\rangle$ we
construct the ratio of the appropriate three-point function with the
pion two-point function with a source at $t_0$ and a sink at $t_s$:
\begin{equation}
  \label{eq:averxratio}
  R^{\pi}(t_s,t_{\rm ins})=
  \frac{G_{\rm 3pt}^{44}({\bf 0},t_s,t_{\rm ins})}{G^{\pi}_\mathrm{2pt}(t_s)}
\end{equation}
with
\begin{equation}
  G_{\rm 3pt}^{44}({\bf 0},t_s,t_{\rm ins})=\sum_{\mathbf{y}}\langle {J}_{\pi^\pm}(t_s)\,
  \mathcal{O}^{44}(t_{\rm ins},\mathbf{y})\,
          { J}_{\pi^\pm}^\dagger(t_0)\rangle\,.
\end{equation}
with $J_{\pi^{+}}(x)=\bar{d}(x)\gamma_5 u(x)$
($J_{\pi^{-}}(x)=\bar{u}(x)\gamma_5 d(x)$) is the interpolating field
of $\pi^+$ ($\pi^-$). As in the case of the nucleon, in the isovector
combination up to lattice artifacts, the disconnected
contributions vanish and will thus
be dropped.

The pion two- and three-point correlators, unlike those of the
nucleon, are evaluated by using a stochastic time-slice source
(Z(2)-noise in both real and imaginary
part)~\cite{Dong:1993pk,Foster:1998vw,McNeile:2006bz} for all color,
spin and spatial indices. This method, which is particularly suited for the pion,  has been first applied to
moments of parton distribution functions in Ref.~\cite{Baron:2007ti}.
The quark propagator $S^b_\beta(y)$ is obtained by solving
\begin{equation}
  \sum_y\,  D^{ab}_{\alpha\beta}(z,y)\,
  S^b_\beta(y)=\xi(\mathbf{z})^a_\alpha\, \delta_{z_0,t_0}\quad
  (\textrm{source at}~ t_0) 
\end{equation}
for $S$.
$\xi(\mathbf{z})^a_\alpha$ is a $Z(2)$ random source satisfying
\begin{equation}
  \begin{split}
    \langle\xi^*(\mathbf{x})^a_\alpha\ \xi(\mathbf{y})^b_\beta\rangle_r&=
    \delta_{\mathbf{x}\mathbf{y}}\delta_{ab}\delta_{\alpha\beta}\,,\\
    \langle\xi(\mathbf{x})^a_\alpha\
    \xi(\mathbf{y})^b_\beta\rangle_r&=0\,,\\
  \end{split}
\end{equation}
where $\langle\cdot\rangle_r$ denotes the average over many random
sources.  Using $S$ we can define a so-called sequential or
generalized propagator $\Sigma^b_\beta(y)$
from~\cite{Martinelli:1987zd}
\begin{equation}
  \sum_y\, D^{ab}_{\alpha\beta}(z,y)\, \Sigma^b_\beta(y)\ =\ \gamma_5\, S^a_\alpha(z)\, \delta_{z_0,t_s}\quad (\textrm{sink at}~ t_s)\,.
\end{equation}
This method represents a generalization of the
one-end-trick~\cite{McNeile:2000hf} to moments of parton distribution
functions.  Its clear advantage is an increased signal to noise ratio
at reduced computational costs at least when it is applied for 
meson observables.  With a point source, 24 inversions
per gauge configuration are needed: 12 (3 colors $\times$ 4 spins) for
the quark propagator and 12 for the generalized propagator.  With the
stochastic source discussed above, only two inversions are needed: one
for the quark propagator and another one for the generalized propagator.
For a comparison of stochastic versus point sources we refer to
Ref.~\cite{Baron:2007ti}.

The stochastic method described above can be adapted to work for other
mesons.  However, for moments of nucleon parton distribution functions
we found no improvement in the signal to noise ratio for a comparable 
computational effort.

For $\langle x\rangle^{\pi^{\pm}}_{u-d}$ it is sufficient to fix $t_{\rm
  s}-t_0 = T/2$, where $T$ is the temporal extent of the lattice.
As in the case of the nucleon, the value of $t_0$ is chosen randomly
on every gauge configuration in order to reduce autocorrelation.

\subsection{Ensuring ground state dominance}

To extract the nucleon matrix element from the ratio defined in
Eq.~(\ref{eq:ratio}) one needs to make sure that the contribution of
the terms due to the excited states in numerator and denominator of
Eq.~(\ref{eq:ratioexc}), the so called  contamination due to excited states,
is negligible. We will employ three methods to check for ground state
dominance, as described below.

In the first method, which we will refer to as the \textit{plateau
  method}, one probes the region for which $\Delta(t_s-t_{\rm ins})\gg
1$ and $\Delta(t_{\rm ins}-t_0)\gg 1$ such that excited state
contributions are much smaller than the contribution of the ground state. Within this time interval the ratio becomes time
independent and the time range where this happens is referred to as
the \textit{plateau region}. Fitting the ratio
\be
R(\Gamma^\lambda,t_s,t_{\rm ins})\xrightarrow[\Delta(t_{\rm
    ins}-t_0)\gg 1]{\Delta(t_s-t_{\rm ins})\gg 1}\Pi(\Gamma^\lambda)
\label{eq:plat}
\ee
over $t_{\rm ins}$ within this plateau region one obtains the plateau value, which is the desired
matrix element $\mathcal{M}$. To ensure excited state suppression one
repeats this procedure for multiple values of $t_s$, checking that the
plateau value does not change. However, the statistical errors grow
exponentially with $t_s$, which means that as the sink-source time
separation increases the signal is lost as compared to the statistical
noise making it difficult to detect any time-dependence.  Increasing
$t_s$ therefore requires a corresponding increase in statistics if
this check is to be useful~\cite{Dinter:2011sg}.

The second approach is to use the \textit{summation method} proposed
some time ago~\cite{Maiani:1987by} and recently applied to the study
of the nucleon axial charge~\cite{Capitani:2012gj}. One sums the ratio
over the time of the insertion,
\be
R^{\rm
  sum}(\Gamma^\lambda)=\sum_{t_{\rm
    ins}=t_0+\tau}^{t_s-\tau}R(\Gamma^\lambda,t_s,t_{\rm ins}),
\label{eq:sumratio}
\ee
with $\tau$ selected such that contact terms are not included,
i.e. $\tau=1$ for local operators and $\tau=2$ for derivative
operators. The sum over the excited state contributions given in
Eq.~(\ref{eq:ratioexc}) is a geometric series and can easily be summed
to yield
\be
R^{\rm sum}(\Gamma^\lambda) \propto \mathcal{C}'+(t_s-t_0)\mathcal{M}+O(e^{-\Delta (t_s-t_0)})
\label{eq:summeth}
\ee
with $\mathcal{C}'$ a constant independent of $t_s$. The advantage
over the plateau method is that excited state contamination is
suppressed by a larger factor [$\Delta (t_s-t_0)$ as opposed to
  $\Delta(t_s-t_{\rm ins})$ or $\Delta (t_{\rm ins}-t_0)$]. However,
the extraction of $\mathcal{M}$ requires a fit to two parameters,
resulting in general in larger statistical uncertainties.
Nevertheless, this method provides a good consistency check of our
results.

A third approach to extract the desired matrix element is to take into
account in the fit the contribution of the first excited state in
Eq.~(\ref{eq:ratioexc}). In this case, we simultaneously fit the two-
and three-point correlation functions obtained from the lattice
including the ground state and the first excited state
contributions. This is done by performing a combined fit to all  sink-source separations and to both
correlation functions with $t_{\rm ins}$ and $t_s$ as independent
variables. Like for the summation method, we exclude the contact terms, i.e. for $t_{\rm ins} \in [t_0+1, t_s-1]$ for the scalar, axial and tensor charges, and $t_{\rm ins} \in [t_0+2, t_s-2]$ for the momentum fraction, polarized moment and
 trasversity moment, which include a derivative. We will refer to this method as the \textit{two-state fit  method}.

In this work, we consider agreement among the above three methods
yielding the same value for $\mathcal{M}$ as our criterion that
excited states are sufficiently damped out.

If one has ground state dominance the nucleon matrix elements of the
scalar, axial and tensor local operators, at zero momentum transfer
and Euclidean time are related to the ratio as follows:

\beq
\Pi_S(\Gamma^4) &=& \frac{g_S}{2} \nonumber \\
\Pi^j_A(\Gamma_k) &=& -i\delta_{jk}\frac{g_A}{2}\nonumber \\
\Pi^{ij}_T(\Gamma_k) &=& \epsilon_{ijk}\frac{g_T}{2}.
\eeq
The corresponding expressions for the vector, axial and tensor
one-derivative operators are:
\beq
\Pi_V^{44}(\Gamma^4) &=& -\frac{3m_N}{4}\langle x \rangle_{u\pm d} \nonumber \\
\Pi_V^{kk}(\Gamma^4) &=&  \frac{m_N}{4}\langle x \rangle_{u\pm d} \nonumber \\
\Pi^{j4}_A(\Gamma_k) &=& -\frac{i}{2}\delta_{jk}m_N\langle x\rangle_{\Delta u\pm\Delta d}\nonumber \\
\hspace*{-0.4cm} \Pi^{\mu\nu\rho}_T(\Gamma_k) &=& i\epsilon_{\mu\nu\rho k}\frac{m_N}{8}(2\delta_{4\rho}-\delta_{4\mu}-\delta_{4\nu})\langle x\rangle_{\delta u\pm \delta d}.\nonumber \\
  & &
\eeq
Note that after symmetrization and subtraction of the trace as
indicated in Eq.~(\ref{eq:derivops}), only one of the two expressions
for $\langle x\rangle_{u\pm d}$ is independent.

For the case of the pion we only present $\langle x
\rangle^{\pi^{\pm}}_{u-d}$. We consider the largest sink-source
separation possible on each lattice, namely $t_s-t_0=T/2$. This is
possible for the pion since its two point function has constant 
signal to noise ratio independently of $t_s-t_0$. Therefore,
we extract the pion momentum fraction using the plateau method at this
single value of the sink-source separation:
\beq
R^{\pi}(t_s,t_{\rm ins})\xrightarrow[\Delta(t_{\rm ins}-t_0)\gg 1]{\Delta(t_s-t_{\rm ins})\gg 1}\Pi^\pi
\eeq
where we use $\Delta$ to generically denote the energy gap between the
energy of the first excited state and the ground state of the hadron
of interest in its rest frame. Given ground state dominance, the pion
momentum fraction is at zero momentum transfer and Eucledian time
obtained from the ratio via:
\beq
\Pi^\pi = \frac{m_\pi}{2}\langle x\rangle^{\pi^\pm}_{u-d}. 
\eeq

\section{Simulation details}
We use the (maximally) twisted mass fermion (TMF) formulation of
lattice QCD~\cite{Frezzotti:2000nk}, which is particularly suited for hadron structure
calculations since it provides automatic ${\cal O}(a)$ improvement
requiring no operator modification~\cite{Frezzotti:2003ni,Jansen:2005cg,Farchioni:2004us,Farchioni:2005bh}. Twisted mass ensembles with two degenerate flavors of
light sea quarks ($N_f$ = 2) 
as well as ensembles including the strange and charm sea quarks ($N_f=2+1+1$) are 
produced by the European Twisted Mass Collaboration (ETMC) and technical
details on the simulations can be found in Refs.~\cite{Boucaud:2007uk,Boucaud:2008xu,Baron:2009wt} and ~\cite{Baron:2010bv} respectively. This work focuses on the analysis
of gauge configurations produced using two degenerate flavors of
twisted mass light sea quarks ($N_f$ = 2) including a clover term. For the gauge action we use the Iwasaki action.  The
parameters of the four ensembles considered in this work are given in
Table~\ref{Table:params}. More details on
the choice of action and the simulations are given in
Refs.~\cite{Abdel-Rehim:2013yaa,Abdel-Rehim:2014nka,Abdel-Rehim:2015pwa}.

\begin{table}[h]
\begin{center}
\caption{Input parameters of our new lattice ensembles used in this
  work. For each ensemble we give the lattice size, the bare quark
  mass (a$\mu$) and the corresponding pion mass ($m_{\pi}$). These
  ensembles use TMF at one value of $\beta$ with a clover term with
  $c_{SW}=1.57551$. The lattice spacing given in the table is determined using the
  nucleon mass as explained in the text.}
\label{Table:params}
\begin{tabular}{c|llll}
\hline\hline
\multicolumn{5}{c}{ $\beta=2.1$, $a=0.093(1)$~fm, ${r_0/a}=5.32(5)$}\\\hline
$24^3\times 48$, $L=2.23$~fm  &$a\mu$ & 0.006        & 0.003    & \\
                              &$m_\pi$ (GeV)   & 0.338(9)     & 0.244(8) & \\
$32^3\times 64$, $L=2.97$~fm  &$a\mu$ & 0.006        &          & \\
                              &$m_\pi$ (GeV)   & 0.335(9)     &          & \\
$48^3\times 96$, $L=4.46$~fm  &$a\mu$ & 0.0009       &          & \\
                              &$m_\pi$ (GeV)   & 0.1312(13)     &          & \\
\hline\hline
\end{tabular}
\end{center}
\vspace*{-.0cm}
\end{table}
\begin{table}[h]
\begin{center}
\caption{  The lattice spacing 
and value of the  scale parameter, $r_0$~\cite{Sommer:1993ce}, 
for the $N_f=2$ and $N_f=2+1+1$ TMF ensembles
as well as for the new  $N_f=2$ TMF ensembles with the clover term,  determined using
  the nucleon mass as explained in the text. The first  error is statistical. The second error is the difference in the value when discarding ensembles with pion mass larger than 300~MeV.}
\label{tab:as}
\begin{tabular}{c|ccc}
\hline\hline
\multicolumn{4}{c}{ $N_f=2$}\\\hline
$\beta$  & 3.9 & 4.05 & 4.2\\
$a$~(fm)   & 0.088(2)(2) & 0.071(2)(1) & 0.056(2)(1)\\
$r_0$~(fm)  &0.458(10)(1)&0.467(12)(7)&0.465(13)(7)\\
\hline
\multicolumn{4}{c}{ $N_f=2+1+1$}\\\hline
$\beta$ & 1.90 & 1.95 & 2.1 \\
$a$~(fm) &  0.094(1)(2) & 0.082(1)(2) & 0.065(1)(1) \\ 
$r_0$~(fm) & 0.501(7)(9) & 0.492(6)(3) & 0.499(6)(5)\\\hline
\hline
\multicolumn{4}{c}{ $N_f=2$ with $c_{SW}=1.57551$}\\\hline
$\beta$ & 2.1 & &  \\
$a$~(fm) & 0.093(1)(0)  & & \\
$r_0$~(fm) & 0.493(5)(0) \\
\hline
\end{tabular}
\end{center}
\end{table}

\begin{figure}
\includegraphics[width=0.9\linewidth]{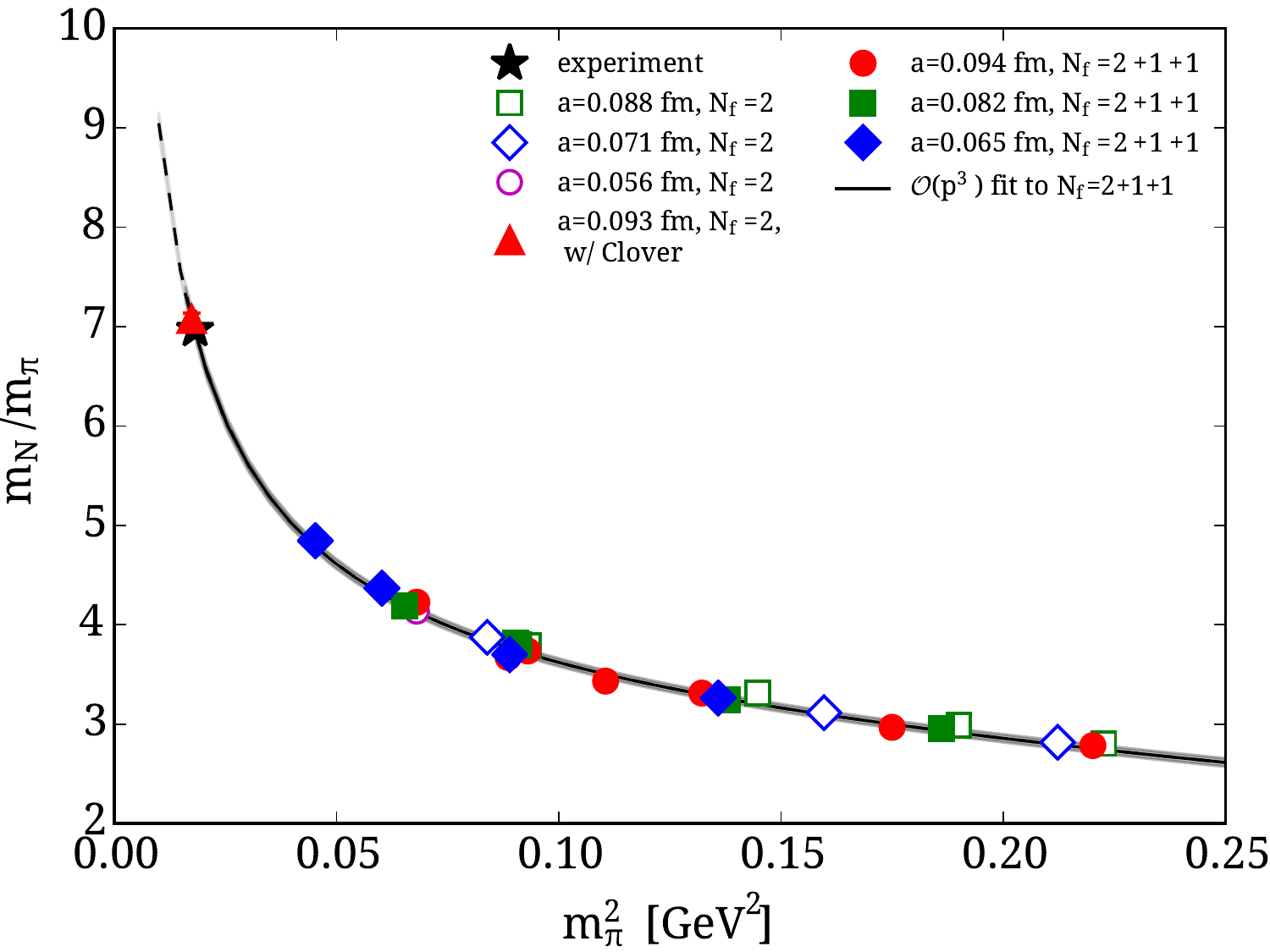}
\caption{The ratio of the nucleon mass to the pion mass as a function
  of the pion mass squared. For determining the pion mass squared the
  scale is set using the nucleon mass at the physical point as
  described in the text. The fit only used the $N_f=2+1+1$ ensembles
  without a clover term (filled circles, diamonds and squares). The plot also shows  the $N_f=2$  TMF results
  (open circles, diamonds and squares) and
  the $N_f=2$ ensemble with a clover term at the physical point
  (filled triangle). For the latter action ($N_f=2$ with a clover term) we restrict our analysis of nucleon observables to the ensemble simulated at a physical value of the 
pion mass only.}
\label{fig:mN}
\end{figure}

For the nucleon structure observables, we analyze the ensemble with
$a\mu=0.0009$.  We will refer to this ensemble as the physical
ensemble and speak in what follows of the {\em physical point}.  For
the case of the pion momentum fraction, we use all four ensembles of
TMF with a clover term.

Although the observables of interest in this work are dimensionless
and do not depend on the lattice spacing, it is useful to study their
dependence on the pion mass, which is a dimensionfull quantity. 
In Ref.~\cite{Abdel-Rehim:2015pwa}, the lattice spacing for the new $N_f=2$ ensembles
 with the clover term was determined using gluonic quantities as well as 
the pion and kaon decay constants. Another determination of the lattice spacing
mentioned in Ref.~\cite{Abdel-Rehim:2015pwa} is via the nucleon mass.

The  physical value of the nucleon mass  was used as an input for
the determination of the lattice spacings in our previous analysis of the $N_f=2$~\cite{Alexandrou:2012xk,Alexandrou:2010hf} and $N_f=2+1+1$~\cite{Alexandrou:2014sha} twisted mass ensembles. Each set of $N_f=2+1+1$ and $N_f=2$ ensembles 
involved 
 three values of the lattice spacing.  Since those simulations 
involved larger than physical light quark masses a chiral extrapolation was needed.
   We used the lowest order heavy
baryon chiral perturbation theory expression, given by~\cite{Gasser:1987rb}
\be am_N = am^0_N-4(c_1/a)(am_\pi)^2-\frac{3g^2_A}{16\pi
  (af_\pi)^2}(am_\pi)^3\, ,
\label{chiral}
\ee 
which is well-established within baryon chiral perturbation theory. $m_N^0$ is the value of the nucleon mass in the chiral limit and $-4c_1$ gives  the $\sigma$-term written in units of the lattice spacing.
The fit was constrained to reproduce the physical nucleon mass, by fixing the value of $c_1$.
 Including an $a^2$-term in
Eq.~(\ref{chiral}) had a negligible effect on the fit showing that
indeed cut-off effects are small~\cite{Alexandrou:2014sha} for lattice spacings smaller than 0.1~fm.
This justified the utilization of  continuum chiral perturbation theory
to  determine the three lattice spacings for each $N_f=2$ or  $N_f=2+1+1$ by
simultaneously fitting  each set of 17 $N_f=2+1+1$ or 11 $N_f=2$ ensembles. 
The values of the nucleon mass used 
are taken from Ref.~\cite{Alexandrou:2014sha} for the $N_f=2+1+1$ ensembles and from Ref.~\cite{Alexandrou:2010hf} for the $N_f=2$  ensembles.
These values of the lattice spacings are used to obtain the pion mass
for these ensembles.

For the physical ensemble using the values 
\be
am_\pi = 0.06196(9) \quad am_N = 0.440(4),
\label{eq:mpi mn}
\ee
and assuming that we are exactly at  the physical point we find
$a=0.0925(8)$~fm where the average nucleon mass $m_N=0.939$~GeV is used as an input. 
With this lattice spacing we find $m_\pi=0.1323(12)$~MeV, where the largest part of the error comes from the error on the lattice spacing. This is about 5\% less than the average physical pion mass. 
Using the values of Eq.~(\ref{eq:mpi mn}) we find for the ratio
$m_N/m_{\pi^\pm}=7.10(6)$ compared to the physical value of
$0.939/0.138=6.8$, which  again differs by less than 5\% from the physical value. 
In order to check what the effect of  a possible small mismatch in the pion mass 
would be on the lattice spacing, we use the fit extracted
from the 
 $N_f=2+1+1$ ensembles to interpolate to the physical value of pion mass.
This is done by  making a combined fit  of the 17 $N_f=2+1+1$ ensembles with their three lattice spacings and the lattice spacing for the physical ensemble as well as  $m^0_N$ as fit parameters. The fit yields $\chi^2/{\rm d.o.f}=1.6$ for ${\rm d.o.f}=12$, which is a reasonable value.
 We find a value of $a=0.093(1)$~fm
for the physical ensemble, consistent with the determination using Eq.~(\ref{eq:mpi mn}), while the lattice spacings for the
$N_f=2+1+1$ remain unchanged compared to the values obtained when the
physical ensemble was not included. 
Using $a=0.093(1)$ we find $m_{\pi^{\pm}}=0.1312(13)$~GeV, which is consistent with the value extracted from Eq.~\ref{eq:mpi mn}.
Excluding from the fit pion masses larger than 300~MeV  yields consistent results for the lattice spacings of the $N_f=2+1+1$ ensembles while it does not change the value of the lattice spacing at the physical point.  We note that if we fit using the  $N_f=2$ ensembles~\cite{Alexandrou:2010hf,Alexandrou:2012xk} instead of the $N_f=2+1+1$ ensembles 
the value of $a=0.093(1)$~fm  is unchanged.
This indicates that the mild interpolation is very robust. 
In
Fig.~\ref{fig:mN} we show the ratio of the nucleon to pion mass
$m_N/m_{\pi^\pm}$, which is a dimensionless observable  determined
purely from lattice quantities.
We note that the values
of the lattice spacings affect only the determination of the pion mass plotted as the x-axis.
  The curve shown in Fig.~\ref{fig:mN}  is the fit to the ratio performed on the  17 $N_f=2+1+1$ ensembles alone.
  The resulting chiral fit using $m_\pi<500$~MeV yields $\chi^2/{\rm d.o.f}=1.4$ and describes very well the
data.  In the figure we also include the values for the ratio for the  $N_f=2$ ensembles, which also fall on the
same curve. 
This can be taken as an
indication that indeed strange and charm sea quark effects are small for the
nucleon sector.  
The consistency of our new result is demonstrated
in Fig.~\ref{fig:mN} by the fact that the ratio $m_N/m_{\pi^\pm}$ for
our physical ensemble falls on the curve determined from fitting the
$N_f=2+1+1$ alone. Fig.~\ref{fig:mN} provides  a nice demonstration of the negligible effect of   lattice artifacts
on the $m_N/m_{\pi^\pm}$ ratio.

We note that the value of the lattice spacing determined from the nucleon mass analysis
is fully consistent with the one determined from  gluonic quantities such as the one  related to the static quark-antiquark potential,
$r_0$, and the ones related to the action density renormalised through the gradient flow. It is, however,  larger by about 1\% as compared to  that extracted using  $f_\pi$~\cite{Abdel-Rehim:2015pwa}.  This was also observed  in our analysis of $N_f=2$
and  $N_f=2+1+1$ TMF ensembles~\cite{Carrasco:2014cwa}.
  In
Table~\ref{tab:as} we collect  the lattice spacings for all the TMF ensembles
determined using the nucleon mass and Eq.~(\ref{chiral}).  We take as a systematic error due to 
the chiral extrapolation  the shift in  the mean value
when discarding  ensembles with pion mass greater  than 300~MeV. 
For completeness we also give the values of $r_0$ determined from the nucleon mass in the same way as the lattice  spacings, although they are not needed in this work.
In what follows we use, for the physical ensemble,  the value $a=0.093(1)$. The lattice spacings given
in Table~\ref{tab:as} are used  to convert the pion mass to physical units. No other physical quantity presented in this work is affected by
the value of the lattice spacings.

\section{Results}

For the nucleon observables we analyze 96 gauge field configurations with 16 randomly chosen
positions for each configuration yielding a total of 1536 measurements.
For the nucleon observables,  we use three sink-source separations for
both the plateau and the summation methods, namely $t_s/a=$10, 12, and
14, corresponding to approximately 0.9~fm, 1.1~fm, and 1.3~fm. For all
separations we have 1536 measurements, by computing the required two-
and three-point correlation functions. First results on these
quantities were presented in
Refs.~\cite{Alexandrou:2014wca,Alexandrou:2013wka}. For the pion we use
the largest possible time separation namely $T/2$.

\vspace*{0.3cm}

\subsection{Renormalization}
\begin{figure}[!h]
  \includegraphics[width=\linewidth]{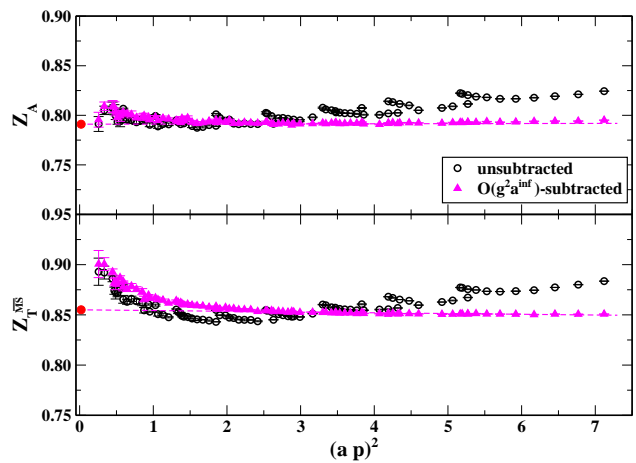}
  \caption{Results for the axial (upper) and tensor(lower) renormalization 
    functions as a function of the momentum square in lattice units. The open (black) circles are the unsubtracted results while the filled triangles (magenta) show the data after ${\cal O}(g^2\,a^\infty)$-terms are subtracted. The filled (red) circle at $(ap)^2=0$ is the value extracted by fitting to the plateau region [2-7] the subtracted data.}
\label{fig:remormalization}
\end{figure}

We determine the renormalization functions for the lattice matrix
elements non-perturbatively, in the RI$'$-MOM scheme employing a
momentum source~\cite{Gockeler:1998ye}. For the computation of the
renormalization functions of the $N_f=2+1+1$ ensembles we employed
$N_f=4$ simulations for at least three different values of the pion
mass taking the chiral limit.  A similar analysis was
performed for the $N_f=2$ TMF ensembles as well as for our new
$N_f=2$ TMF ensembles that include the clover term
using the ensembles with $a\mu=0.006, 0.003$ and 0.0009, the latter being at the physical pion point. 
 In
Refs.~\cite{Constantinou:2009tr,Alexandrou:2010me} we carried out a
perturbative subtraction of ${\cal O}(a^2)$ terms that subtracts the
leading cut-off effects yielding only a very weak dependence of the
renormalization factors on $(ap)^2$ for which the $(ap)^2\rightarrow
0$ limit can be reliably taken. In this work, we reduce even further
the ${\cal O}(a^2)$ contributions by subtracting lattice artifacts
computed perturbatively to one-loop and to all orders in the lattice
spacing, ${\cal O}(g^2\,a^\infty)$, so that we eliminate a large part
of the cut-off effects. In fig.~\ref{fig:remormalization}
we show the results on the axial and tensor renormalization functions after subtraction. As can be seen, lattice artifacts are practially removed 
allowing a robust extrapolation to $(ap)^2=0$. Due to the good quality of the plateaus after the subtraction of the ${\cal O}(g^2\,a^\infty)$ any choice for the fit within the non-perturbative region $(a p)^2 \in (2-7)$ yields consistent results.
Details on this computation can be found in
Ref.~\cite{Alexandrou:2015sea}. 

Our previous
chiral extrapolations have shown that for all renormalization
functions except $Z_P$ the pion mass dependence is very weak. For the
physical ensemble we compute  $Z_P$ for three pion masses corresponding to
$a\mu=0.006\,,0.003$ and 0.0009 and performed the pole subtraction.
Our value is given in Table~\ref{tab1}~\cite{Alexandrou:2015sea}.  The
scheme and scale dependent renormalization functions are converted in
the $\overline{\rm MS}$-scheme at a scale of 2~GeV, using the
intermediate RGI scheme. 

We collect the values of all relevant renormalization functions in
Table~\ref{tab1}, converting the scale dependent renormalization
function in the $\overline{\rm MS}$ scheme at a scale $ \mu=2$~GeV,
which is applicable to all except $Z_A$. The systematic error
is computed by varying the interval for the continuum extrapolation $(ap)^2 \rightarrow 0$. The values  of $Z_P$
 for the $N_f=2+1+1$ ensembles
 are taken from
Refs.~\cite{Carrasco:2014cwa,Constantinou:2010gr} where the pole subtraction
was performed, while
 for the new ensembles with the clover term
we use the result of this paper.  The values
given in Table~\ref{tab1} are used to renormalize the lattice matrix
elements studied in this work. More details are reported in Ref.~\cite{Alexandrou:2015sea}.
For the $N_f=2$ TMF ensembles without the clover term we do not calculate the scalar charge and transversity and therefore
the renormalization functions are not given.

\begin{widetext}
\begin{center}
\begin{table}[!h]
\caption{Renormalization functions for the ensembles used in this work. They are given in the twisted basis.  They are the same in the physical basis except for $Z_P$, which renormalizes the scalar operator in the physical basis. The renormalization functions for the local axial-vector, scalar and tensor operators are given in columns two, three and four, respectively. The three last columns give the renormalization functions for the derivative vector, axial vector and tensoroperators. The first error is statistical and the second error systematic. }
\label{tab1}
\begin{tabular}{ccccccc}
\hline
\hline
$\beta$	&$Z_{\rm A}$	&$Z^{\overline{\rm MS}}_{\rm P}$  &$Z^{\overline{\rm MS}}_{\rm T}$ &$Z^{\overline{\rm MS}}_{\rm DV}$	&$Z^{\overline{\rm MS}}_{\rm DA}$	&$Z^{\overline{\rm MS}}_{\rm DT}$ \\
\hline
\hline
	&	&	&$N_f{=}2$	&	&    & \\
\hline
\hline
3.90	&0.769(2)(1)	&$\,$	&0.758(2)(4)	&1.028(2)(6)	&1.102(5)(7)	& \\
4.05	&0.787(1)(1)	&$\,$	&0.796(1)(3)	&1.080(2)(11)	&1.161(4)(13)	& \\
4.20	&0.791(1)(1)	&$\,$	&0.814(1)(3)	&1.087(3)(12)	&1.164(3)(6)	& \\
\hline
\hline
	&	&	&$N_f{=}4$	&	&  &  \\
\hline
\hline
1.90	&0.7474(6)(4) 	&0.529(7)(45) 	&0.7154(6)(6)	&1.0268(26)(103)	&1.1170(54)(223)	&1.0965(90)(278) \\
1.95	&0.7556(5)(85)	&0.509(4)(37)	&0.7483(6)(94)	&1.0624(108)(33)	&1.1555(36)(289)	&1.1727(121)(73) \\
2.10	&0.7744(7)(31)	&0.516(2)(29)	&0.7875(9)(15)	&1.0991(29)(55)	        &1.1819(47)(147)	&1.1822(59)(118) \\
\hline
\hline
	&	&	&$N_f{=}2+c_{\rm SW}$	&	&    & \\
\hline
\hline
2.10	&0.7910(4)(5)	&0.5012(75)(258)   &0.8551(2)(15)	&1.1251(27)(17)	        &1.1357(20)(205)	&1.1472(121)(48) \\
\hline
\hline
\end{tabular}
\end{table}
\end{center}
\end{widetext}


The renormalization functions are given for the twisted basis. Going
from the twisted to the physical basis affects only the
renormalization function for the scalar charge, which, in the twisted
basis, is renormalized with $Z_P$.  Furthermore, since disconnected
contributions are neglected, the isovector and isoscalar are
renormalized using the same renormalization functions.  All our results on the scalar and
tensor charges and on the moments of PDFs are given in the
${\overline{\rm MS}}$ scheme at an energy scale of $2$~GeV.

\subsection{Nucleon  scalar, axial and tensor charges}

\begin{figure}[!h]
  \includegraphics[width=\linewidth]{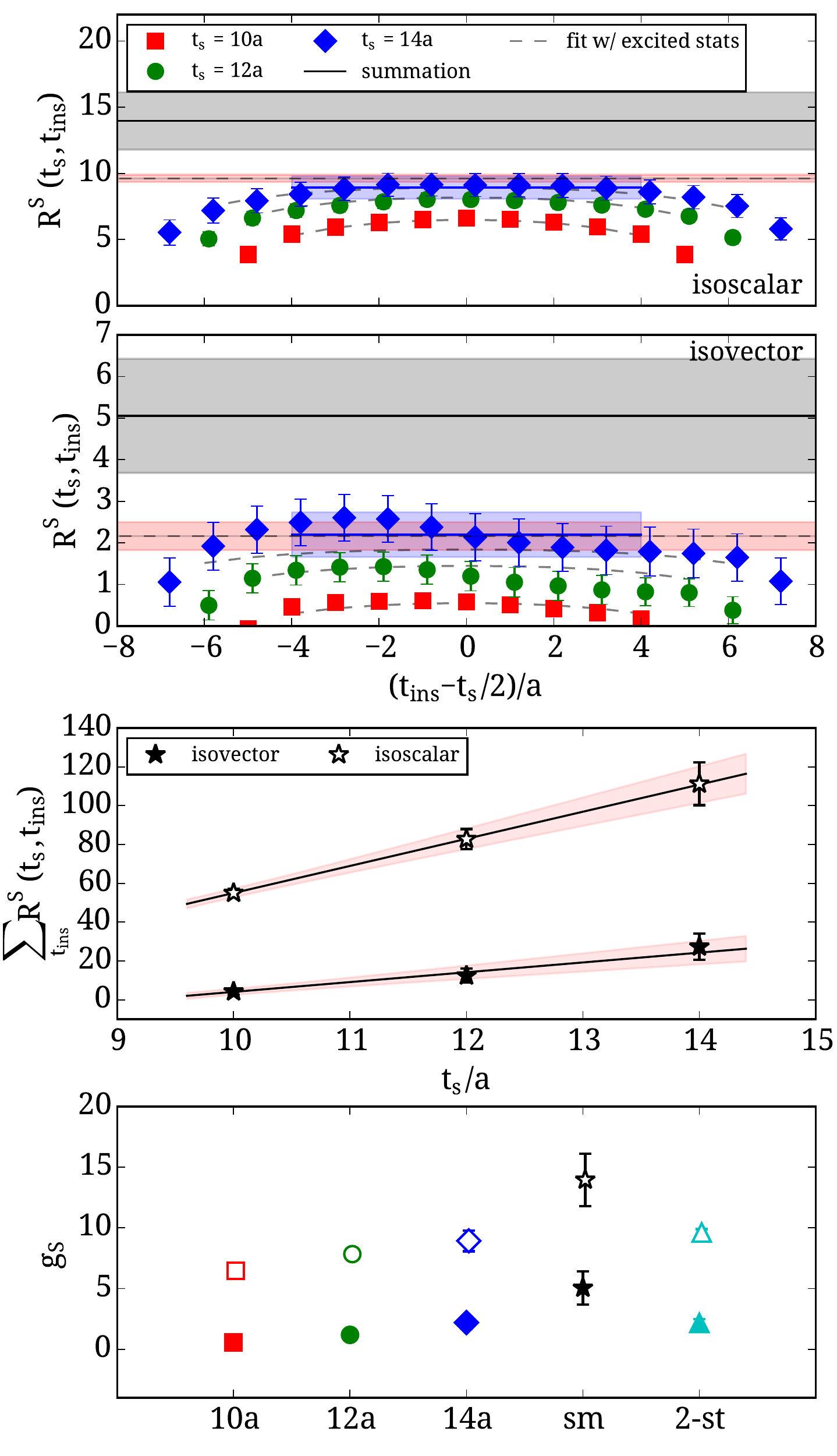}
  \caption{Results for the isovector and isoscalar nucleon scalar
    charge: Upper two panels is the ratio from which $g_S$
    is extracted as a function of $t_{\rm ins}-t_s/2$ for the
    isoscalar (upper) and the isovector (lower). The blue
    bands spanning from $(t_{\rm ins}-t_s/2)/a$=-4 to 4 are fits to
    the ratio for $t_s/a=14$. The dashed lines show the result
    of the two-state fit method. The dashed (solid) line spanning the entire
    x-range show the value obtained via the two-state (summation)  method,
     with the band indicating the
    corresponding statistical error. In the third panel, the summed
    ratio is shown for the isovector (filled symbols) and isoscalar
    (open symbols) case. The line shows the result of a linear fit,
    while the bands show the statistical error based on the jackknife
    error of the fitted parameters. In the bottom panel, we show the
    result for $g_S$ when using the plateau method with $t_s/a=10, 12$ and $14 $ (squares, circles, and
    rhombuses respectively), as well as when using the summation
    method denoted by ``sm'' (asterisks) and the two-state fit ``2-st.''
    (triangles).}
  \label{fig:gS}
\end{figure}

In what follows we will use the same format to present our results for
a given observable in four plots unless otherwise mentioned. Our
presentation is illustrated in Fig.~\ref{fig:gS}. In the two upper
panels we present the ratio of Eq.~(\ref{eq:ratio}), as a function of
the insertion time ($t_{\rm ins}$), shifted by half the sink-source
separation, i.e. $t_{\rm ins}-t_s/2$. This way, the midpoint time of
the ratio coincides for all sink-source separations at $t_{\rm
  ins}-t_s/2=0$. In what follows all times are measured relative to
$t_0$ and thus we drop the reference to $t_0$. In the third panel we
show the summed ratio as a function of the sink time, as obtained by
Eq.~(\ref{eq:sumratio}) and in the bottom panel we compare results
from the summation method and from the two-state fit method with those
obtained by the plateau method.

\begin{figure}[!h]
\includegraphics[width=\linewidth]{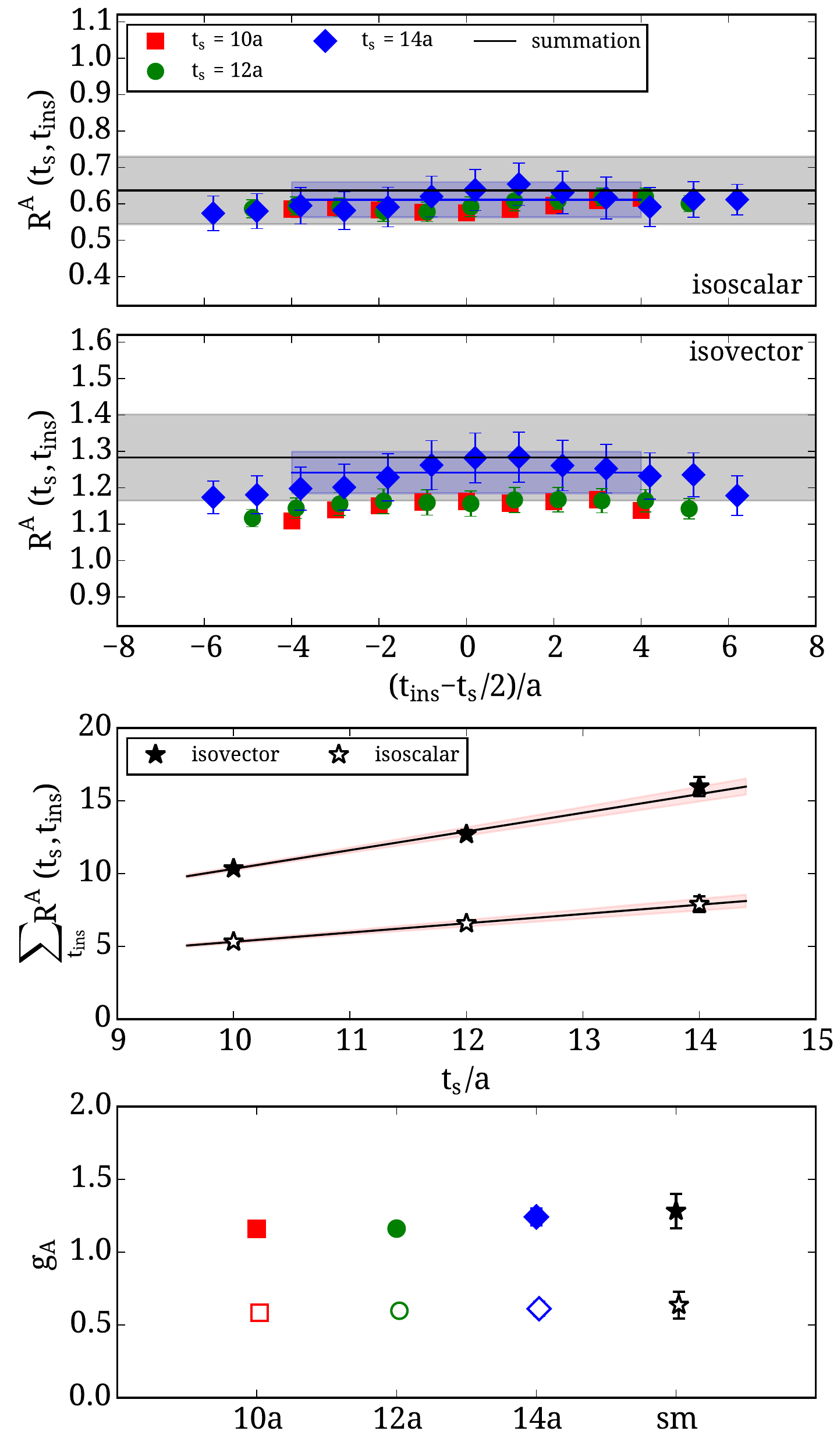}
\caption{Results for the axial charge. The notation is the same as
  that in Fig.~\ref{fig:gS}.}
\label{fig:gA}
\end{figure}

\begin{figure}[!h]
\includegraphics[width=\linewidth]{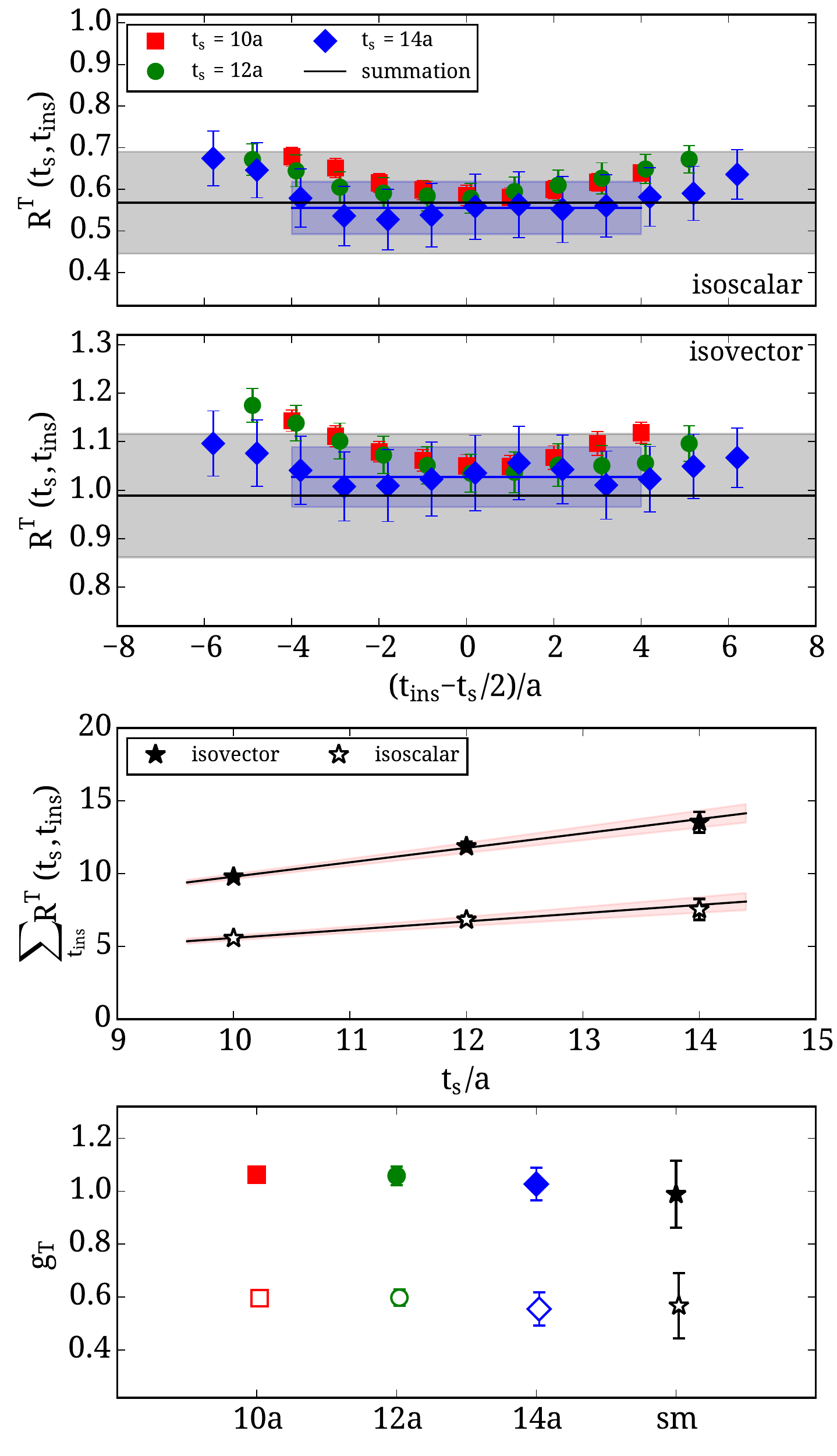}
\caption{Results for the tensor charge. The notation is the same as
  that in Fig.~\ref{fig:gS}.}
\label{fig:gT}
\end{figure}

Let us first discuss the results for the scalar charge shown in
Fig.~\ref{fig:gS}. In the two upper panels we show the ratio for the
isoscalar and isovector scalar charges, for the three sink-source
separations considered. As explained in the previous section, when the
time separations $\Delta t_{\rm ins}\gg 1$ and $\Delta(t_s-t_{\rm
  ins})\gg 1$, the ratio becomes time-independent. Fitting in the
plateau region to a constant value, which we refer to as the plateau
value, we obtain $g_S$, as in Eq.~(\ref{eq:plat}). This is shown by
the blue band in Fig.~\ref{fig:gS} for $t_s=14a$.  One observes an
increasing trend for the plateau value and a clear curvature,
especially for the isoscalar, indicating dependence on excited
states. Carrying out a two-state fit yields the dashed lines. As in the case of the summation method, the contact points 
$t_{\rm ins}=t_0$ and $t_{\rm ins}=t_s$ are omitted. 
The value for
$g_S$ obtained by the two-state fit is given by the dashed line that
spans the entire x-range of the figure, with the red band indicating
the statistical error, while the result of the summation method is
shown with the solid line and  gray band indicating the
error. As can be seen, within errors the two-state fit is consistent
with the plateau fit, but not with the value from the summation method,
which, however, carries a very large error indicating the need to
increase statistics in order to have a better assessment of the result
of this method. 
In the third panel of Fig.~\ref{fig:gS} we show the
summed ratio for the scalar charge as a function of the sink time, as
obtained by Eq.~(\ref{eq:sumratio}), for the isovector and the
isoscalar cases. Fitting to a linear dependence with respect to $t_s$,
one extracts the desired matrix element from the slope
(Eq.~(\ref{eq:summeth})), which is the result shown by the gray band in the
two upper graphs of Fig.~\ref{fig:gS}. The width of the
bands is obtained by a jack-knife re-sampling of the summed ratio to
obtain jack-knife errors for the slope, $\mathcal{M}$, and
intersection $\mathcal{C}'$ of Eq.~(\ref{eq:summeth}). The values for
$g_S$ from the summation method and those obtained by the plateau
method and the two-state fit method are shown in the bottom panel of
the figure.  One clearly observes the increasing trend of the plateau
values with increasing $t_s/a$ as well as the larger values of the
summation method shown by the asterisks. This study shows that both
larger sink-source time separations as well as larger statistics are
needed in order to obtain a meaningful convergence of all methods. 
This corroborates our findings of our high statistics analysis
of the $N_f=2+1+1$ TMF ensemble with pion mass 373~MeV, referred to as  B55.32
 ensemble,  where we showed that  $t_s\sim 1.5$~fm is needed~\cite{Alexandrou:2014wca}. Our current
statistics do not allow to use such a large sink-source separation for the physical ensemble.  We note that as a check of the robustness of the two-state fit, we omit more points besides the time slice of the source and the sink. In the case of the scalar charge, taking the fit range $t_{\rm ins} \in [t_0+2, t_s-2]$ and $t_{\rm ins} \in [t_0+3, t_s-3]$ we obtain: 2.18(34) and 2.22(33) respectively for the isovector case and 9.68(26) and 9.75(24) for the isoscalar, which are consistent with  2.16(34) and 9.62(27) extracted when just omitting the source and the sink. Thus for the scalar charge, the fluctuation of the central value when changing the fit-range is within the statistical error and of the order of 2\%. 

In Fig.~\ref{fig:gA} we show results for the axial charge following
the same notation as  that in Fig.~\ref{fig:gS}. For the axial charge, one
observes a milder dependence on $t_s$ showing that excited states
contributions are suppressed for this observable. Because of this
weaker dependence a two-state fit does not yield a meaningful result
for these values of $t_s/a$, at least within the statistical accuracy
of 1536 measurements. We therefore only show results for the plateau
and summation methods. The values from the plateau method do not vary
as a function of $t_s$ and are in agreement with the value extracted
from the summation method, within the large statistical uncertainties
of the latter.

\begin{figure}[!h]
\includegraphics[width=\linewidth]{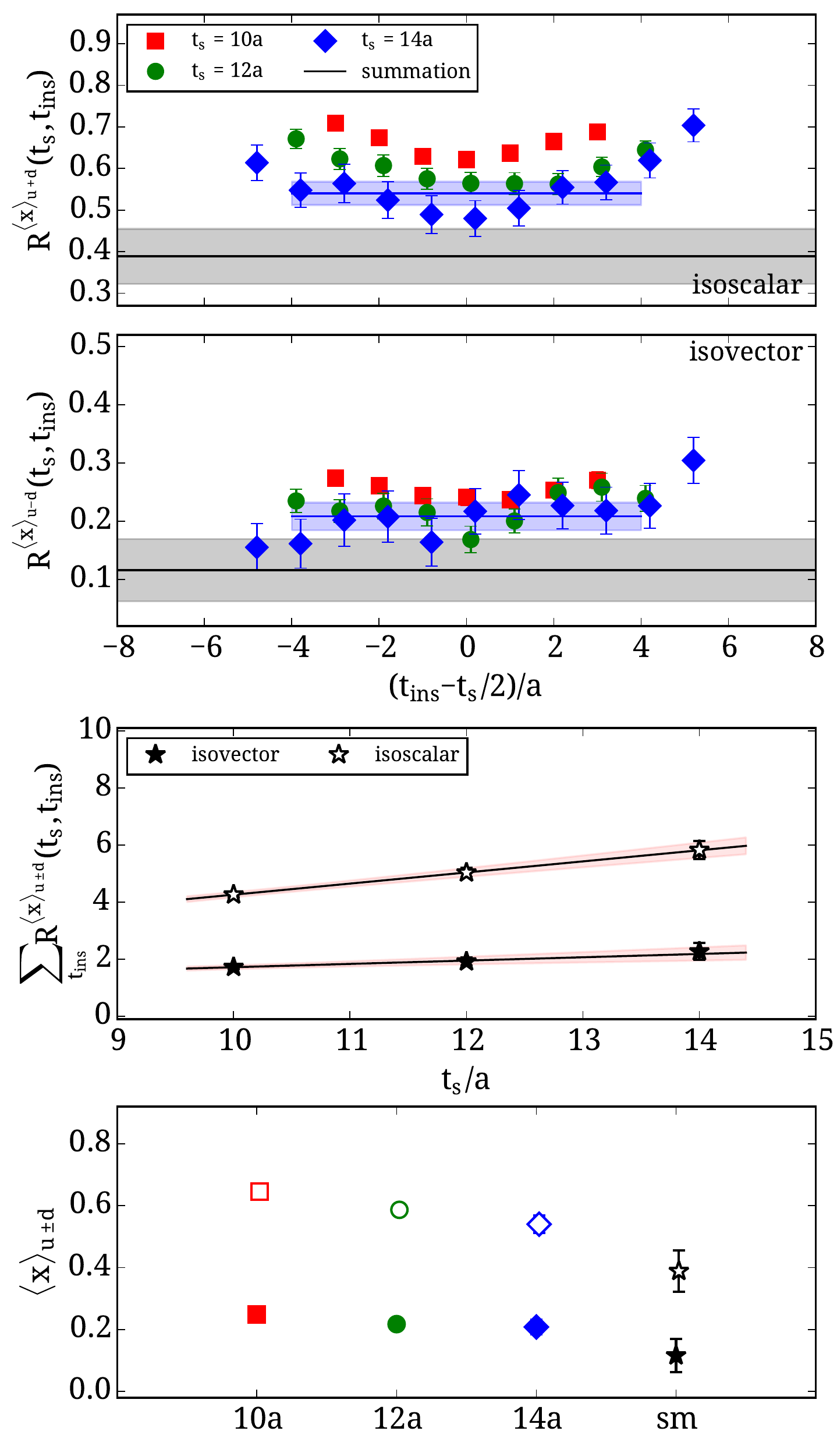}
\caption{Results for the momentum fraction. The notation is the same as that  in Fig.~\ref{fig:gS}.}
\label{fig:xq}
\end{figure}

\begin{figure}[!h]
\includegraphics[width=\linewidth]{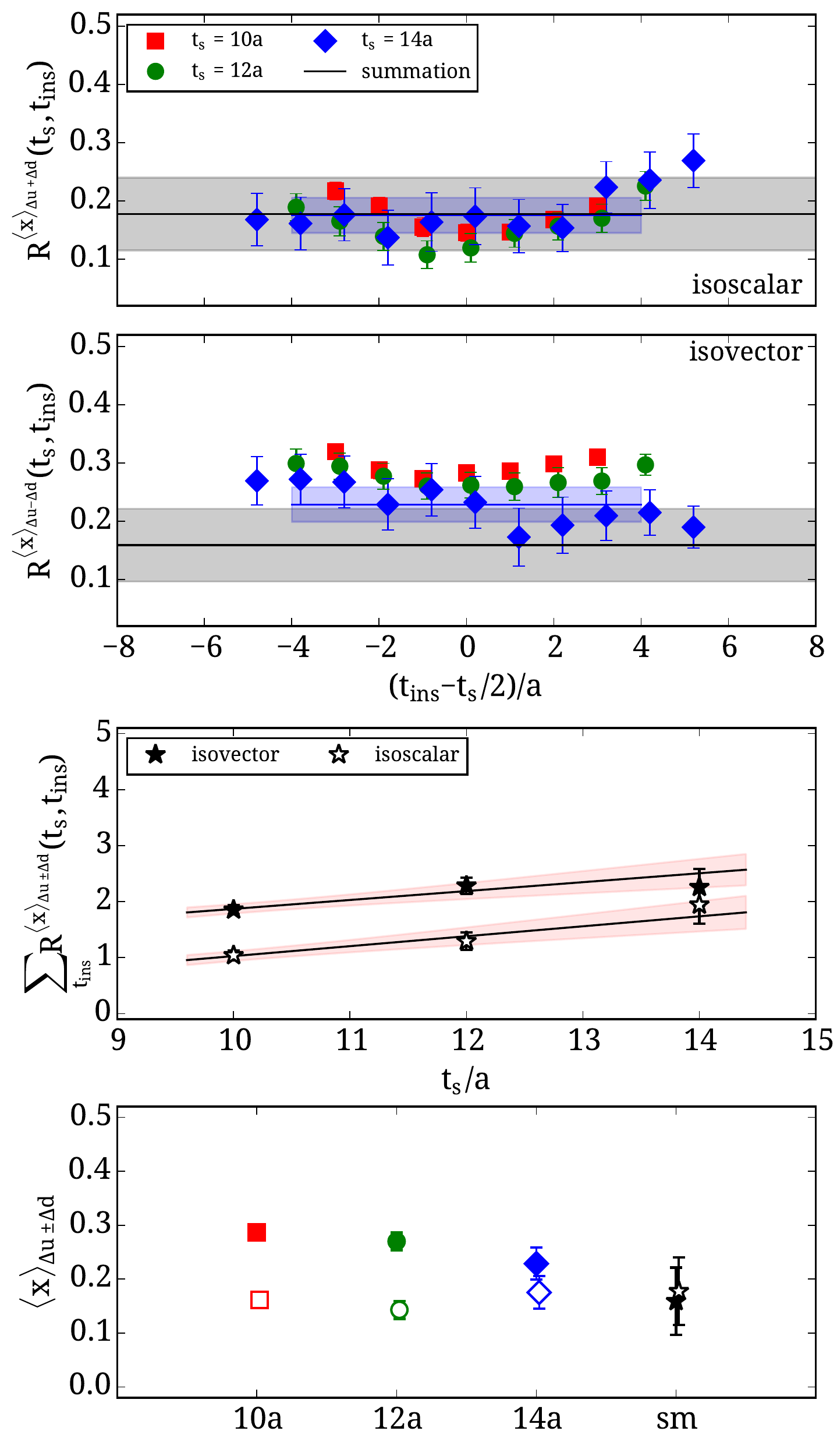}
\caption{Results for the nucleon helicity moment. The notation is the same as that  in Fig.~\ref{fig:gS}.}
\label{fig:xDq}
\end{figure}

\begin{figure}[!h]
\includegraphics[width=\linewidth]{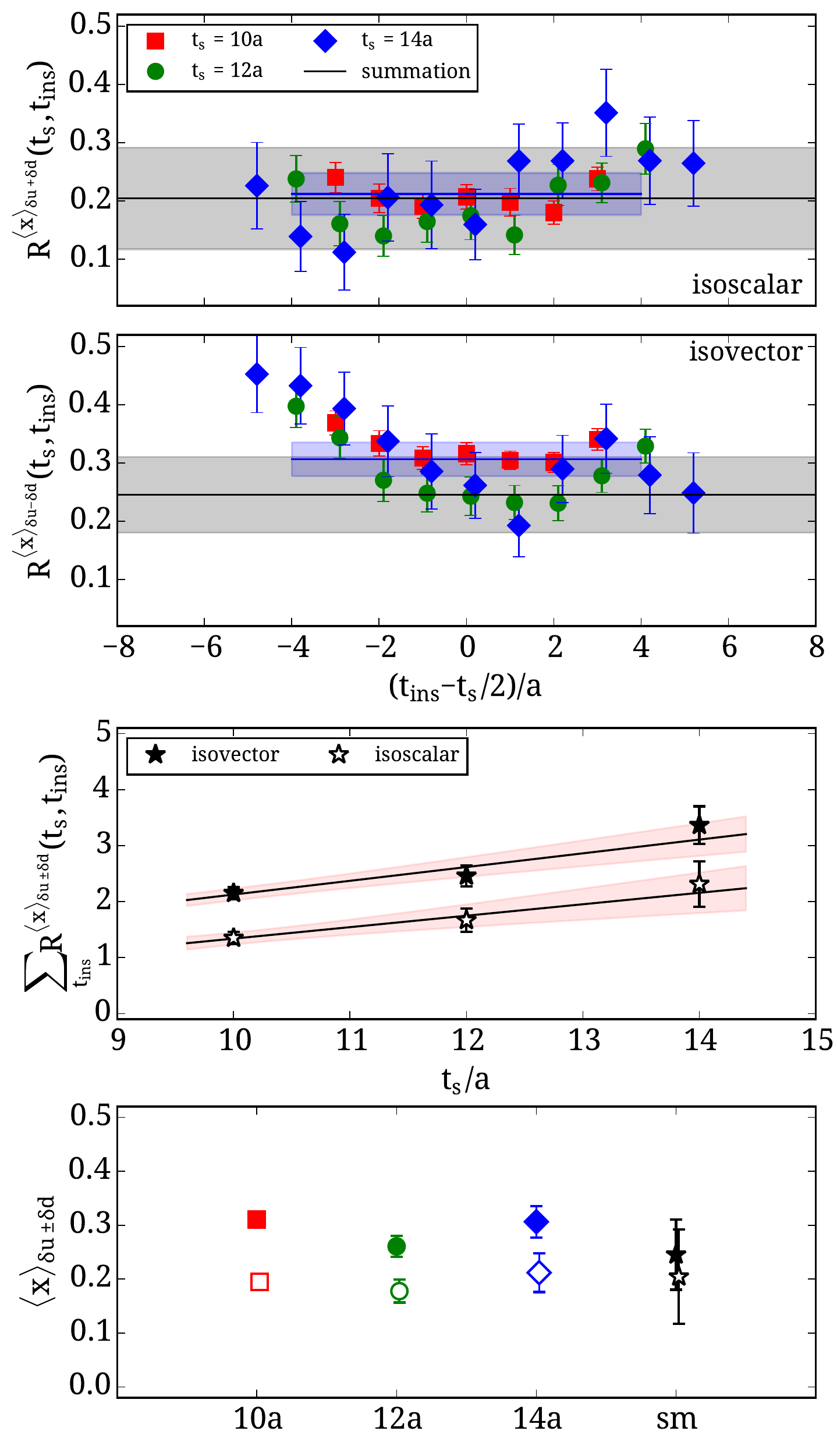}
\caption{Results for the transversity moment. The notation is the same as that  in Fig.~\ref{fig:gS}.}
\label{fig:xtq}
\end{figure}

Our results for the tensor charge ($g_T$) are shown in
Fig.~\ref{fig:gT}.  The dependence of $g_T$ on
$t_s$ is similar to that observed in the case of $g_A$ and thus a
two-state fit fails to accurately resolve the excited states. Thus we
only compare the results extracted using the summation and plateau
methods in the bottom panel. As can be seen, $g_T$ exhibits no
dependence on the sink-source separation within the current
statistical uncertainties, evident by the same values extracted by
fitting to the plateau for the three different sink-source
separations. The value extracted from the summation method is in
agreement but carries a much larger error and thus  does not  provide a
stringent check.

\subsection{ Nucleon momentum fraction, helicity  and transversity moments}

The momentum fraction is shown in Fig.~\ref{fig:xq} for the connected
isoscalar $\langle x\rangle_{u+d}$ and isovector $\langle
x\rangle_{u-d}$ combinations. Both isoscalar and isovector channels
exhibit excited state contributions, especially for the isoscalar
channel.  A two-state fit is performed  using $t_{\rm ins} \in [t_0+2, t_s-2]$, which however yields too
large errors to include in the plots (see
Table~\ref{table:charges-two-state}). Using $t_{\rm ins} \in [t_0+3, t_s-3]$, we find  a  value of 0.48(19) for the isoscalar, which
is consistent with the value given in Table~\ref{table:charges-two-state}. For the isovector both fit ranges yield consistent results
albeit with a large error that does not allow to access the sensitivity on the fit range.
Furthermore,  as can be seen in the lowest panel of panel of Fig.~\ref{fig:xq},  a
decreasing trend is observed as the sink-source separation is
increased from 10$a$ to 14$a$, showing that elimination of excited state effects is
responsible for reducing the value of this matrix element. The
summation method yields a value that is even lower but with a large
statistical uncertainty. Like for the case of the scalar charge a
larger value of $t_s/a$ and increased statistics will be needed to
reach consistency among the various methods with meaningful errors.

The helicity moment $\langle x \rangle_{\Delta u\pm \Delta d}$ and
transversity moment $\langle x \rangle_{\delta u\pm \delta d}$ are
shown in Figs.~\ref{fig:xDq} and ~\ref{fig:xtq}. For both isoscalar
observables, a milder dependence on the sink-source separation is
observed. This is also true for the isovector transversity moment. On
the other hand, the isovector helicity moment shows a decreasing
trend similar to that observed in the case of the isovector momentum
fraction. The value obtained using the summation method is consistent
in all cases with the plateau value when $t_s/a=14$, albeit with a
large statistical uncertainty.

\begin{table}
  \centering
  \caption{Results for the nucleon scalar charge and momentum fraction when employing the two-state fit.}
  \label{table:charges-two-state}
  \begin{tabular}{ccccc}
    \hline\hline
    Moment                      & isovector & isoscalar \\
    \hline
    $g_S$                       & 2.16(34)  & 9.62(27)  \\ 
    $\langle x\rangle_q$          & 0.19(24)  & 0.50(20)  \\ 
    \hline\hline
  \end{tabular}
\end{table}

\begin{table}
  \centering
\caption{Results for the nucleon charges and first moments computed with the
physical ensemble. The first column denotes the observable, with $u-d$
  indicating the isovector contribution and $u+d$ the connected
  isoscalar. Results extracted using the  plateau method are given
  for $t_s/a= 10, 12$ and $t_s/a=14$ in the second, third, and fourth
  columns. The value extracted using the summation method is given in the
  last column. The errors are obtained using jack-knife.}
  \label{table:charges}
  \begin{tabular}{ccccc}
    \hline\hline
    \multirow{2}{*}{Moment}              &\multicolumn{3}{c}{Plateau}        & Summation \\
                                         & 10$a$    & 12$a$    & 14$a$       & Method    \\
    \hline
    $g^{u+d}_S$                           & 6.46(27)  & 7.84(48)  & 8.93(86)  & 14.0(2.2) \\
    $g^{u-d}_S$                           & 0.55(18)  & 1.18(34)  & 2.20(54)  &  5.0(1.4) \\
    $g^{u+d}_A$                           & 0.583(14) & 0.597(23) & 0.611(48) & 0.637(92) \\
    $g^{u-d}_A$                           & 1.158(16) & 1.162(30) & 1.242(57) & 1.28(12)  \\
    $g^{u+d}_T$                           & 0.596(21) & 0.598(31) & 0.555(63) & 0.57(12)  \\
    $g^{u-d}_T$                           & 1.062(21) & 1.058(35) & 1.027(62) & 0.99(13)  \\
    $\langle x\rangle_{u+d}$              & 0.645(13) & 0.587(18) & 0.540(28) & 0.389(66) \\
    $\langle x\rangle_{u-d}$              & 0.248(09) & 0.218(15) & 0.208(24) & 0.116(54) \\
    $\langle x\rangle_{\Delta u+\Delta d}$  & 0.161(12) & 0.143(17) & 0.175(30) & 0.177(63) \\
    $\langle x\rangle_{\Delta u-\Delta d}$  & 0.286(11) & 0.270(16) & 0.229(33) & 0.159(62) \\
    $\langle x\rangle_{\delta u+\delta d}$  & 0.195(15) & 0.178(21) & 0.212(36) & 0.204(87) \\
    $\langle x\rangle_{\delta u-\delta d}$  & 0.311(13) & 0.261(19) & 0.306(29) & 0.245(65) \\
    \hline\hline
  \end{tabular}
\end{table}

\begin{table}
  \begin{center}
    \caption{Results for the nucleon axial and tensor charges and
      first moments of parton distributions. In the first column we
      give the observable, in the second the isovector combination and
      in the third and fourth the $u$ and $d$ values neglecting
      disconnected contributions.}
  \label{tab:results}
\begin{tabular}{lr@{.}lr@{.}lr@{.}l}
 \hline\hline
  & \multicolumn{2}{c}{isovector} 
  & \multicolumn{2}{c}{up} 
  & \multicolumn{2}{c}{down} \\
 \hline
 $g_A$                         & 1&242(57) & 0&926(47) & -0&315(24) \\
 $g_T$                         & 1&027(62) & 0&791(53) & -0&236(33) \\
 $\langle x\rangle_q$          & 0&208(24) & 0&373(22) &  0&166(13) \\
 $\langle x\rangle_{\Delta q}$ & 0&229(30) & 0&202(26) & -0&027(16) \\
 $\langle x\rangle_{\delta q}$ & 0&306(29) & 0&264(25) & -0&045(21) \\
 \hline\hline  
\end{tabular}
\end{center}
\end{table}
\begin{figure}[!h]
\includegraphics[width=\linewidth]{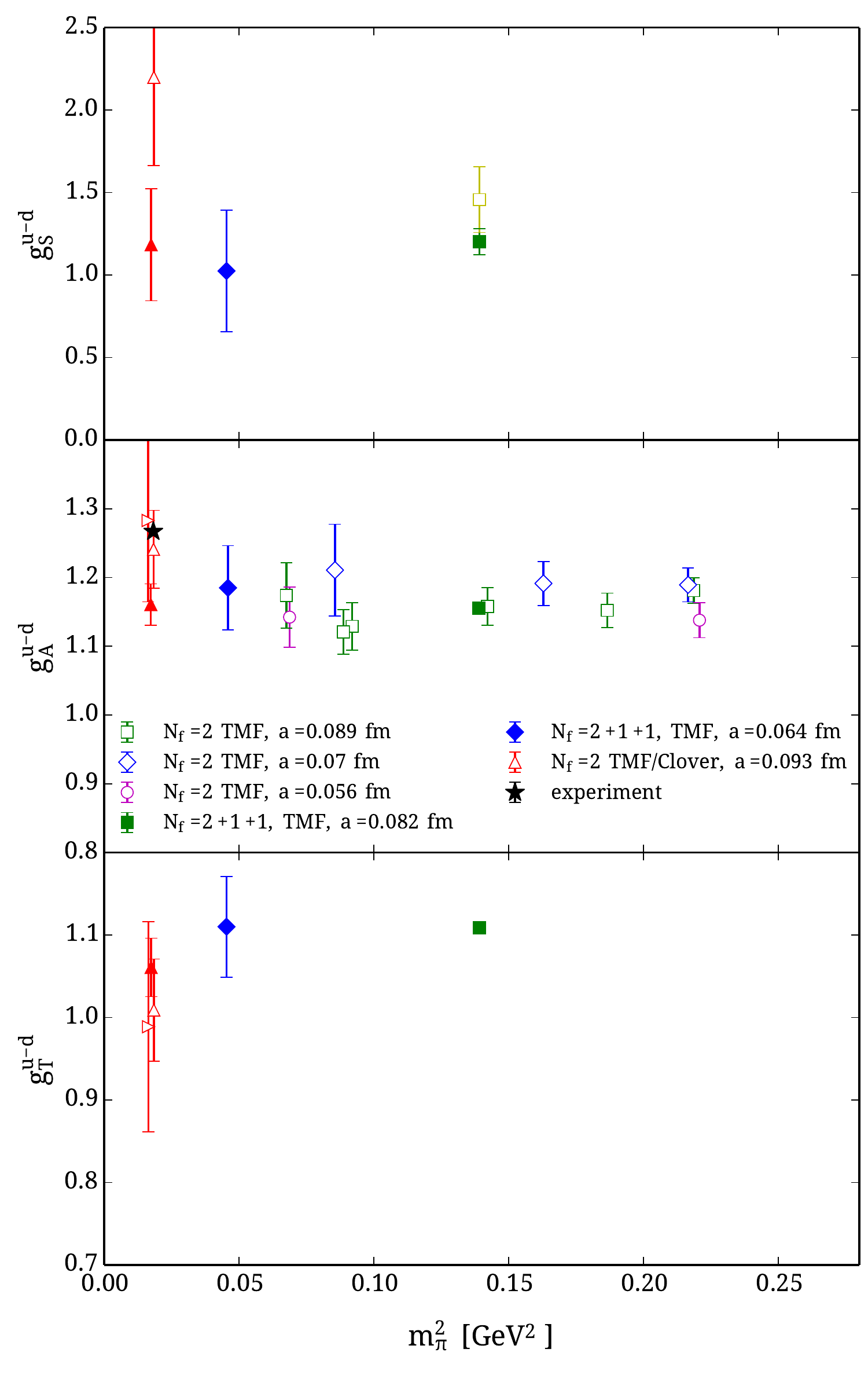}
\caption{Isovector nucleon scalar charge $g_S^{u-d}$ (upper), axial
  charge $g_A^{u-d}$ (middle), and tensor charge $g_T^{u-d}$ (lower)
  using the values of Table~\ref{tab:results}. Twisted mass fermion
  (TMF) results are shown for i) $N_f=2$, $a=0.089$~fm (open green
  squares), $a=0.07$~fm (open blue diamonds) and $a=0.056$~fm (open
  magenta circles); ii) $N_f=2+1+1$, $a=0.082$~fm
  (filled green squares), $a=0.064$~fm (filled blue diamonds) and iii)
  $N_f=2$ TMF clover-improved $a=0.093$~fm (physical ensemble),
  $t_s/a=12$ (filled red triangle), $t_s/a=14$ (open red triangle),
  summation method (open right triangle). The physical value is shown
  with the black asterisk. For the scalar charge we show with the open
  yellow square the value when $t_s\sim 1.5$~fm.}
\label{fig:g ETMC}
\end{figure}

\begin{figure}[!h]
\includegraphics[width=\linewidth]{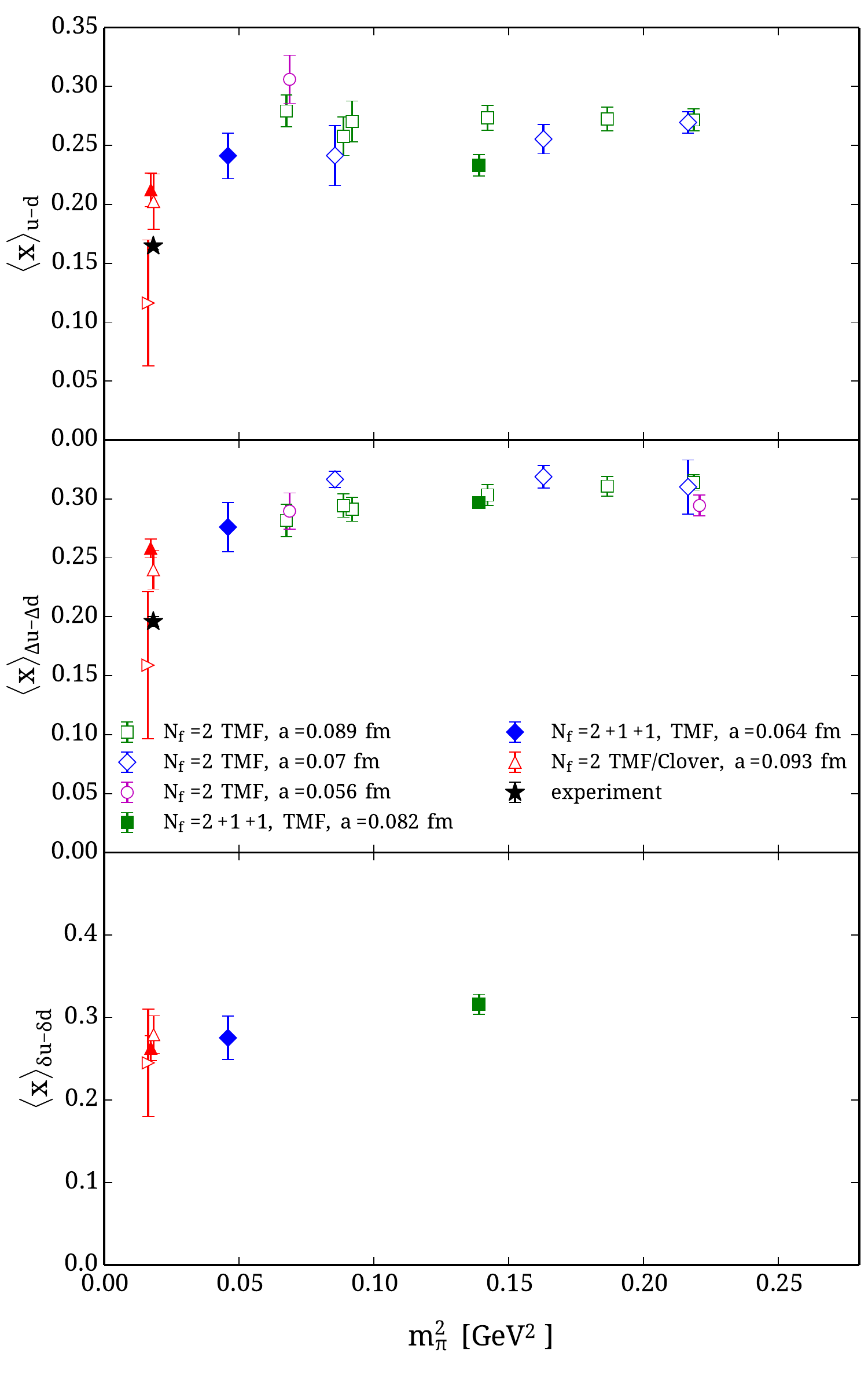}
\caption{ Isovector nucleon momentum fraction $\langle x \rangle_q$
  (upper), helicity $\langle x \rangle_{\Delta q}$ (middle), and
  transversity $\langle x \rangle_{\delta q}$ (lower). The notation is
  the same as that in Fig.~\ref{fig:g ETMC}}
\label{fig:A20 ETMC}
\end{figure}

The nucleon results presented in Figs.~\ref{fig:gS} to~\ref{fig:gT}
and Figs.~\ref{fig:xq} to~\ref{fig:xtq} are summarized in
Table~\ref{table:charges}, where we give the values obtained when
using the plateau method and the summation method. Our results with
sink-source time separation 12$a$, 14$a$, and the summation method
agree within one to two standard deviations. The errors exhibited by
the summation method, however, are still large, which is explained by
the fact that this method relies on a two parameter fit, contrary to
the plateau method which is a fit to a constant. Nevertheless, within
our current statistics, the summation method can provide an additional
check of excited state effects.

We observe that the scalar charge $g_S$ and
momentum fraction $\langle x \rangle_{u\pm d}$ exhibit non-negligible
excited state effects when increasing the sink-source separation. For
these cases we show in Table~\ref{table:charges-two-state} the results
when employing the two-state fit method. For both observables, the
two-state fit result agrees with the plateau method for $t_s/a=14$.

In Table~\ref{tab:results} we give our results for the isovector
quantities as determined from the plateau method using $t_s \sim
1.3$~fm, as well as for the up- and down-quark contributions
neglecting disconnected diagrams. We note that for the up- and
down-quark contributions of these quantities we carry out the complete
analysis with jack-knife resampling starting from three-point
correlation functions with only an up- or down-quark
insertion. Alternatively one can form linear combinations of the final
isovector and isoscalar results of Table~\ref{table:charges}, which
will give consistent up- and down-quark contributions within
statistical errors. Except for the scalar and the momentum fraction,
the results of Table~\ref{tab:results} are in agreement with the value
obtained using the summation method. For the scalar and the momentum
fraction they are consistent with the result extracted using the
two-state fit albeit with large statistical error especially for the
momentum fraction. Since for the scalar there are large differences
still between the results at different values of $t_s$ as well as from
the value extracted using the summation method we do not include a
single value in the table.

Our TMF results for the three isovector charges $g_S^{u-d}, g_A^{u-d}$
and $g_T^{u-d}$ and moments $\langle x \rangle _{u-d}$, $\langle x
\rangle _{\Delta u-\Delta d}$ and $\langle x \rangle _{\delta u-\delta
  d}$ are collected in Figs.~\ref{fig:g ETMC} and ~\ref{fig:A20 ETMC}.
 In the Appendix we give our updated
 results for our high statistics analysis of the B55.32 ensemble for several
sink-source separations in Tables~\ref{tab:B55 results} and \ref{tab:B55 results 2} as well as for one $N_f=2+1+1$ ensemble at a finer lattice spacing~\cite{Alexandrou:2013joa} in Table~\ref{tab:updated results}. These results use the 
new renormalization functions given in Table~\ref{tab1}. For $g_A^{u-d}$, $\langle x \rangle
_{u-d}$ and $\langle x \rangle _{\Delta u-\Delta d}$ we include
results using $N_f=2$ at three lattice spacings and, for one mass, at
two different volumes~\cite{Alexandrou:2010hf,Alexandrou:2011nr}. These are
given with the new renormalization functions in Table~\ref{tab:updated results 2} of the Appendix.
 These results show that cut-off effects are small for
lattice spacings smaller than 0.1~fm. Finite volume effects are not
visible within our statistical accuracy when comparing results for two
ensembles simulated at a pion mass of about 300~MeV and $Lm_\pi=3.3$
and $Lm_\pi=4.6$. For the physical ensemble we show results for
$t_s=12a\simeq 1.1$~fm and $t_s=14a\simeq1.3$~fm and from  using the
summation method. We expect $\langle x \rangle _{u-d}$ to have moderate
excited states contamination as revealed by our high-statistics
investigation of $\langle x \rangle _{u-d}$ for the $N_f=2+1+1$
ensemble with $m_\pi=373$~MeV that showed that its value decreases
with increasing $t_s$~\cite{Dinter:2011sg}. Our current results at the
physical point with 1536 statistics have much larger errors as
compared to what was achieved in Ref.~\cite{Dinter:2011sg} but there
is a clear decreasing trend as we increase $t_s$. For the scalar
charge the excited state contributions are large, as can be seen both
from the results obtained with the $N_f=2+1+1$ ensemble with 373~MeV
where a high statistics analysis is carried out but also for the
physical ensemble where no convergence is achieved with the current
sink-source separations and statistics. Although for $g_A^{u-d}$, $\langle x \rangle _{u-d}$ and $\langle x
\rangle _{\Delta u-\Delta d}$ our results from the summation method are
in agreement with the experimental values, the errors are still too large
and must be reduced by a factor of at least two to draw a safe conclusion.
However, we stress that our value for $g_A^{u-d}$ from the plateau method 
using $t_s\sim 1.3$~fm agrees with the experimental value. To our knowledge,
this is the first computation for which the value of the axial charge is reproduced 
from the plateau method, without any chiral extrapolation.



\subsection{Pion momentum fraction}
In this section we present results on the isovector pion momentum fraction. Three  $N_f=2$ TMF ensembles with the clover term are analyzed with
heavier than physical pion masses, two of which with spatial lattice
size 2.23~fm and one with spatial lattice
size 2.98~fm. A fourth ensemble that includes the clover term 
is simulated using  the physical
value of the pion mass and spatial lattice extent of 4.46~fm.  This is the
ensemble 
used for the nucleon observables and the ensemble details can be found
in Table~\ref{Table:params}. The number of
 measurements, which are well separated
in the number of HMC trajectories, is given in Table~\ref{table:pionx}.
\begin{table}
  \centering
  \caption{Results for the renormalized $\langle x\rangle_{u-d}^{\pi^{\pm}}$ in $\overline{\rm MS}$ at 2~GeV for
    the four TMF clover-improved ensembles considered in this
    work. In the first column we give the bare twisted quark mass,
    and in the second column the spacial extent in lattice units. In
    the fourth column we provide the  value of $\langle
    x\rangle_{u-d}^{\pi^\pm}$ of the pion with its statistical and
    systematic uncertainty computed as explained in the text. The last
    column gives the number of measurements for each ensemble.}
  \label{table:pionx}
\begin{tabular}{cccc}
   \hline\hline
    $a\mu$ & $L/a$& $\langle x\rangle_{u-d}^{\pi^{\pm}}$ in $\overline{\rm MS}$ at 2~GeV& $N_\mathrm{meas}$ \\
   \hline\hline
    $0.006$ & $24$ & $0.249(7)(^{+3}_{-3})$   & $210$ \\
    $0.006$ & $32$ & $0.259(4)(^{+1}_{-1})$   & $240$ \\
    $0.003$ & $24$ & $0.219(13)(^{+6}_{-5})$  & $276$ \\
    $0.0009$& $48$ & $0.214(15)(^{+12}_{-9})$ & $309$ \\
   \hline\hline
 \end{tabular}
\end{table}
\begin{figure}[h!]
  \centering
  \includegraphics[width=\linewidth]{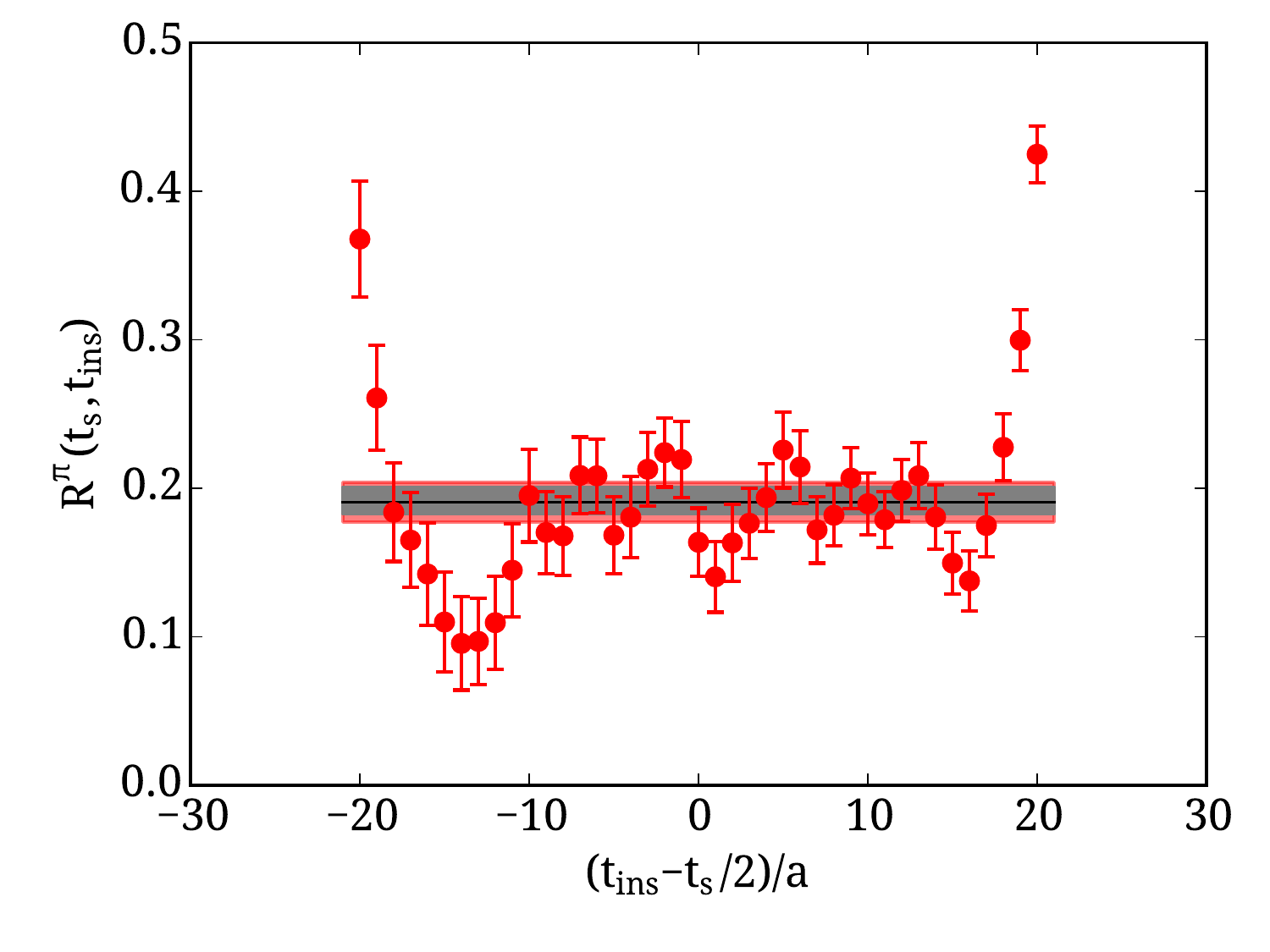}
  \caption{The ratio for $\langle x\rangle_{u-d}^{\pi^\pm}$ using the
    $a\mu=0.0009$ ensemble. We show the weighted median over the
    different fit-ranges as a solid black line, the statistical error
    as the red band and the systematic errors as the gray band.}
  \label{fig:averxplateau}
\end{figure}

\begin{figure}[h!]
  \centering
  \includegraphics[width=\linewidth]{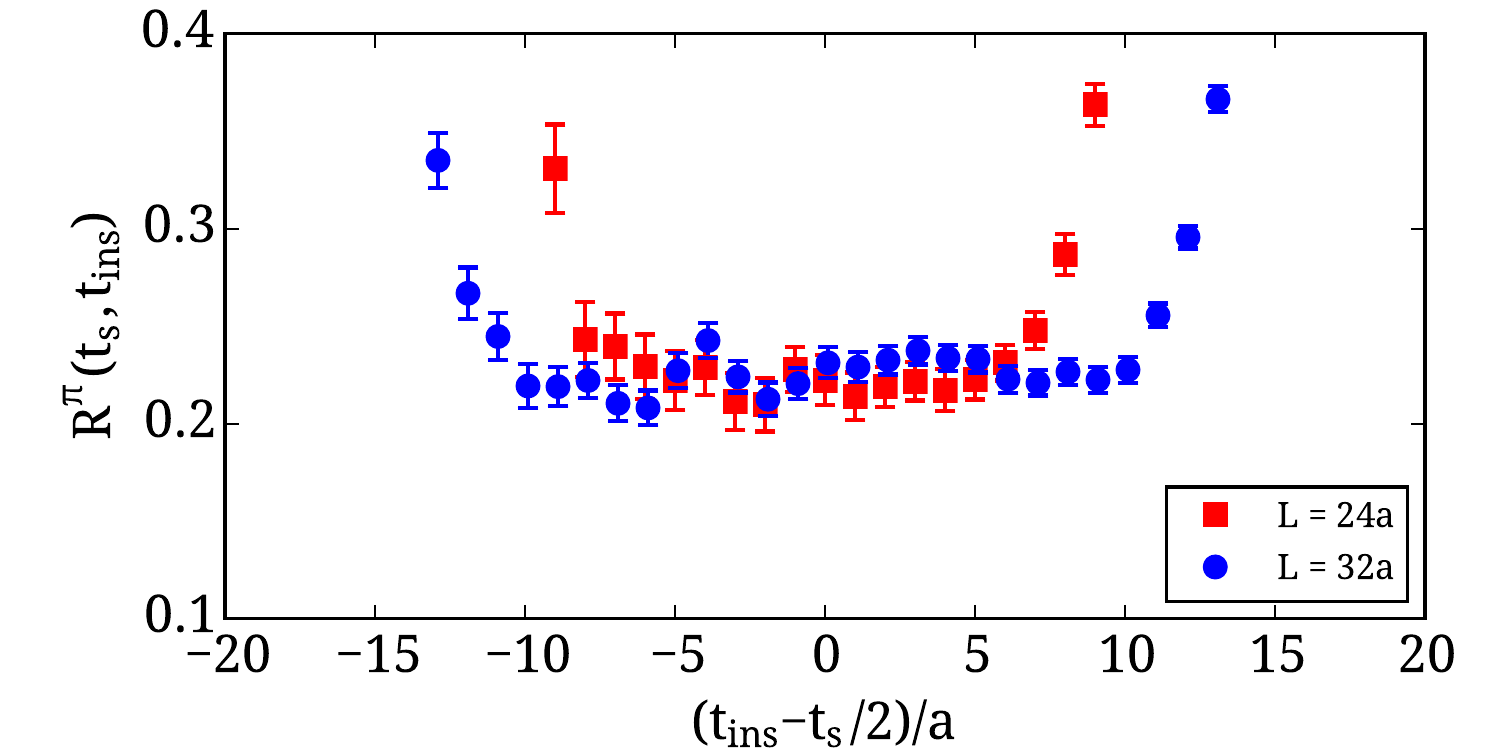}
  \caption{The ratio for $\langle x\rangle_{u-d}^{\pi^{\pm}}$ for $a\mu=0.006$,
    comparing results obtained for lattices of two sizes: $L=24a$ (red squares) and $L=32a$ (blue circles).}
  \label{fig:averxplateau06}
\end{figure}

\begin{figure}[h!]
  \centering
  \includegraphics[width=\linewidth]{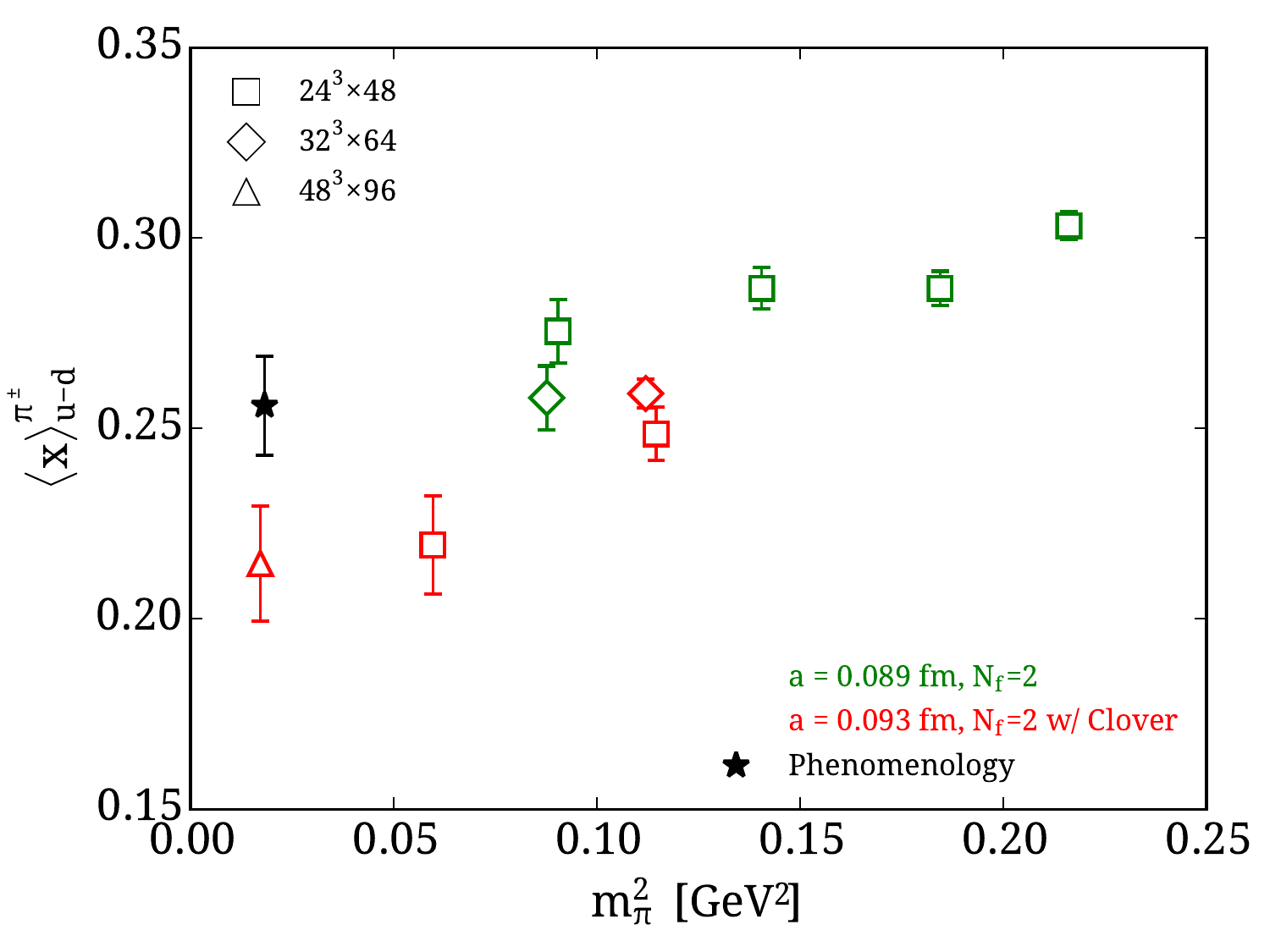}
  \caption{The renormalised momentum fraction of the pion $\langle
    x\rangle^{\pi^{\pm}}_{u-d}$ as a function of the squared pion mass
    at renormalization scale $\mu=2\ \mathrm{GeV}$ in the
    $\overline{\mathrm{MS}}$ scheme. The results of this work (red
    symbols) are shown together with previous results obtained using
    $N_f=2$ ensembles (green symbols)~\cite{Baron:2007ti}
    compared to the phenomenological value (black star) from
    Ref.~\cite{Wijesooriya:2005ir}. Results at two values of the pion mass but different lattice volumes are shown by the open squares ($24^3\times 48$) and  diamonds ($32^3\times 64$).}
  \label{fig:averx}
\end{figure}

In Fig.~\ref{fig:averxplateau} we show the ratio
Eq.~(\ref{eq:averxratio}) as a function of $t_{\rm ins}/a$ for the
physical ensemble. The black horizontal line represents the value
quoted in Table~\ref{table:pionx}. The statistical accuracy of the
pion correlation functions allows for a more careful assessment of
systematic uncertainties in the plateau fit as compared to the case of
the nucleon. Namely, we obtain the plateau value by performing
constant fits to the data with all possible fit ranges with degrees of
freedom larger than $5$. For each of these fits a weight factor
\[
w = \left(\frac{1-2|p - 0.5|}{W}\right)^2\left(\frac{1-2|p_{m_\pi} - 0.5|}{W_{m_\pi}}\right)^2\,
\]
is computed, where $p$ ($p_{m_\pi}$) is the p-value of the fit and
W ($W_{m_\pi}$) the statistical error of the fit parameter
(of $m_\pi$) determined using $1500$ bootstrap samples. The pion mass
itself is also determined for a large number of fit-ranges. The final
result is determined as the weighted median over all combinations of
fit-ranges. The 68.54\% confidence interval of the weighted
distribution is quoted as the systematic uncertainty.

For the twisted mass value $a\mu=0.006$ we have two spatial lattice
sizes available, namely $L/a=24$ and $L/a=32$. Within errors, the
result for $\langle x\rangle^{\pi^{\pm}}_{u-d}$ agrees between these two ensembles. We
show the $\langle x\rangle^{\pi^{\pm}}_{u-d}$ results for $L/a=24$ and $L/a=32$ in
Fig.~\ref{fig:averxplateau06} as a function of $t_{\rm ins}/a$. 

We compare our $N_f=2$ clover-improved results with results obtained
for $N_f=2$ twisted mass ensembles without the clover term published
in Ref.~\cite{Baron:2007ti}. These ensembles were simulated using
$\beta=3.90$ with four values of the bare quark mass: $a\mu=0.004$,
0.0064, 0.0085 and 0.0100. The lattice spacing for these ensembles is
0.089(1)~fm, which is similar to the clover-improved ensembles. For
$a\mu=0.004$ we have again two spatial lattice extents $L/a=24$ and
$L/a=32$ available. As for the clover-improved ensembles the results
for $\langle x\rangle_{u-d}^{\pi^\pm}$ agree between the two volumes.

From the comparison of the different available spatial lattice sizes
we conclude that within the current statistical uncertainties we
cannot detect significant finite volume effects for $m_\pi L \leq 3.2$
realized for the $\beta=3.90, a\mu=0.004, L/a=24$ ensemble. The
physical ensemble has slightly smaller $m_\pi L=2.97$, and the clover
ensemble with $a\mu=0.003$ has $m_\pi L=2.77$. Therefore, we cannot
 completely  exclude finite size effects for these two ensembles. All other
ensembles have $m_\pi L> 3.2$. Note that we are currently generating a
physical ensemble with $L/a=64$, which will allow us to check for
finite size effects. We expect pion observables to be more sensitive
to finite size effects than nucleon observables; having a larger
volume will enable us to confirm this expectation.

The results for the renormalized isovector momentum fraction of the pion  $\langle
x\rangle_{u-d}^{\pi^\pm}$  are summarized in Table~\ref{table:pionx}.  The
renormalization factors used are given in Table~\ref{tab1}  at
$2\ \mathrm{GeV}$ in the $\overline{\mathrm{MS}}$ scheme.
The results are also displayed in Fig.~\ref{fig:averx}  as a function of the squared pion mass.  In
Fig.~\ref{fig:averx} one observes that there is agreement between the
clover-improved and non-clover improved ensembles within errors. Note
that systematic uncertainties are not displayed.

In Fig.~\ref{fig:averx} we compare with the latest phenomenological
value for $\langle x\rangle^{\pi^{\pm}}_{u-d}$ which can be found in
Ref.~\cite{Wijesooriya:2005ir} and reads
\[
\langle x\rangle^{\pi^{\pm}}_{u-d} = 0.256(13)\,.
\]
Note that the result  given in Ref.~\cite{Wijesooriya:2005ir} is at
$\mu=5.2\ \mathrm{GeV}$ renormalization scale and we have translated
it to $\mu=2\ \mathrm{GeV}$ using three loop perturbation theory. The
phenomenological value  is compatible with the value of $0.214(15)(^{+12}_{-9})$
  computed for the
physical ensemble.

The results presented here can be compared to
Ref.~\cite{Bali:2013gya}, where $N_f=2$ non-perturbatively clover
improved Wilson fermions have been used, including two ensembles with
pion mass values around $150\ \mathrm{MeV}$. Two values of the lattice
spacing are investigated, $a=0.06$ and $a=0.07\ \mathrm{fm}$,
respectivley. In that reference a bending of the momentum fraction
values towards small pion mass values is observed, while the agreement
at $m_\pi>300\ \mathrm{MeV}$ to the results presented here is
reasonable. 

\subsection{Comparison of nucleon observables with different fermion actions}

In this section we compare our results on nucleon observables with
other recent results obtained using simulations with similar
parameters.
\begin{figure}[!h]
\includegraphics[width=\linewidth]{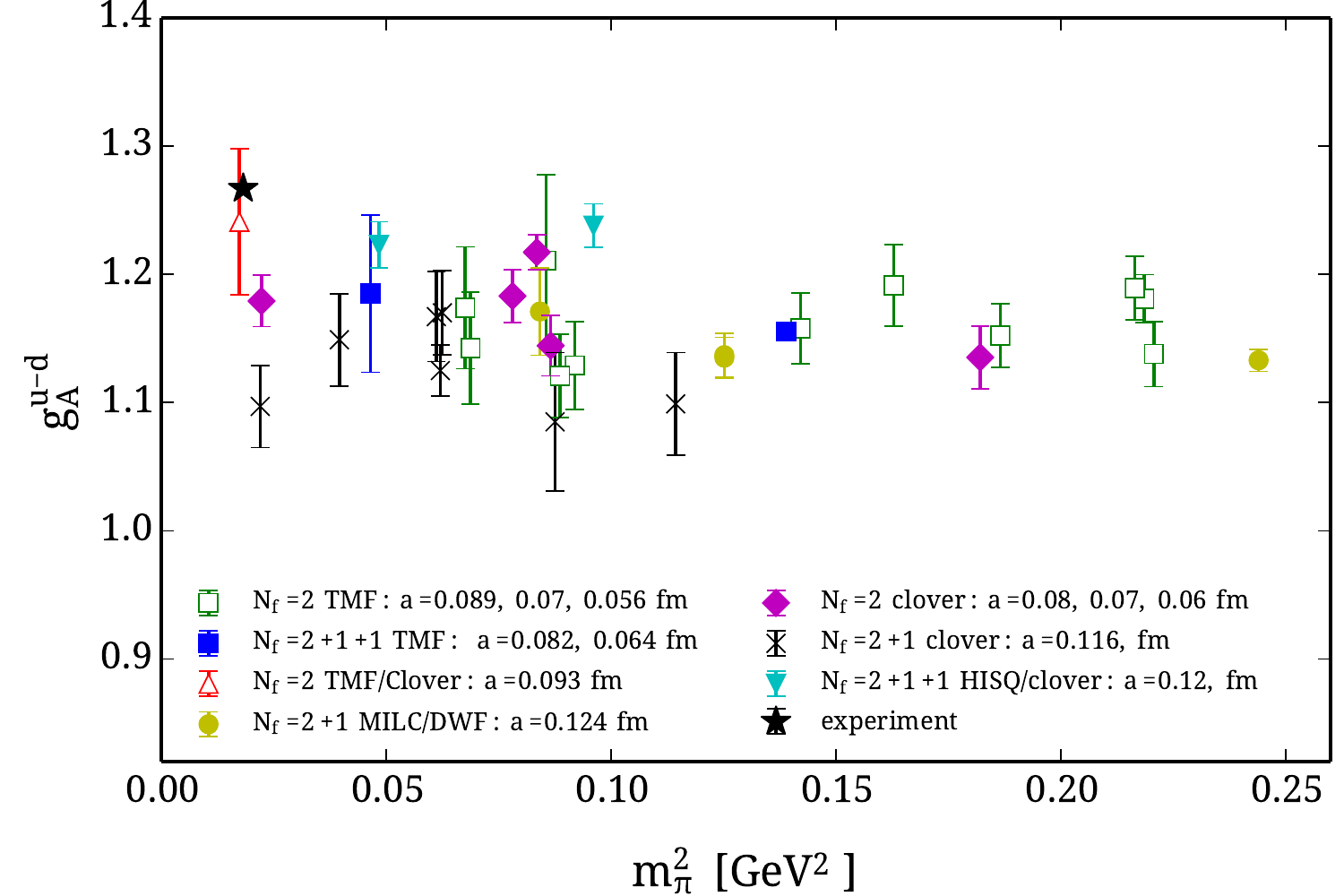}\\
\caption{Results for the nucleon axial charge for different fermion
  actions.  Twisted mass fermion results are shown with open green
  squares for $N_f=2$ ensembles~\cite{Alexandrou:2010hf}, with filled
  blue squares for $N_f=2+1+1$~\cite{Alexandrou:2013joa} and with the
  open red triangle for the physical ensemble using the plateau value
  at $t_s/a=14$ (see Table~\ref{tab:results}). Results are also shown
  using $N_f=2$ clover fermions (filled purple
  diamonds)~\cite{Bali:2014nma}; $N_f=2+1+1$ staggered sea and clover
  valence quarks (filled light blue inverted
  triangles)~\cite{Bhattacharya:2013ehc}; $N_f=2+1$ with DWF on a
  staggered sea (filled yellow circles)~\cite{Bratt:2010jn}; and
  $N_f=2+1$ clover (black x-symbols)~\cite{Green:2012ud}.}
\label{fig:gA comparison}
\end{figure}

\begin{figure}[!h]
\includegraphics[width=\linewidth]{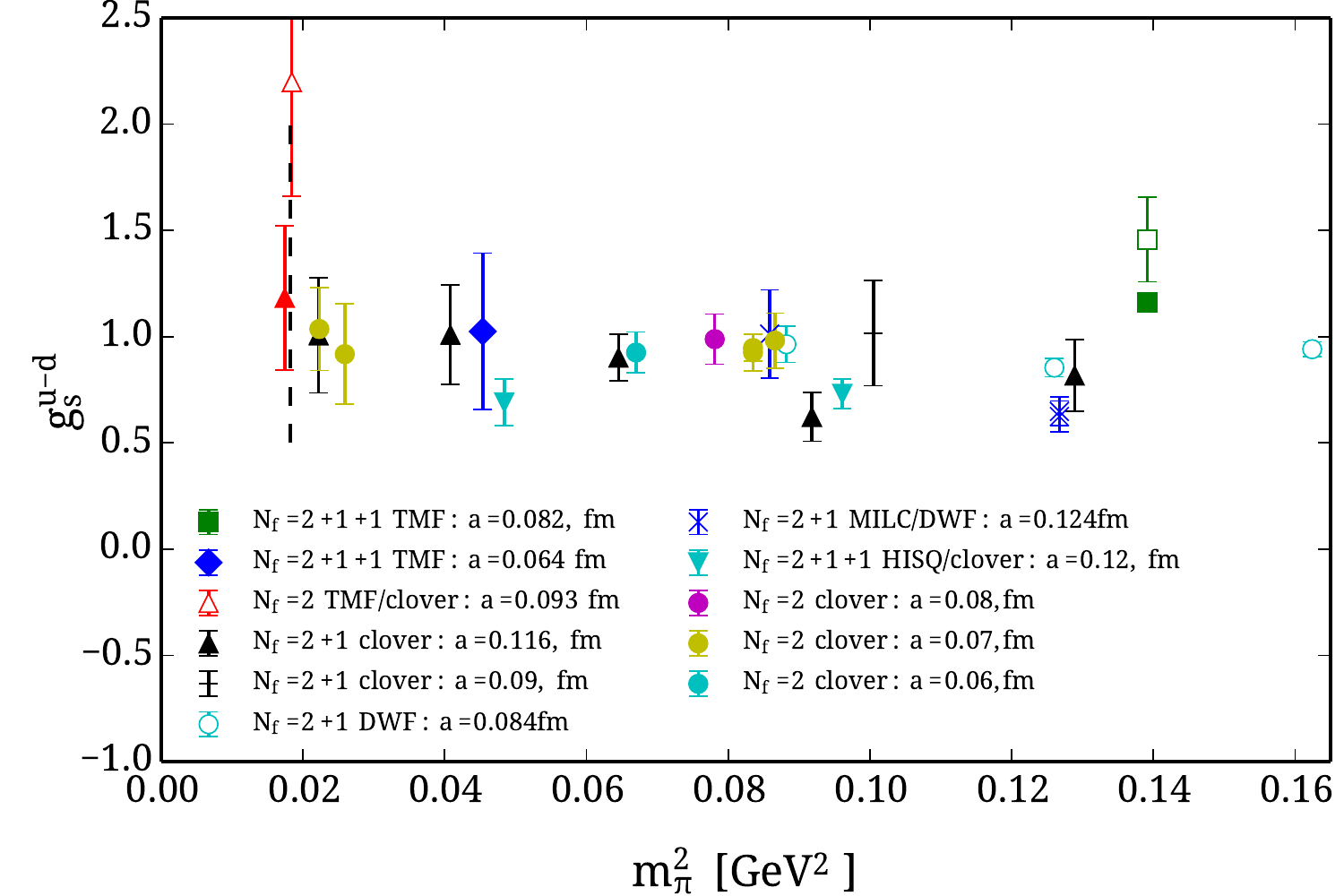}\\
\includegraphics[width=\linewidth]{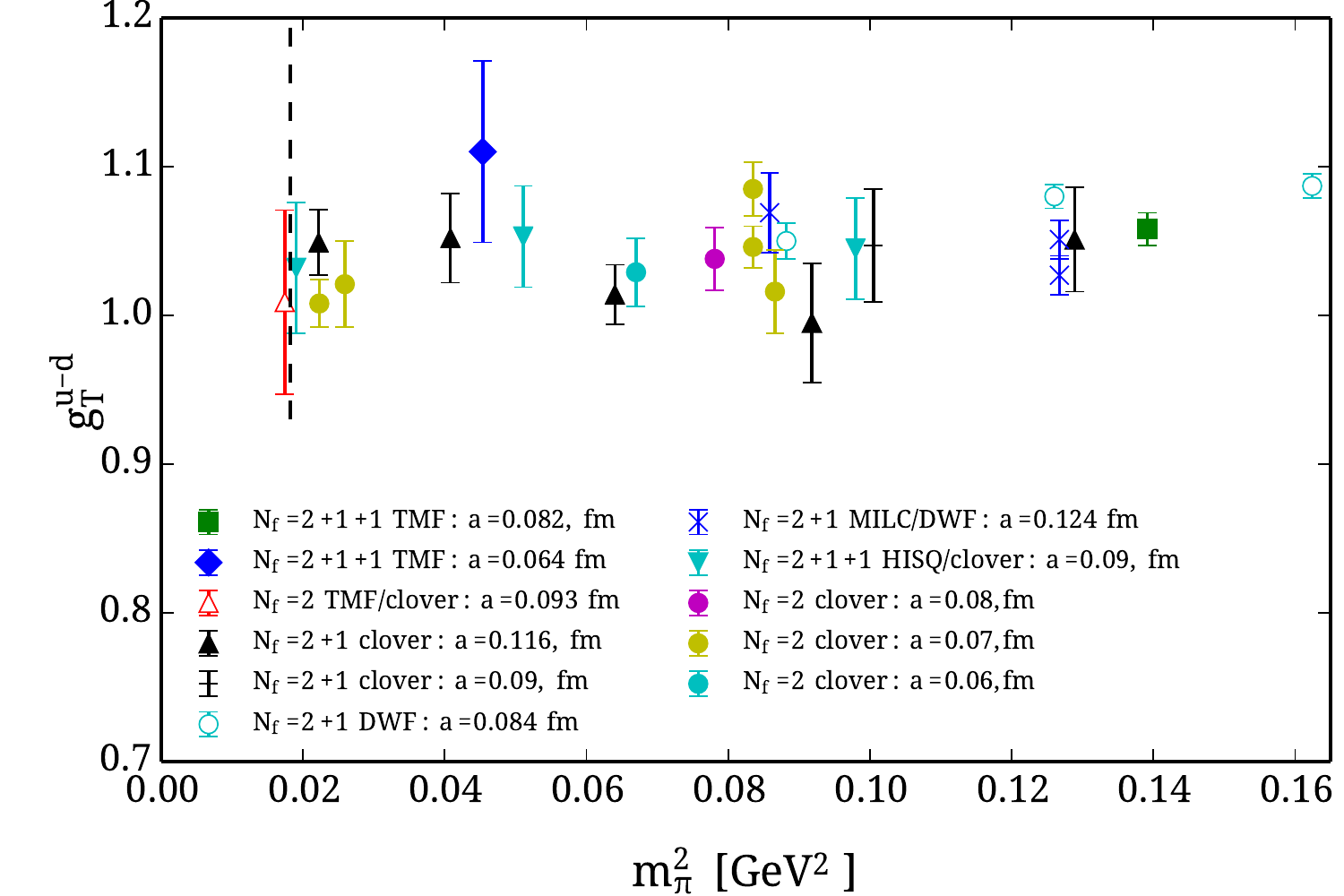}
\caption{ Isovector nucleon scalar charge $g_S^{u-d}$ (upper) and
  tensor charge $g_T^{u-d}$ (lower) versus $m_\pi^2$. Twisted mass
  fermion (TMF) results are shown for two ensembles of $N_f=2+1+1$
  fermions (filled green square for $t_s\sim 1.15$~fm and open green square for $t_s \sim 1.5$~fm, and filled blue diamond) and for the
  physical ensemble (filled red triangle for $t_s\sim 1.1$~fm and open red triangle for $t_s\sim 1.3$~fm).  Results are also shown using:
  clover fermions on $N_f=2+1+1$ staggered sea from
  Ref.~\cite{Bhattacharya:2013ehc,Gupta:2014dla} for $g_S^{u-d}$ and from
  Ref.~\cite{Bhattacharya:2015wna}, for $g_T^{u-d}$ (filled light blue downwards
  triangles); $N_f=2+1$ clover (black filled triangles and crosses),
  $N_f=2+1$ domain wall fermions (open light blue circles) and hybrid
  (blue crosses)~\cite{Green:2012ej}; $N_f=2$ Clover fermions for
  three values of the lattice spacing (filled magenta, yellow and
  light blue circles crosses)~\cite{Bali:2014nma}. All results were
  extracted using the plateau method except those from
  Ref.~\cite{Bhattacharya:2013ehc,Bhattacharya:2015wna}, which used a two-state fit. }
\label{fig:gSgT comparison}
\end{figure}

\begin{figure}[h!]
\includegraphics[width=\linewidth]{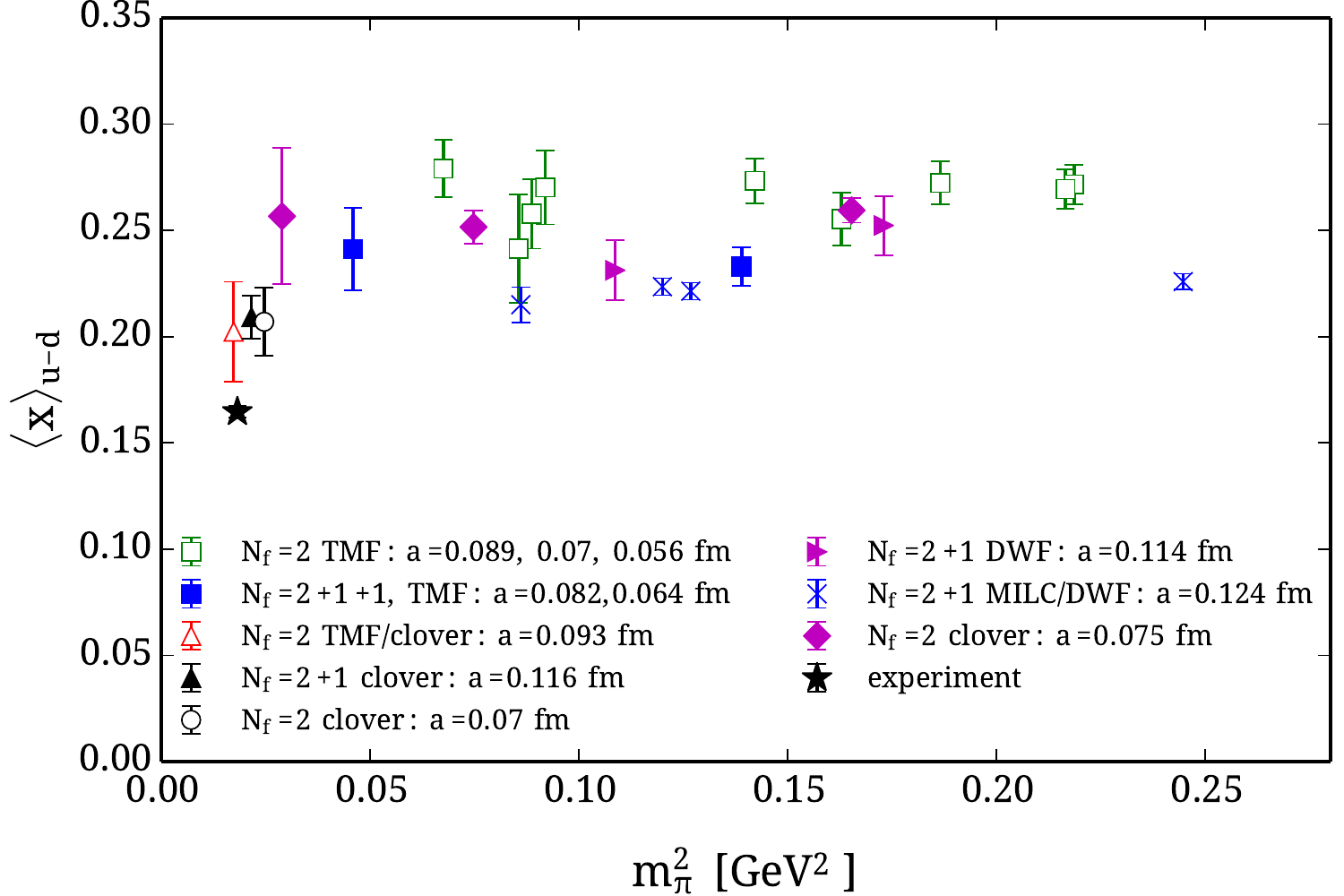}\\
\includegraphics[width=\linewidth]{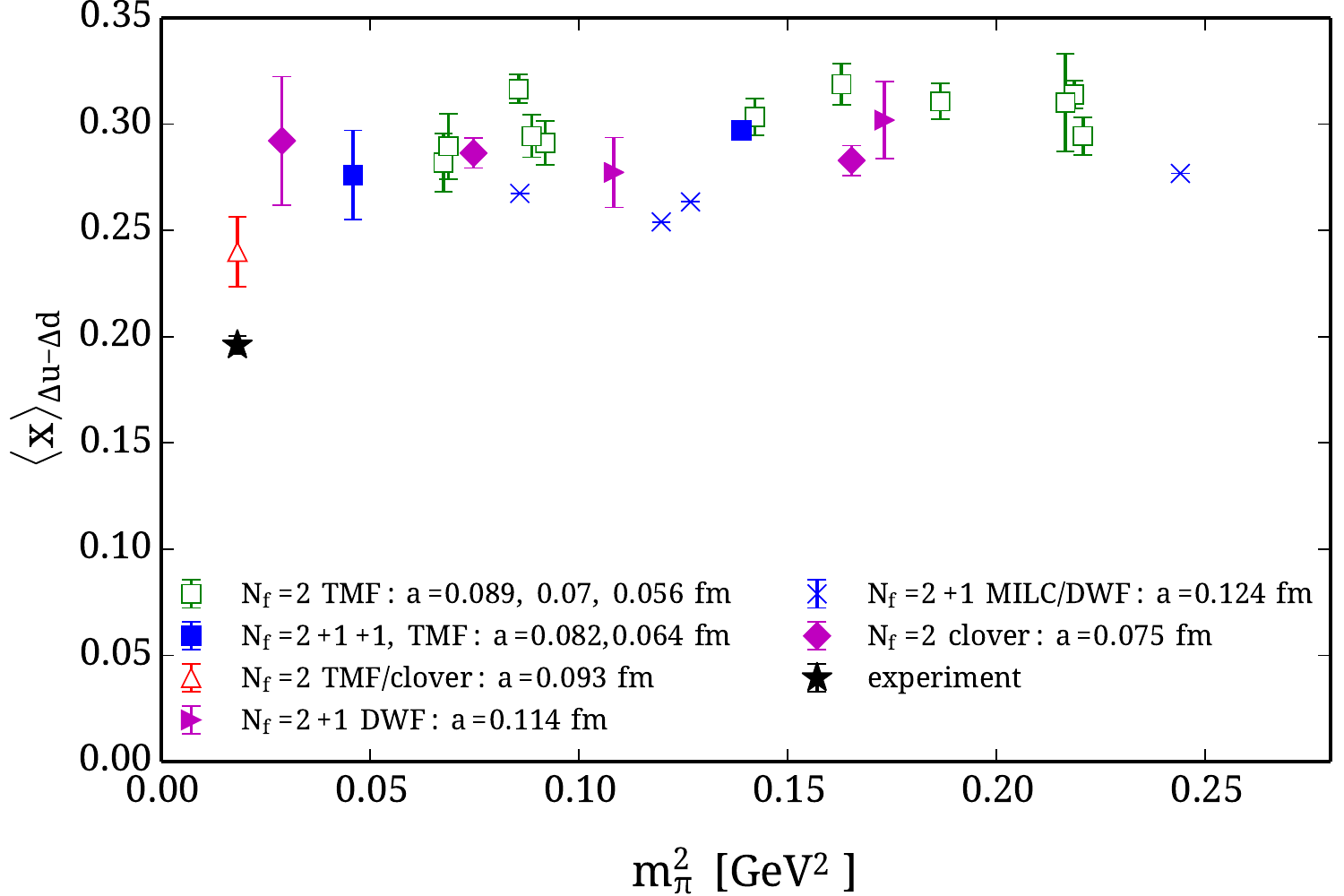}
\caption{Isovector nucleon momentum fraction
  $\langle x \rangle_{u-d}$ and helicity $\langle x \rangle_{\Delta u
    - \Delta d}$. Twisted mass fermion results are shown for $N_f=2$
  ensembles (open green squares), for two $N_f=2+1+1$ ensembles (blue
  filled square) and for the physical ensemble with a clover term
  (open red triangle) taken from Table~\ref{tab:results}.  Also shown
  are results from RBC-UKQCD using $N_f=2+1$ DWF (magenta right
  pointing triangle)~\cite{Aoki:2010xg}, from LHPC using DWF on
  $N_f=2+1$ staggered sea (blue crosses)~\cite{Bratt:2010jn} and QCDSF/UKQCD using $N_f=2$ clover
  fermions (filled magenta diamond)~\cite{Pleiter:2011gw}. For
  $\langle x \rangle_{u-d}$ we also show results from LHPC using
  $N_f=2+1$ clover with 2-HEX smearing (filled black
  triangles)~\cite{Green:2012ud} and $N_f=2$ clover (open black
  circle)~\cite{Bali:2014gha}. All values are extracted using the
  plateau method and $t_s\sim (1-1.2)$~fm, except our result at the
  physical point for which $t_s\sim 1.3$~fm was used. The experimental value for $\langle x\rangle_{u-d}$ is taken from Ref.~\cite{Alekhin:2012ig} and for $\langle x\rangle_{\Delta u-\Delta d}$ from Ref.~\cite{Blumlein:2010rn}.}
\label{fig:A20 comparison}
\end{figure}
 
A number of lattice QCD collaborations are investigating
$g_A$ since, as emphasized already, this is considered a benchmark quantity
for lattice QCD. In Fig.~\ref{fig:gA comparison} we show results for
$N_f=2$~\cite{Alexandrou:2010hf} and
$N_f=2+1+1$~\cite{Alexandrou:2013joa} twisted mass fermions obtained
in previous analyses using simulations with pion masses in the range
of 450~MeV to 210~MeV for various volumes always satisfying the
condition $L m_\pi> 3$. For $N_f=2$ ensembles, three values of the
lattice spacing were analyzed, namely $a=0.089$~fm, 0.07~fm and
0.056~fm and, at one pion mass of about 300~MeV, for two different
volumes.  As already pointed out, the consistency among these results
indicates small cut-off and finite volume effects.  The $N_f=2$ values
are consistent with the values extracted using two $N_f=2+1+1$
ensembles with lattice spacing $a=0.082$~fm and $a=0.064$~fm, showing
that there are no visible strange and charm sea quark effects on these
quantities at least to the accuracy we now have. This allows a
comparison with results using different fermion discretization schemes
even before the continuum extrapolation is performed. In
Fig.~\ref{fig:gA comparison} we include results obtained using clover
improved fermions from two collaborations: In Ref.~\cite{Bali:2014nma}
results were obtained using $N_f=2$ clover fermions with smallest pion
mass of 150~MeV and three lattice spacings $a=0.08$~fm, 0.07~fm and
0.06~fm as well as several volumes. These results are in agreement
with ours. The LHPC analyzed $N_f=2+1$ tree-level clover-improved
Wilson fermions with 2-HEX stout smeared gauge links provided
by the BMW collaboration using smallest pion mass of 149~MeV at one
lattice spacing $a=0.116$~fm~\cite{Green:2012ud}. These tend in
general to have lower values. This is particularly severe close to
the physical pion mass. LHPC also computed the axial charge in a
mixed action approach that uses DWF on staggered sea quarks by
LHPC~\cite{Bratt:2010jn} with $a=0.124$~fm where high accuracy results
were produced for heavier pion masses. These results are in good
agreement with ours.  We note that both TMF and clover-improved
results are extracted using the plateau method with sink-source time
separation of about 1~fm to 1.2~fm with the exception of our result
for the physical ensemble where we used time separations of up to
1.3~fm. In addition, results at two pion masses using a hybrid action
with clover valence on $N_f=2+1+1$ staggered fermions used a two-state
fit~\cite{Bhattacharya:2013ehc} and are included here for comparison.
They tend to be higher than other results although they are compatible
with our $N_f=2+1+1$ value at pion mass of about 210~MeV, which albeit
carries a large error. The general conclusion is that the lattice
results for pion mass higher than about 300~MeV, which are more
accurate as compared to those at smaller pion masses, are in
agreement. This is an indication that lattice systematics are under
control in this pion mass range. Results for pion masses smaller than
about 300~MeV, in general, have larger statistical errors and
agreement among them is harder to assess. They clearly indicate the
need for more precise values and a reliable assessment of systematic
uncertainties. This is particularly relevant for the results close to
the physical point where we observe a disagreement between clover
results from LHPC at pion mass of 149~MeV and from
Ref.~\cite{Bali:2014nma} at similar pion mass. This discrepancy was
claimed to be due to excited states, which were shown to be suppressed
with improved smearing in Ref.~\cite{Bali:2014nma}. This needs to be
further investigated with a dedicated precision calculation with a
complete assessment of systematic uncertainties. Other results using
clover-improved fermions not shown here are those by the CLS
collaboration~\cite{Capitani:2012gj}, which extracted their values
from the summation method. A complete set of the results on $g_A$ can
be found in Ref.~\cite{Constantinou:2014tga}. The result of this work
is shown with the open triangle obtained for $t_s=1.3$~fm.  This value
is in agreement with the experimental value with, however admittedly a
rather large error.

The calculation of the scalar and tensor charges has received more
attention recently due to their relevance for searches of new scalar
and tensor couplings beyond the familiar weak interactions of the
Standard Model in the decay of ultra-cold neutrons. We compare our TMF
results in Fig.~\ref{fig:gSgT comparison} with those obtained by three
groups whose results on the nucleon axial charge were also included in
Fig.~\ref{fig:gA comparison}.  The first set of results are from
Ref.~\cite{Bali:2014nma} using $N_f=2$ clover fermions at three
lattice spacings. The second set is from the LHPC group which in
Ref.~\cite{Green:2012ej} used $N_f=2+1$ clover with 2-HEX stout
smeared gauge links at lattice spacings $a=0.116$~fm and $a=0.09$~fm,
  $N_f=2+1$ DWF with $a=0.084$~fm and a hybrid
action of DWF on staggered sea with $a=0.124$~fm. Both these groups
used the plateau method and sink-source time separation within 1 to
1.2~fm. The third set of results are from
Ref.~\cite{Bhattacharya:2013ehc} at one lattice spacing for the scalar
and from Ref.~\cite{Bhattacharya:2015wna} at three lattice spacings for the tensor
obtained using a hybrid action of DWF on $N_f=2+1+1$ staggered
fermions and employing a two-state fit.
For the case of the scalar charge shown in the upper panel of
Fig.~\ref{fig:gSgT comparison}, we observe overall a good agreement
among all lattice results obtained with similar sink-source
separation. However, our high-statistics analysis using $N_f=2+1+1$ TMF
at pion mass 373~MeV revealed excited states contamination, which only
become negligible when $t_s\sim 1.5$~fm, increasing the value of
$g_S$. The value extracted when $t_s=1.48$~fm is shown in
Fig.~\ref{fig:gSgT comparison} by the open green square.  A similar
analysis for the physical ensemble also reveals large contributions
from excited states for $g_S^{u-d}$. Comparing results obtained for
$t_s\sim 1.1$~fm and $1.3$~fm we confirm an increasing value as we
increase $t_s$. However, the statistical error is large despite the
fact that we have 1536 measurements as compared to 1200 used for the
ensemble at $m_\pi=373$~MeV. This demonstrates that, obtaining the
same accuracy at the physical point for $t_s\sim 1.5$~fm, which may be
needed to suppress excited states, requires more than an order of
magnitude increase in statistics.

Results on the isovector tensor charge are compared in the lower panel of
Fig~\ref{fig:gSgT comparison}. Our TMF results shown in
Fig.~\ref{fig:g ETMC} show that excited state contributions are less
severe for $g_T^{u-d}$ and that the values at $t_s/a=12$ and
$t_s/a=14$ are consistent.  Indeed our value at the physical point
obtained using $t_s\sim 1.3$~fm is in very good agreement with other
lattice results providing a prediction for this important quantity
directly at the physical point.

Recent lattice QCD results have also been obtained for the isovector
momentum fraction and helicity.  A comparison of our results for
$\langle x \rangle _{u-d}$ and $\langle x \rangle _{\Delta u-\Delta
  d}$ with other collaborations is shown in Fig.~\ref{fig:A20
  comparison}. We only show results extracted using the plateau
method. Most of the analyses employed a sink-source separation of 1 to
1.2~fm including our TMF $N_f=2$ ensembles. As shown in
Ref.~\cite{Dinter:2011sg} where $\langle x \rangle _{u-d}$ was
computed using our $N_f=2+1+1$ ensemble at pion mass of 373~MeV and
high-statistics, excited states may not be negligible for this
observable. Indeed, this is confirmed by our current analysis for the
physical ensemble where there is a decreasing trend as $t_s$
increases. Our value at the physical point is in agreement with the
other lattice values extracted close to the physical point. These are
from Ref.~\cite{Bali:2014gha}, which it is an update of Ref.~\cite{Bali:2012av} for $m_\pi\sim 160$~MeV and from LHPC at
$m_\pi\sim 150$~MeV using $N_f=2+1$ clover fermions with 2-HEX smeared
gauge action~\cite{Green:2012ud}.  Our value at $t_s\sim 1.3$~fm is
still larger than the experimental value. We are currently performing
a high statistics analysis for our physical ensemble using larger
values of $t_s$ to investigate contamination due to excited states,
which tend to decrease the value of $\langle x \rangle _{u-d}$.
 For larger pion masses
we show results using $N_f=2+1$ DWF from the RBC-UKQCD
collaborations~\cite{Aoki:2010xg}, from LHPC~\cite{Bratt:2010jn} using
DWF on $N_f=2+1$ staggered sea and from the QCDSF collaboration using
$N_f=2$ clover fermions~\cite{Pleiter:2011gw}. Results from LHPC used
perturbative renormalization which could explain the fact that these
are in general lower than other lattice results. For the case of
$\langle x \rangle _{\Delta u-\Delta d}$ the situation is similar and
our value using the physical ensemble is still larger than its
experimental value. As for the momentum fraction there is a decreasing
trend as $t_s$ increases. In fact the summation method yields a value
that is consistent with the experimental value as can be seen in
Fig.~\ref{fig:A20 ETMC}. However, the error is too large and our goal in
a future analysis is to reduce it by a factor of two so as to confirm
agreement with the experimental value. Resolving these discrepancies
will give more confidence on our prediction for the transversity
moment.

\section{Conclusions} 
In this work we present results on the pion momentum fraction and key nucleon 
observables using
lattice QCD simulations at the physical value of the up and down
quarks. Our analysis of the isovector pion momentum fraction uses $N_f=2$
ensembles with the clover term simulated at three different values of
the light quark mass. We find 
a value of  $\langle x\rangle_{u-d}^{\pi^\pm}=0.214(15)(^{+12}_{-9})$ in the $\overline{\rm MS}$ at 2~GeV at the physical point.

For the nucleon system, we compute 
 the three local and three one-derivative isovector and isoscalar
 matrix elements  at zero momentum transfer.
 In our calculation we analyze three sink-source time
separations, which allows us to investigate excited state effects by
observing the dependence of the extracted nucleon matrix elements on
this separation. For all observables we compare the plateau method
with the summation method.  In some cases the sensitivity on the sink time $t_s$ is
good enough so that a two-state fit can also be applied as a
third method to detect excited state contaminations. Employing these
different methods to ensure that contamination from excited states is
suppressed is crucial in obtaining reliable results. However, for this
study to be meaningful one needs large statistics in particular for
large sink-source time separations and for the summation method. For
the pion momentum fraction where statistical errors are smaller one
extracts the relevant matrix element using the largest possible time
separation ensuring ground state dominance. Our results for the
nucleon axial charge and isovector pion momentum fraction are in
agreement with their experimental values, that constitutes a very
important conclusion of this study. Since the tensor charge is found
to behave similarly to the axial charge as far as ground state
dominance is concerned we can predict its value at the physical point
to be $g_T^{u-d}=1.027(62)$ in the
$\overline{\text{MS}}$ scheme at $2$ GeV. Assuming that disconnected contributions
remain as small at the physical point as were found at a pion mass of
373~MeV where they were shown to be
negligible~~\cite{Abdel-Rehim:2013wlz,Alexandrou:2013wca}, we can give
a direct prediction for the individual up- and down-quark tensor
charges. We find $g_T^u=0.791(53)$ and $g_T^d= -0.236(33)$ (see
Table~\ref{tab:results}).

Thus, this first lattice study of nucleon and pion structure at the
physical values of the light quark masses is very promising for future
precision calculations of these key quantities directly at the
physical point. Ongoing plans include an analysis with increased
statistics for general momentum transfer and new $N_f=2+1+1$ simulations with
 their mass fixed to physical values, combined with larger volumes and improved
algorithms for noise reduction such as multiple right-hand-side
solvers. After reproduction of benchmark quantities such as $g_A$ for
the nucleon and the pion, lattice QCD is in a position to turn to
quantities more difficult to obtain experimentally such as the scalar
and tensor charges $g_S$ and $g_T$. Such charges are of high interest
in phenomenology and experiments since these enter in couplings of
protons to super-symmetric candidate particles. Their precise
determination can therefore be used to exclude regions in dark matter
searches and influence future experimental set-ups for new physics
searches.

\section*{Acknowledgments}

We would like to thank the members of the ETMC for a most enjoyable collaboration.
This work was supported by a grant from the Swiss National Supercomputing Centre (CSCS) under project ID s540 and in addition used computational resources  from
the John von Neumann-Institute for Computing on the Juropa system and
the BlueGene/Q system Juqueen at the research center in J\"ulich, and
Cy-Tera at the Cyprus Institute. We also acknowledge PRACE for
awarding us access to the Tier-0 computing resources Curie, Fermi and
SuperMUC based in CEA, France, Cineca, Italy and LRZ, Germany. 
We
thank the staff members at all sites for their kind and sustained
support. This work is supported in part by the DFG Sino-German
collaborative research centre CRC 110 and
Sonder\-for\-schungs\-be\-reich/ Trans\-regio SFB/TR9 and by funding
received from the Cyprus Research Promotion Foundation under contracts
NEA Y$\Pi$O$\Delta$OMH/$\Sigma$TPATH/0308/31) co-financed by the
European Regional Development Fund and
TECHNOLOGY/$\Theta$E$\Pi$I$\Sigma$/0308(BE)/17. K.H. and Ch. K. acknowledge
support from the Cyprus Research Promotion Foundation under contract
T$\Pi$E/$\Pi\Lambda$HPO/0311(BIE)/09.  B.K. gratefully acknowledges full financial support under AFR PhD grant
2773315 of the National Research Fund, Luxembourg.

\section*{ Appendix}

In this Appendix we give the ETMC results for the $N_f=2$ and the $N_f=2+1+1$ ensembles published in Refs.~\cite{Alexandrou:2010hf,Alexandrou:2013joa} respectively. These results are updated using the new renormalization functions 
given in Table~\ref{tab1}.  In Tables~\ref{tab:B55 results} and \ref{tab:B55 results 2} we collect the results for the B55.32 ensemble for which a high statistics analysis is carried out for several sink-source separations. In Figs.~\ref{fig:B55 isov plateaus} and \ref{fig:B55 isos plateaus} we show the ratios from which the isovector and isoscalar connected charges and first moments of PDFs
are extracted. The ratios are computed for a total of five sink-source time separations spanning a time range of about 0.8~fm to 1.5~fm enabling us to apply the summation method as check for . This high statistics analysis
allows us to perform a two-state fit for all quantities except the axial charge where the excited state contamination is the mildest. As our final values we take
the plateau value that is in agreement with the value extracted from the summation method and two-state fit when possible.
 Finally in Tables~\ref{tab:updated results} and \ref{tab:updated results 2}  we give the results for all our other ensembles where only one sink-source separation was employed.

\begin{widetext}

\begin{table}[h!]
\begin{center}
\caption{Updated results for the $N_f=2+1+1$ B55.32 ensemble for the nucleon charges and first moments of parton distribution functions. For the isoscalar combination we give only the connected contribution.}
\label{tab:B55 results}
\begin{tabular}{cccccccc}
\hline\hline
\multicolumn{6}{c}{Plateau method}                             &   summation   & two-state  \\
$t_s/a:$  &   10    &   12    &   14    &   16     &   18      &    method     &    fit     \\
\hline
\multicolumn{8}{c}{   $g_S$ }\\
 isovector& 1.08(3) & 1.12(3) & 1.16(4) & 1.08(10) & 1.46(20)  &   1.19(10)    &   1.23(10) \\
 isoscalar& 5.07(4) & 5.45(4) & 5.74(6) & 5.95(13) & 6.33(28)  &   6.46(15)    &   6.81(23) \\  
 stat.    & 2429    & 4396    & 4396    &  2018    &  1200     &               &            \\
\hline
\multicolumn{8}{c}{   $g_A$ }\\
 isovector&1.143(4) &1.152(5)  &1.155(8)  &1.174(20) &         &   1.184(19)   &            \\
 isoscalar&0.596(3) &0.591(4)  &0.589(7)  &0.605(16) &         &   0.583(16)   &            \\
 stat.    &  2429   &  4396    & 4396     &  2018    &         &               &            \\
\hline
\multicolumn{8}{c}{   $g_T$ }\\
 isovector&1.119(6) & 1.087(7) &1.058(11) & 1.080(30)&         &   1.023(27)   &   1.053(21)\\ 
 isoscalar&0.680(5) & 0.666(6) &0.660(9)  & 0.663(21)&         &   0.624(22)   &   0.646(9) \\
 stat.    &2278     & 4040     &4040      &1762      &         &               &            \\
\hline
\end{tabular}
\end{center}
\end{table}

\begin{table}[!h]
\begin{center}
\caption{Updated results for the $N_f=2+1+1$ B55.32 ensemble for the nucleon first moments of parton distributions. For the isoscalar combination we give only the connected contribution.}
\label{tab:B55 results 2}
\begin{tabular}{cccccccc}
\hline\hline
\multicolumn{6}{c}{Plateau method}                             &   summation   & two-state  \\
$t_s/a:$  &   10    &   12    &   14    &   16     &   18      &    method     &    fit     \\
\multicolumn{8}{c}{   $\langle x\rangle_q$ }\\
 isovector&0.290(4) & 0.270(3) & 0.252(4) & 0.233(9) &0.252(19)&   0.223(9)    & 0.214(13)  \\
 isoscalar&0.720(8) & 0.677(5) & 0.639(6) & 0.607(11)&0.616(21)&   0.554(15)   & 0.558(19)  \\
 stat.    &2429     & 4396     &4396      &2018      &1200     &               &            \\
\hline
\multicolumn{8}{c}{   $\langle x\rangle_{\Delta q}$ }\\
 isovector&0.328(3) & 0.312(3) & 0.297(3) & 0.298(8) &         &   0.270(8)    &  0.286(9)  \\
 isoscalar&0.207(3) & 0.198(2) & 0.189(3) & 0.193(8) &         &   0.172(7)    &  0.184(7)  \\
 stat.    &2429     & 4396     & 4396     & 2018     &         &               &            \\
\hline
\multicolumn{8}{c}{   $\langle x\rangle_{\delta q}$ }\\
 isovector&0.372(5) & 0.349(4) & 0.322(5) & 0.316(12)&         &   0.283(14)   &   0.284(17) \\
 isoscalar&0.254(4) & 0.239(4) & 0.219(6) & 0.215(15)&         &   0.178(14)   &   0.183(21) \\
 stat.    &2278     & 4041     & 4041     & 1763     &         &               &             \\
\hline           
\end{tabular}
\end{center}
\end{table}

\begin{figure}[h!]
  \centering
  \includegraphics[width=\linewidth]{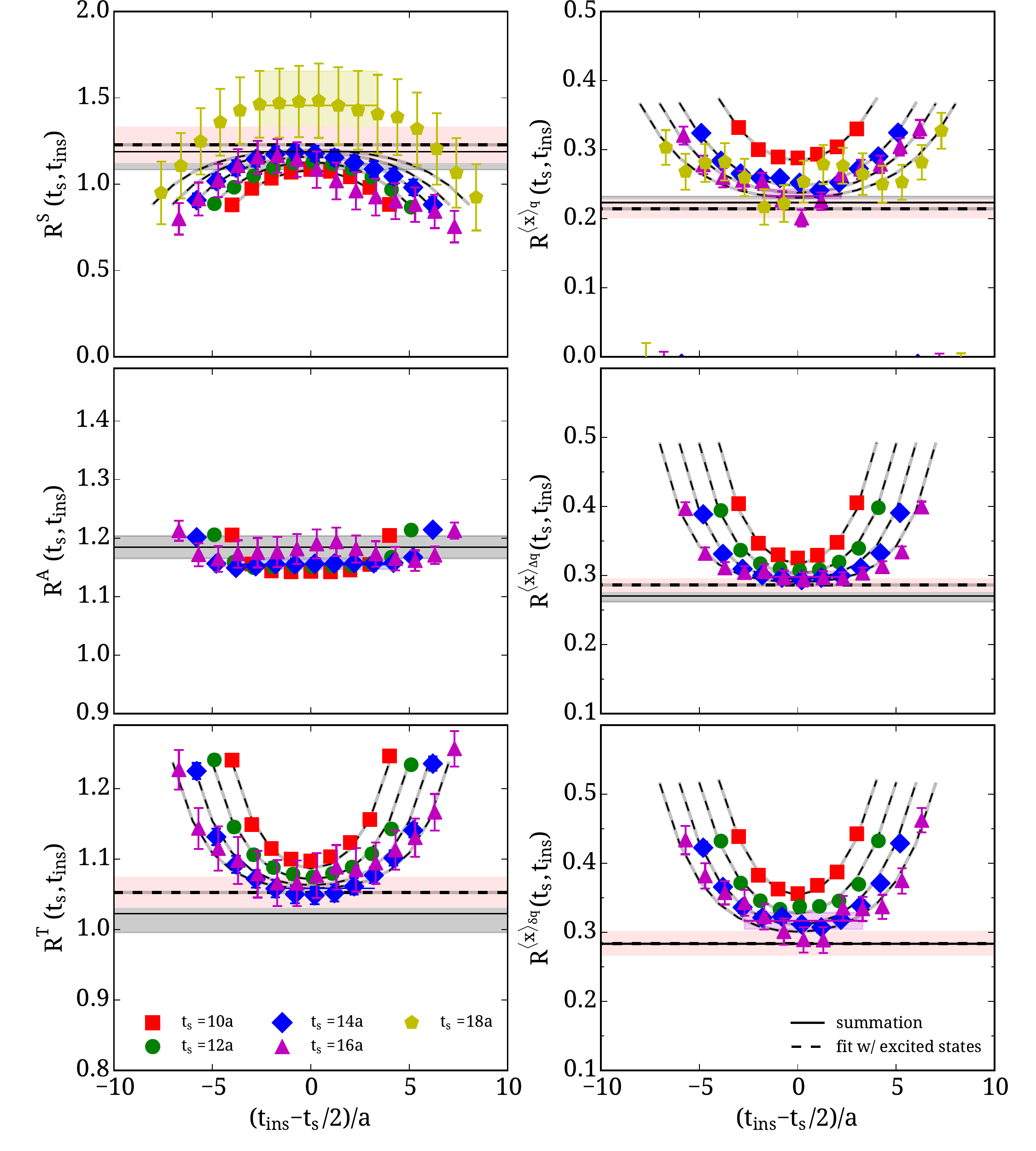}
  \caption{The ratios from which the isovector charges (left) and the first moments of PDFs (right) are extracted as a function of the insertion-source time separation
for the  B55.32 ensemble. The statistics used are given in Tables~\ref{tab:B55 results} and \ref{tab:B55 results 2}.}
  \label{fig:B55 isov plateaus}
\end{figure}\FloatBarrier
\begin{figure}[h!]
  \centering
  \includegraphics[width=\linewidth]{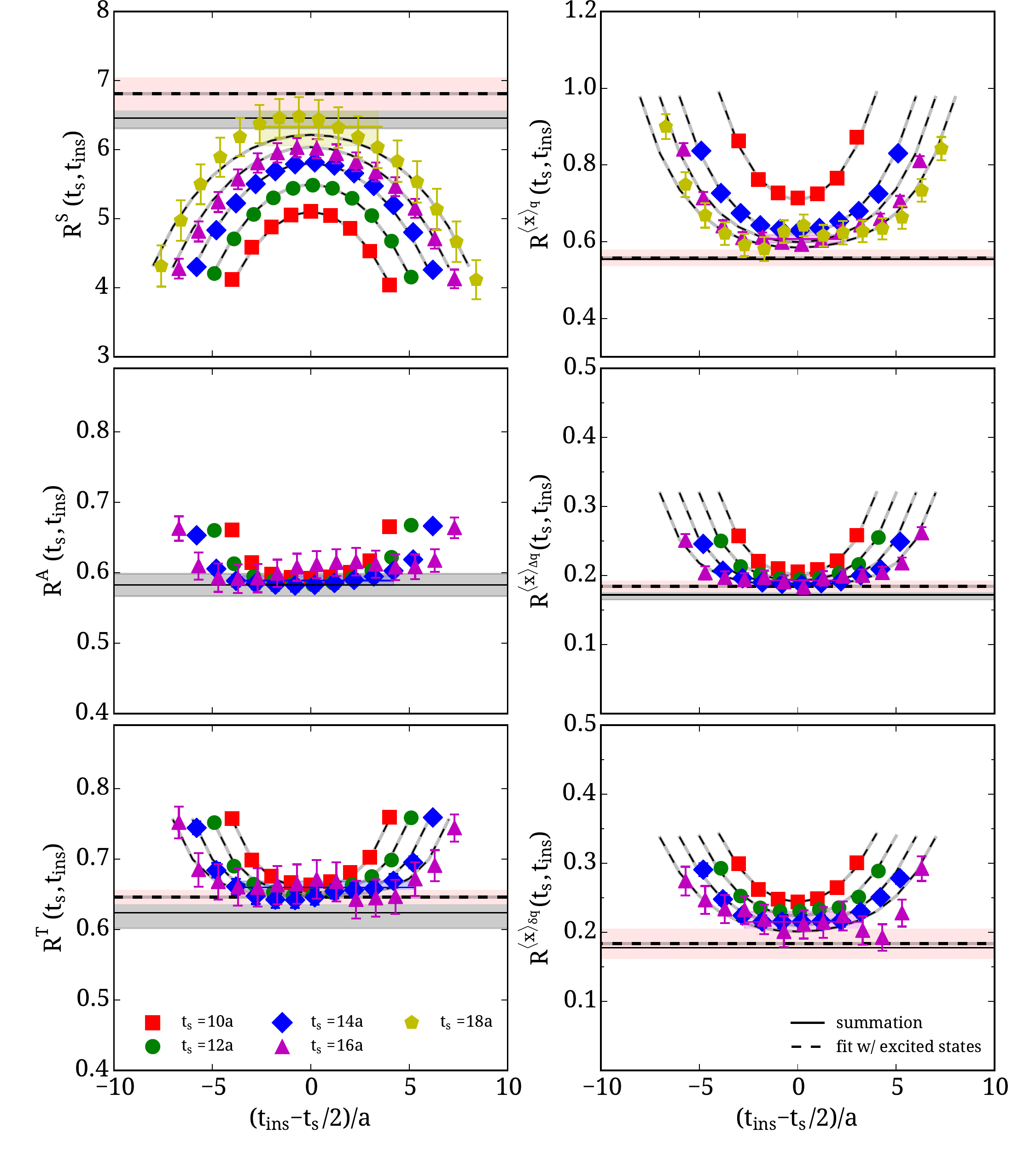}
  \caption{The ratios from which the isoscalar charges (left) and the first moments (right) are extracted as a function of the insertion-source time separation
for the  B55.32 ensemble. Only connected contributions are included. The statistics used are given in Tables~\ref{tab:B55 results} and \ref{tab:B55 results 2}.}
  \label{fig:B55 isos plateaus}
\end{figure}\FloatBarrier

\begin{table}[h!]
\begin{center}
\caption{Updated results for the $N_f=2+1+1$ TMF ensemble with $m_\pi=213$~MeV and $a=0.064$~fm. In the last column we give the number of measurements for all observables. }
\label{tab:updated results}
\begin{tabular}{cccccccccc}
\hline
\hline
$\beta$ ($L^3\times T$)	&$m_\pi$ (GeV) &$g_A^{u-d}$	&$g_S^{u-d}$  &$g_T^{u-d}$  &$\langle x \rangle_{u-d}$	&$\langle x \rangle_{\Delta u-\Delta d}$	&$\langle x \rangle_{\delta u-\delta d}$   &$t_s/a$ &Statistics \\
\hline
\hline
    &	&	&	&  &$N_f{=}2+1+1$	&	&  &  &\\
\hline
\hline
2.10 (48$^3\times$96)	&$\,\,\,$0.2134(6)	&$\,\,\,$1.185(61)	&$\,\,\,$1.024(368)	&$\,\,\,$1.110(61)	&$\,\,\,$0.241(19)       &$\,\,\,$0.276(21)	&$\,\,\,$0.275(26) &$\,\,\,$18  &$\,\,\,$900\\
\hline
\hline
\end{tabular}
\end{center}
\end{table}

\begin{table}[!h]
\begin{center}
\caption{Updated results for the $N_f=2$ TMF ensembles.}
\label{tab:updated results 2}
\begin{tabular}{ccccccc}
\hline
\hline
$\beta$ ($L^3\times T$)	&$m_\pi$ (GeV) &$g_A^{u-d}$	 &$\langle x \rangle_{u-d}$	&$\langle x \rangle_{\Delta u-\Delta d}$  &$t_s/a$ &Statistics\\
\hline
\hline
&	&  &$N_f{=}2$	&	&   &\\
\hline
\hline
3.90 (24$^3\times$48)	&$\,\,\,$0.3032(16)	&$\,\,\,$1.129(34)	 &$\,\,\,$0.270(17)     &$\,\,\,$0.291(10)     &$\,\,\,$12    &$\,\,\,$943\\
	                &$\,\,\,$0.3770(9)	&$\,\,\,$1.158(28)	 &$\,\,\,$0.273(11)     &$\,\,\,$0.303(9)      &$\,\,\,$      &$\,\,\,$553\\
	                &$\,\,\,$0.4319(12)	&$\,\,\,$1.152(25)	 &$\,\,\,$0.272(10)     &$\,\,\,$0.311(8)      &$\,\,\,$      &$\,\,\,$365\\
	                &$\,\,\,$0.4675(12)	&$\,\,\,$1.181(19)	 &$\,\,\,$0.272(9)      &$\,\,\,$0.314(6)      &$\,\,\,$      &$\,\,\,$477\\
3.90 (32$^3\times$64)	&$\,\,\,$0.2600(9)	&$\,\,\,$1.174(48)	 &$\,\,\,$0.279(13)     &$\,\,\,$0.282(14)     &$\,\,\,$      &$\,\,\,$667\\
	                &$\,\,\,$0.2978(6)	&$\,\,\,$1.121(32)	 &$\,\,\,$0.258(16)     &$\,\,\,$0.294(10)     &$\,\,\,$      &$\,\,\,$351\\
4.05 (32$^3\times$64)	&$\,\,\,$0.2925(18)	&$\,\,\,$1.211(67)       &$\,\,\,$0.241(25)     &$\,\,\,$0.317(24)     &$\,\,\,$16    &$\,\,\,$447\\
	                &$\,\,\,$0.4035(18)	&$\,\,\,$1.191(32)       &$\,\,\,$0.255(12)     &$\,\,\,$0.319(10)     &$\,\,\,$      &$\,\,\,$326\\
	                &$\,\,\,$0.4653(15)	&$\,\,\,$1.189(25)       &$\,\,\,$0.269(9)      &$\,\,\,$0.310(7)      &$\,\,\,$      &$\,\,\,$419\\
4.20 (32$^3\times$64)	&$\,\,\,$0.4698(18)	&$\,\,\,$1.138(25)       &$\,\,\,$0.250(13)     &$\,\,\,$0.294(9)      &$\,\,\,$18    &$\,\,\,$357\\
4.20 (48$^3\times$96)	&$\,\,\,$0.2622(11)	&$\,\,\,$1.142(44)	 &$\,\,\,$0.274(20)     &$\,\,\,$0.290(15)     &$\,\,\,$      &$\,\,\,$245\\
\hline
\hline
\end{tabular}
\end{center}
\end{table}

\end{widetext}

\newpage

\bibliography{refs}

\begin{thebibliography}{77}
\expandafter\ifx\csname natexlab\endcsname\relax\def\natexlab#1{#1}\fi
\expandafter\ifx\csname bibnamefont\endcsname\relax
  \def\bibnamefont#1{#1}\fi
\expandafter\ifx\csname bibfnamefont\endcsname\relax
  \def\bibfnamefont#1{#1}\fi
\expandafter\ifx\csname citenamefont\endcsname\relax
  \def\citenamefont#1{#1}\fi
\expandafter\ifx\csname url\endcsname\relax
  \def\url#1{\texttt{#1}}\fi
\expandafter\ifx\csname urlprefix\endcsname\relax\def\urlprefix{URL }\fi
\providecommand{\bibinfo}[2]{#2}
\providecommand{\eprint}[2][]{\url{#2}}

\bibitem[{\citenamefont{Bernard et~al.}(1995)\citenamefont{Bernard, Kaiser, and
  Meissner}}]{Bernard:1995dp}
\bibinfo{author}{\bibfnamefont{V.}~\bibnamefont{Bernard}},
  \bibinfo{author}{\bibfnamefont{N.}~\bibnamefont{Kaiser}}, \bibnamefont{and}
  \bibinfo{author}{\bibfnamefont{U.-G.} \bibnamefont{Meissner}},
  \bibinfo{journal}{Int.J.Mod.Phys.} \textbf{\bibinfo{volume}{E4}},
  \bibinfo{pages}{193} (\bibinfo{year}{1995}), \eprint{hep-ph/9501384}.

\bibitem[{\citenamefont{Hemmert et~al.}(2003)\citenamefont{Hemmert, Procura,
  and Weise}}]{Hemmert:2003cb}
\bibinfo{author}{\bibfnamefont{T.~R.} \bibnamefont{Hemmert}},
  \bibinfo{author}{\bibfnamefont{M.}~\bibnamefont{Procura}}, \bibnamefont{and}
  \bibinfo{author}{\bibfnamefont{W.}~\bibnamefont{Weise}},
  \bibinfo{journal}{Phys.Rev.} \textbf{\bibinfo{volume}{D68}},
  \bibinfo{pages}{075009} (\bibinfo{year}{2003}), \eprint{hep-lat/0303002}.

\bibitem[{\citenamefont{Alexandrou}(2010)}]{Alexandrou:2010cm}
\bibinfo{author}{\bibfnamefont{C.}~\bibnamefont{Alexandrou}},
  \bibinfo{journal}{PoS} \textbf{\bibinfo{volume}{LATTICE2010}},
  \bibinfo{pages}{001} (\bibinfo{year}{2010}), \eprint{1011.3660}.

\bibitem[{\citenamefont{Constantinou}(2014)}]{Constantinou:2014tga}
\bibinfo{author}{\bibfnamefont{M.}~\bibnamefont{Constantinou}},
  \bibinfo{journal}{PoS} \textbf{\bibinfo{volume}{LATTICE2014}},
  \bibinfo{pages}{001} (\bibinfo{year}{2014}), \eprint{1411.0078}.

\bibitem[{\citenamefont{Alexandrou
  et~al.}(2013{\natexlab{a}})\citenamefont{Alexandrou, Constantinou, Drach,
  Hatziyiannakou, Jansen et~al.}}]{Alexandrou:2013cda}
\bibinfo{author}{\bibfnamefont{C.}~\bibnamefont{Alexandrou}},
  \bibinfo{author}{\bibfnamefont{M.}~\bibnamefont{Constantinou}},
  \bibinfo{author}{\bibfnamefont{V.}~\bibnamefont{Drach}},
  \bibinfo{author}{\bibfnamefont{K.}~\bibnamefont{Hatziyiannakou}},
  \bibinfo{author}{\bibfnamefont{K.}~\bibnamefont{Jansen}},
  \bibnamefont{et~al.}, \bibinfo{journal}{Nuovo Cim.}
  \textbf{\bibinfo{volume}{C036}}, \bibinfo{pages}{111}
  (\bibinfo{year}{2013}{\natexlab{a}}), \eprint{1303.6818}.

\bibitem[{\citenamefont{Alexandrou}(2014)}]{Alexandrou:2014yha}
\bibinfo{author}{\bibfnamefont{C.}~\bibnamefont{Alexandrou}}
  (\bibinfo{year}{2014}), \eprint{1404.5213}.

\bibitem[{\citenamefont{Syritsyn}(2014)}]{Syritsyn:2014saa}
\bibinfo{author}{\bibfnamefont{S.}~\bibnamefont{Syritsyn}},
  \bibinfo{journal}{PoS} \textbf{\bibinfo{volume}{LATTICE2013}},
  \bibinfo{pages}{009} (\bibinfo{year}{2014}), \eprint{1403.4686}.

\bibitem[{\citenamefont{Bali et~al.}(2014{\natexlab{a}})\citenamefont{Bali,
  Collins, Glässle, Göckeler, Javadi-Motaghi et~al.}}]{Bali:2013gya}
\bibinfo{author}{\bibfnamefont{G.}~\bibnamefont{Bali}},
  \bibinfo{author}{\bibfnamefont{S.}~\bibnamefont{Collins}},
  \bibinfo{author}{\bibfnamefont{B.}~\bibnamefont{Glässle}},
  \bibinfo{author}{\bibfnamefont{M.}~\bibnamefont{Göckeler}},
  \bibinfo{author}{\bibfnamefont{N.}~\bibnamefont{Javadi-Motaghi}},
  \bibnamefont{et~al.}, \bibinfo{journal}{PoS}
  \textbf{\bibinfo{volume}{LATTICE2013}}, \bibinfo{pages}{447}
  (\bibinfo{year}{2014}{\natexlab{a}}), \eprint{1311.7639}.

\bibitem[{\citenamefont{Horsley et~al.}(2014)\citenamefont{Horsley, Nakamura,
  Nobile, Rakow, Schierholz et~al.}}]{Horsley:2013ayv}
\bibinfo{author}{\bibfnamefont{R.}~\bibnamefont{Horsley}},
  \bibinfo{author}{\bibfnamefont{Y.}~\bibnamefont{Nakamura}},
  \bibinfo{author}{\bibfnamefont{A.}~\bibnamefont{Nobile}},
  \bibinfo{author}{\bibfnamefont{P.}~\bibnamefont{Rakow}},
  \bibinfo{author}{\bibfnamefont{G.}~\bibnamefont{Schierholz}},
  \bibnamefont{et~al.}, \bibinfo{journal}{Phys.Lett.}
  \textbf{\bibinfo{volume}{B732}}, \bibinfo{pages}{41} (\bibinfo{year}{2014}),
  \eprint{1302.2233}.

\bibitem[{\citenamefont{Bhattacharya et~al.}(2012)\citenamefont{Bhattacharya,
  Cirigliano, Cohen, Filipuzzi, Gonzalez-Alonso et~al.}}]{Bhattacharya:2011qm}
\bibinfo{author}{\bibfnamefont{T.}~\bibnamefont{Bhattacharya}},
  \bibinfo{author}{\bibfnamefont{V.}~\bibnamefont{Cirigliano}},
  \bibinfo{author}{\bibfnamefont{S.~D.} \bibnamefont{Cohen}},
  \bibinfo{author}{\bibfnamefont{A.}~\bibnamefont{Filipuzzi}},
  \bibinfo{author}{\bibfnamefont{M.}~\bibnamefont{Gonzalez-Alonso}},
  \bibnamefont{et~al.}, \bibinfo{journal}{Phys.Rev.}
  \textbf{\bibinfo{volume}{D85}}, \bibinfo{pages}{054512}
  (\bibinfo{year}{2012}), \eprint{1110.6448}.

\bibitem[{\citenamefont{Gao et~al.}(2011)\citenamefont{Gao, Gamberg, Chen,
  Qian, Qiang et~al.}}]{Gao:2010av}
\bibinfo{author}{\bibfnamefont{H.}~\bibnamefont{Gao}},
  \bibinfo{author}{\bibfnamefont{L.}~\bibnamefont{Gamberg}},
  \bibinfo{author}{\bibfnamefont{J.}~\bibnamefont{Chen}},
  \bibinfo{author}{\bibfnamefont{X.}~\bibnamefont{Qian}},
  \bibinfo{author}{\bibfnamefont{Y.}~\bibnamefont{Qiang}},
  \bibnamefont{et~al.}, \bibinfo{journal}{Eur.Phys.J.Plus}
  \textbf{\bibinfo{volume}{126}}, \bibinfo{pages}{2} (\bibinfo{year}{2011}),
  \eprint{1009.3803}.

\bibitem[{\citenamefont{Ellis and Olive}(2010)}]{Ellis:2010kf}
\bibinfo{author}{\bibfnamefont{J.}~\bibnamefont{Ellis}} \bibnamefont{and}
  \bibinfo{author}{\bibfnamefont{K.~A.} \bibnamefont{Olive}}
  (\bibinfo{year}{2010}), \eprint{1001.3651}.

\bibitem[{\citenamefont{Servant and Tait}(2002)}]{Servant:2002hb}
\bibinfo{author}{\bibfnamefont{G.}~\bibnamefont{Servant}} \bibnamefont{and}
  \bibinfo{author}{\bibfnamefont{T.~M.} \bibnamefont{Tait}},
  \bibinfo{journal}{New J.Phys.} \textbf{\bibinfo{volume}{4}},
  \bibinfo{pages}{99} (\bibinfo{year}{2002}), \eprint{hep-ph/0209262}.

\bibitem[{\citenamefont{Bertone et~al.}(2011)\citenamefont{Bertone, Kong,
  de~Austri, and Trotta}}]{Bertone:2010ww}
\bibinfo{author}{\bibfnamefont{G.}~\bibnamefont{Bertone}},
  \bibinfo{author}{\bibfnamefont{K.}~\bibnamefont{Kong}},
  \bibinfo{author}{\bibfnamefont{R.~R.} \bibnamefont{de~Austri}},
  \bibnamefont{and} \bibinfo{author}{\bibfnamefont{R.}~\bibnamefont{Trotta}},
  \bibinfo{journal}{Phys.Rev.} \textbf{\bibinfo{volume}{D83}},
  \bibinfo{pages}{036008} (\bibinfo{year}{2011}), \eprint{1010.2023}.

\bibitem[{\citenamefont{Collins et~al.}(1985)\citenamefont{Collins, Soper, and
  Sterman}}]{Collins:1985ue}
\bibinfo{author}{\bibfnamefont{J.~C.} \bibnamefont{Collins}},
  \bibinfo{author}{\bibfnamefont{D.~E.} \bibnamefont{Soper}}, \bibnamefont{and}
  \bibinfo{author}{\bibfnamefont{G.~F.} \bibnamefont{Sterman}},
  \bibinfo{journal}{Nucl.Phys.} \textbf{\bibinfo{volume}{B261}},
  \bibinfo{pages}{104} (\bibinfo{year}{1985}).

\bibitem[{\citenamefont{Brock et~al.}(1995)}]{Brock:1993sz}
\bibinfo{author}{\bibfnamefont{R.}~\bibnamefont{Brock}} \bibnamefont{et~al.}
  (\bibinfo{collaboration}{CTEQ Collaboration}),
  \bibinfo{journal}{Rev.Mod.Phys.} \textbf{\bibinfo{volume}{67}},
  \bibinfo{pages}{157} (\bibinfo{year}{1995}).

\bibitem[{\citenamefont{Ji}(2013)}]{Ji:2013dva}
\bibinfo{author}{\bibfnamefont{X.}~\bibnamefont{Ji}},
  \bibinfo{journal}{Phys.Rev.Lett.} \textbf{\bibinfo{volume}{110}},
  \bibinfo{pages}{262002} (\bibinfo{year}{2013}), \eprint{1305.1539}.

\bibitem[{\citenamefont{Xiong et~al.}(2014)\citenamefont{Xiong, Ji, Zhang, and
  Zhao}}]{Xiong:2013bka}
\bibinfo{author}{\bibfnamefont{X.}~\bibnamefont{Xiong}},
  \bibinfo{author}{\bibfnamefont{X.}~\bibnamefont{Ji}},
  \bibinfo{author}{\bibfnamefont{J.-H.} \bibnamefont{Zhang}}, \bibnamefont{and}
  \bibinfo{author}{\bibfnamefont{Y.}~\bibnamefont{Zhao}},
  \bibinfo{journal}{Phys.Rev.} \textbf{\bibinfo{volume}{D90}},
  \bibinfo{pages}{014051} (\bibinfo{year}{2014}), \eprint{1310.7471}.

\bibitem[{\citenamefont{Lin et~al.}(2014)\citenamefont{Lin, Chen, Cohen, and
  Ji}}]{Lin:2014zya}
\bibinfo{author}{\bibfnamefont{H.-W.} \bibnamefont{Lin}},
  \bibinfo{author}{\bibfnamefont{J.-W.} \bibnamefont{Chen}},
  \bibinfo{author}{\bibfnamefont{S.~D.} \bibnamefont{Cohen}}, \bibnamefont{and}
  \bibinfo{author}{\bibfnamefont{X.}~\bibnamefont{Ji}} (\bibinfo{year}{2014}),
  \eprint{1402.1462}.

\bibitem[{\citenamefont{Alexandrou
  et~al.}(2014{\natexlab{a}})\citenamefont{Alexandrou, Cichy, Drach,
  Garcia-Ramos, Hadjiyiannakou et~al.}}]{Alexandrou:2014pna}
\bibinfo{author}{\bibfnamefont{C.}~\bibnamefont{Alexandrou}},
  \bibinfo{author}{\bibfnamefont{K.}~\bibnamefont{Cichy}},
  \bibinfo{author}{\bibfnamefont{V.}~\bibnamefont{Drach}},
  \bibinfo{author}{\bibfnamefont{E.}~\bibnamefont{Garcia-Ramos}},
  \bibinfo{author}{\bibfnamefont{K.}~\bibnamefont{Hadjiyiannakou}},
  \bibnamefont{et~al.}, \bibinfo{journal}{PoS}
  \textbf{\bibinfo{volume}{LATTICE2014}}, \bibinfo{pages}{135}
  (\bibinfo{year}{2014}{\natexlab{a}}), \eprint{1411.0891}.

\bibitem[{\citenamefont{Abdel-Rehim et~al.}(2015)}]{Abdel-Rehim:2015pwa}
\bibinfo{author}{\bibfnamefont{A.}~\bibnamefont{Abdel-Rehim}}
  \bibnamefont{et~al.} (\bibinfo{collaboration}{ETM}) (\bibinfo{year}{2015}),
  \eprint{1507.05068}.

\bibitem[{\citenamefont{Alexandrou
  et~al.}(2011{\natexlab{a}})}]{Alexandrou:2010hf}
\bibinfo{author}{\bibfnamefont{C.}~\bibnamefont{Alexandrou}}
  \bibnamefont{et~al.} (\bibinfo{collaboration}{ETM Collaboration}),
  \bibinfo{journal}{Phys.Rev.} \textbf{\bibinfo{volume}{D83}},
  \bibinfo{pages}{045010} (\bibinfo{year}{2011}{\natexlab{a}}),
  \eprint{1012.0857}.

\bibitem[{\citenamefont{Alexandrou
  et~al.}(2011{\natexlab{b}})\citenamefont{Alexandrou, Carbonell, Constantinou,
  Harraud, Guichon et~al.}}]{Alexandrou:2011nr}
\bibinfo{author}{\bibfnamefont{C.}~\bibnamefont{Alexandrou}},
  \bibinfo{author}{\bibfnamefont{J.}~\bibnamefont{Carbonell}},
  \bibinfo{author}{\bibfnamefont{M.}~\bibnamefont{Constantinou}},
  \bibinfo{author}{\bibfnamefont{P.}~\bibnamefont{Harraud}},
  \bibinfo{author}{\bibfnamefont{P.}~\bibnamefont{Guichon}},
  \bibnamefont{et~al.}, \bibinfo{journal}{Phys.Rev.}
  \textbf{\bibinfo{volume}{D83}}, \bibinfo{pages}{114513}
  (\bibinfo{year}{2011}{\natexlab{b}}), \eprint{1104.1600}.

\bibitem[{\citenamefont{Alexandrou
  et~al.}(2011{\natexlab{c}})\citenamefont{Alexandrou, Brinet, Carbonell,
  Constantinou, Harraud et~al.}}]{Alexandrou:2011db}
\bibinfo{author}{\bibfnamefont{C.}~\bibnamefont{Alexandrou}},
  \bibinfo{author}{\bibfnamefont{M.}~\bibnamefont{Brinet}},
  \bibinfo{author}{\bibfnamefont{J.}~\bibnamefont{Carbonell}},
  \bibinfo{author}{\bibfnamefont{M.}~\bibnamefont{Constantinou}},
  \bibinfo{author}{\bibfnamefont{P.}~\bibnamefont{Harraud}},
  \bibnamefont{et~al.}, \bibinfo{journal}{Phys.Rev.}
  \textbf{\bibinfo{volume}{D83}}, \bibinfo{pages}{094502}
  (\bibinfo{year}{2011}{\natexlab{c}}), \eprint{1102.2208}.

\bibitem[{\citenamefont{Dinter et~al.}(2011)\citenamefont{Dinter, Alexandrou,
  Constantinou, Drach, Jansen et~al.}}]{Dinter:2011sg}
\bibinfo{author}{\bibfnamefont{S.}~\bibnamefont{Dinter}},
  \bibinfo{author}{\bibfnamefont{C.}~\bibnamefont{Alexandrou}},
  \bibinfo{author}{\bibfnamefont{M.}~\bibnamefont{Constantinou}},
  \bibinfo{author}{\bibfnamefont{V.}~\bibnamefont{Drach}},
  \bibinfo{author}{\bibfnamefont{K.}~\bibnamefont{Jansen}},
  \bibnamefont{et~al.}, \bibinfo{journal}{Phys.Lett.}
  \textbf{\bibinfo{volume}{B704}}, \bibinfo{pages}{89} (\bibinfo{year}{2011}),
  \eprint{1108.1076}.

\bibitem[{\citenamefont{Alexandrou
  et~al.}(2013{\natexlab{b}})\citenamefont{Alexandrou, Constantinou, Dinter,
  Drach, Jansen et~al.}}]{Alexandrou:2013joa}
\bibinfo{author}{\bibfnamefont{C.}~\bibnamefont{Alexandrou}},
  \bibinfo{author}{\bibfnamefont{M.}~\bibnamefont{Constantinou}},
  \bibinfo{author}{\bibfnamefont{S.}~\bibnamefont{Dinter}},
  \bibinfo{author}{\bibfnamefont{V.}~\bibnamefont{Drach}},
  \bibinfo{author}{\bibfnamefont{K.}~\bibnamefont{Jansen}},
  \bibnamefont{et~al.}, \bibinfo{journal}{Phys.Rev.}
  \textbf{\bibinfo{volume}{D88}}, \bibinfo{pages}{014509}
  (\bibinfo{year}{2013}{\natexlab{b}}), \eprint{1303.5979}.

\bibitem[{\citenamefont{Abdel-Rehim
  et~al.}(2014{\natexlab{a}})\citenamefont{Abdel-Rehim, Alexandrou,
  Constantinou, Drach, Hadjiyiannakou et~al.}}]{Abdel-Rehim:2013wlz}
\bibinfo{author}{\bibfnamefont{A.}~\bibnamefont{Abdel-Rehim}},
  \bibinfo{author}{\bibfnamefont{C.}~\bibnamefont{Alexandrou}},
  \bibinfo{author}{\bibfnamefont{M.}~\bibnamefont{Constantinou}},
  \bibinfo{author}{\bibfnamefont{V.}~\bibnamefont{Drach}},
  \bibinfo{author}{\bibfnamefont{K.}~\bibnamefont{Hadjiyiannakou}},
  \bibnamefont{et~al.}, \bibinfo{journal}{Phys.Rev.}
  \textbf{\bibinfo{volume}{D89}}, \bibinfo{pages}{034501}
  (\bibinfo{year}{2014}{\natexlab{a}}), \eprint{1310.6339}.

\bibitem[{\citenamefont{Alexandrou
  et~al.}(2014{\natexlab{b}})\citenamefont{Alexandrou, Constantinou, Drach,
  Hadjiyiannakou, Jansen et~al.}}]{Alexandrou:2013wca}
\bibinfo{author}{\bibfnamefont{C.}~\bibnamefont{Alexandrou}},
  \bibinfo{author}{\bibfnamefont{M.}~\bibnamefont{Constantinou}},
  \bibinfo{author}{\bibfnamefont{V.}~\bibnamefont{Drach}},
  \bibinfo{author}{\bibfnamefont{K.}~\bibnamefont{Hadjiyiannakou}},
  \bibinfo{author}{\bibfnamefont{K.}~\bibnamefont{Jansen}},
  \bibnamefont{et~al.}, \bibinfo{journal}{Comput.Phys.Commun.}
  \textbf{\bibinfo{volume}{185}}, \bibinfo{pages}{1370}
  (\bibinfo{year}{2014}{\natexlab{b}}), \eprint{1309.2256}.

\bibitem[{\citenamefont{Alexandrou et~al.}(1994)\citenamefont{Alexandrou,
  Gusken, Jegerlehner, Schilling, and Sommer}}]{Alexandrou:1992ti}
\bibinfo{author}{\bibfnamefont{C.}~\bibnamefont{Alexandrou}},
  \bibinfo{author}{\bibfnamefont{S.}~\bibnamefont{Gusken}},
  \bibinfo{author}{\bibfnamefont{F.}~\bibnamefont{Jegerlehner}},
  \bibinfo{author}{\bibfnamefont{K.}~\bibnamefont{Schilling}},
  \bibnamefont{and} \bibinfo{author}{\bibfnamefont{R.}~\bibnamefont{Sommer}},
  \bibinfo{journal}{Nucl.Phys.} \textbf{\bibinfo{volume}{B414}},
  \bibinfo{pages}{815} (\bibinfo{year}{1994}), \eprint{hep-lat/9211042}.

\bibitem[{\citenamefont{Gusken et~al.}(1989)\citenamefont{Gusken, Low, Mutter,
  Sommer, Patel et~al.}}]{Gusken:1989ad}
\bibinfo{author}{\bibfnamefont{S.}~\bibnamefont{Gusken}},
  \bibinfo{author}{\bibfnamefont{U.}~\bibnamefont{Low}},
  \bibinfo{author}{\bibfnamefont{K.}~\bibnamefont{Mutter}},
  \bibinfo{author}{\bibfnamefont{R.}~\bibnamefont{Sommer}},
  \bibinfo{author}{\bibfnamefont{A.}~\bibnamefont{Patel}},
  \bibnamefont{et~al.}, \bibinfo{journal}{Phys.Lett.}
  \textbf{\bibinfo{volume}{B227}}, \bibinfo{pages}{266} (\bibinfo{year}{1989}).

\bibitem[{\citenamefont{Alexandrou et~al.}(2008)}]{Alexandrou:2008tn}
\bibinfo{author}{\bibfnamefont{C.}~\bibnamefont{Alexandrou}}
  \bibnamefont{et~al.} (\bibinfo{collaboration}{ETM Collaboration})
  (\bibinfo{year}{2008}), \eprint{0803.3190}.

\bibitem[{\citenamefont{Dolgov et~al.}(2002)}]{Dolgov:2002zm}
\bibinfo{author}{\bibfnamefont{D.}~\bibnamefont{Dolgov}} \bibnamefont{et~al.}
  (\bibinfo{collaboration}{LHPC collaboration, TXL Collaboration}),
  \bibinfo{journal}{Phys.Rev.} \textbf{\bibinfo{volume}{D66}},
  \bibinfo{pages}{034506} (\bibinfo{year}{2002}), \eprint{hep-lat/0201021}.

\bibitem[{\citenamefont{Dong and Liu}(1994)}]{Dong:1993pk}
\bibinfo{author}{\bibfnamefont{S.-J.} \bibnamefont{Dong}} \bibnamefont{and}
  \bibinfo{author}{\bibfnamefont{K.-F.} \bibnamefont{Liu}},
  \bibinfo{journal}{Phys.Lett.} \textbf{\bibinfo{volume}{B328}},
  \bibinfo{pages}{130} (\bibinfo{year}{1994}), \eprint{hep-lat/9308015}.

\bibitem[{\citenamefont{Foster and Michael}(1999)}]{Foster:1998vw}
\bibinfo{author}{\bibfnamefont{M.}~\bibnamefont{Foster}} \bibnamefont{and}
  \bibinfo{author}{\bibfnamefont{C.}~\bibnamefont{Michael}}
  (\bibinfo{collaboration}{UKQCD}), \bibinfo{journal}{Phys. Rev.}
  \textbf{\bibinfo{volume}{D59}}, \bibinfo{pages}{074503}
  (\bibinfo{year}{1999}), \eprint{hep-lat/9810021}.

\bibitem[{\citenamefont{McNeile and Michael}(2006)}]{McNeile:2006bz}
\bibinfo{author}{\bibfnamefont{C.}~\bibnamefont{McNeile}} \bibnamefont{and}
  \bibinfo{author}{\bibfnamefont{C.}~\bibnamefont{Michael}}
  (\bibinfo{collaboration}{UKQCD}), \bibinfo{journal}{Phys. Rev.}
  \textbf{\bibinfo{volume}{D73}}, \bibinfo{pages}{074506}
  (\bibinfo{year}{2006}), \eprint{hep-lat/0603007}.

\bibitem[{\citenamefont{Baron et~al.}(2007)}]{Baron:2007ti}
\bibinfo{author}{\bibfnamefont{R.}~\bibnamefont{Baron}} \bibnamefont{et~al.}
  (\bibinfo{collaboration}{ETM Collaboration}), \bibinfo{journal}{PoS}
  \textbf{\bibinfo{volume}{LAT2007}}, \bibinfo{pages}{153}
  (\bibinfo{year}{2007}), \eprint{0710.1580}.

\bibitem[{\citenamefont{Martinelli and Sachrajda}(1987)}]{Martinelli:1987zd}
\bibinfo{author}{\bibfnamefont{G.}~\bibnamefont{Martinelli}} \bibnamefont{and}
  \bibinfo{author}{\bibfnamefont{C.~T.} \bibnamefont{Sachrajda}},
  \bibinfo{journal}{Phys.Lett.} \textbf{\bibinfo{volume}{B196}},
  \bibinfo{pages}{184} (\bibinfo{year}{1987}).

\bibitem[{\citenamefont{McNeile and Michael}(2000)}]{McNeile:2000hf}
\bibinfo{author}{\bibfnamefont{C.}~\bibnamefont{McNeile}} \bibnamefont{and}
  \bibinfo{author}{\bibfnamefont{C.}~\bibnamefont{Michael}}
  (\bibinfo{collaboration}{UKQCD}), \bibinfo{journal}{Phys. Lett.}
  \textbf{\bibinfo{volume}{B491}}, \bibinfo{pages}{123} (\bibinfo{year}{2000}),
  \eprint{hep-lat/0006020}.

\bibitem[{\citenamefont{Maiani et~al.}(1987)\citenamefont{Maiani, Martinelli,
  Paciello, and Taglienti}}]{Maiani:1987by}
\bibinfo{author}{\bibfnamefont{L.}~\bibnamefont{Maiani}},
  \bibinfo{author}{\bibfnamefont{G.}~\bibnamefont{Martinelli}},
  \bibinfo{author}{\bibfnamefont{M.}~\bibnamefont{Paciello}}, \bibnamefont{and}
  \bibinfo{author}{\bibfnamefont{B.}~\bibnamefont{Taglienti}},
  \bibinfo{journal}{Nucl.Phys.} \textbf{\bibinfo{volume}{B293}},
  \bibinfo{pages}{420} (\bibinfo{year}{1987}).

\bibitem[{\citenamefont{Capitani et~al.}(2012)\citenamefont{Capitani,
  Della~Morte, von Hippel, Jager, Juttner et~al.}}]{Capitani:2012gj}
\bibinfo{author}{\bibfnamefont{S.}~\bibnamefont{Capitani}},
  \bibinfo{author}{\bibfnamefont{M.}~\bibnamefont{Della~Morte}},
  \bibinfo{author}{\bibfnamefont{G.}~\bibnamefont{von Hippel}},
  \bibinfo{author}{\bibfnamefont{B.}~\bibnamefont{Jager}},
  \bibinfo{author}{\bibfnamefont{A.}~\bibnamefont{Juttner}},
  \bibnamefont{et~al.}, \bibinfo{journal}{Phys.Rev.}
  \textbf{\bibinfo{volume}{D86}}, \bibinfo{pages}{074502}
  (\bibinfo{year}{2012}), \eprint{1205.0180}.

\bibitem[{\citenamefont{Frezzotti et~al.}(2001)\citenamefont{Frezzotti, Grassi,
  Sint, and Weisz}}]{Frezzotti:2000nk}
\bibinfo{author}{\bibfnamefont{R.}~\bibnamefont{Frezzotti}},
  \bibinfo{author}{\bibfnamefont{P.~A.} \bibnamefont{Grassi}},
  \bibinfo{author}{\bibfnamefont{S.}~\bibnamefont{Sint}}, \bibnamefont{and}
  \bibinfo{author}{\bibfnamefont{P.}~\bibnamefont{Weisz}}
  (\bibinfo{collaboration}{ALPHA}), \bibinfo{journal}{JHEP}
  \textbf{\bibinfo{volume}{08}}, \bibinfo{pages}{058} (\bibinfo{year}{2001}),
  \eprint{hep-lat/0101001}.

\bibitem[{\citenamefont{Frezzotti and Rossi}(2004)}]{Frezzotti:2003ni}
\bibinfo{author}{\bibfnamefont{R.}~\bibnamefont{Frezzotti}} \bibnamefont{and}
  \bibinfo{author}{\bibfnamefont{G.~C.} \bibnamefont{Rossi}},
  \bibinfo{journal}{JHEP} \textbf{\bibinfo{volume}{08}}, \bibinfo{pages}{007}
  (\bibinfo{year}{2004}), \eprint{hep-lat/0306014}.

\bibitem[{\citenamefont{Jansen et~al.}(2005)}]{Jansen:2005cg}
\bibinfo{author}{\bibfnamefont{K.}~\bibnamefont{Jansen}} \bibnamefont{et~al.},
  \bibinfo{journal}{Phys. Lett.} \textbf{\bibinfo{volume}{B624}},
  \bibinfo{pages}{334} (\bibinfo{year}{2005}), \eprint{hep-lat/0507032}.

\bibitem[{\citenamefont{Farchioni et~al.}(2005)}]{Farchioni:2004us}
\bibinfo{author}{\bibfnamefont{F.}~\bibnamefont{Farchioni}}
  \bibnamefont{et~al.}, \bibinfo{journal}{Eur. Phys. J.}
  \textbf{\bibinfo{volume}{C39}}, \bibinfo{pages}{421} (\bibinfo{year}{2005}),
  \eprint{hep-lat/0406039}.

\bibitem[{\citenamefont{Farchioni et~al.}(2006)\citenamefont{Farchioni,
  Hofmann, Jansen, Montvay, Munster et~al.}}]{Farchioni:2005bh}
\bibinfo{author}{\bibfnamefont{F.}~\bibnamefont{Farchioni}},
  \bibinfo{author}{\bibfnamefont{P.}~\bibnamefont{Hofmann}},
  \bibinfo{author}{\bibfnamefont{K.}~\bibnamefont{Jansen}},
  \bibinfo{author}{\bibfnamefont{I.}~\bibnamefont{Montvay}},
  \bibinfo{author}{\bibfnamefont{G.}~\bibnamefont{Munster}},
  \bibnamefont{et~al.}, \bibinfo{journal}{Eur.Phys.J.}
  \textbf{\bibinfo{volume}{C47}}, \bibinfo{pages}{453} (\bibinfo{year}{2006}),
  \eprint{hep-lat/0512017}.

\bibitem[{\citenamefont{Boucaud et~al.}(2007)}]{Boucaud:2007uk}
\bibinfo{author}{\bibfnamefont{P.}~\bibnamefont{Boucaud}} \bibnamefont{et~al.}
  (\bibinfo{collaboration}{ETM}), \bibinfo{journal}{Phys.Lett.}
  \textbf{\bibinfo{volume}{B650}}, \bibinfo{pages}{304} (\bibinfo{year}{2007}),
  \eprint{hep-lat/0701012}.

\bibitem[{\citenamefont{Boucaud et~al.}(2008)}]{Boucaud:2008xu}
\bibinfo{author}{\bibfnamefont{P.}~\bibnamefont{Boucaud}} \bibnamefont{et~al.}
  (\bibinfo{collaboration}{ETM}), \bibinfo{journal}{Comput.Phys.Commun.}
  \textbf{\bibinfo{volume}{179}}, \bibinfo{pages}{695} (\bibinfo{year}{2008}),
  \eprint{0803.0224}.

\bibitem[{\citenamefont{Baron et~al.}(2010{\natexlab{a}})}]{Baron:2009wt}
\bibinfo{author}{\bibfnamefont{R.}~\bibnamefont{Baron}} \bibnamefont{et~al.}
  (\bibinfo{collaboration}{ETM}), \bibinfo{journal}{JHEP}
  \textbf{\bibinfo{volume}{1008}}, \bibinfo{pages}{097}
  (\bibinfo{year}{2010}{\natexlab{a}}), \eprint{0911.5061}.

\bibitem[{\citenamefont{Baron et~al.}(2010{\natexlab{b}})}]{Baron:2010bv}
\bibinfo{author}{\bibfnamefont{R.}~\bibnamefont{Baron}} \bibnamefont{et~al.},
  \bibinfo{journal}{JHEP} \textbf{\bibinfo{volume}{06}}, \bibinfo{pages}{111}
  (\bibinfo{year}{2010}{\natexlab{b}}), \eprint{1004.5284}.

\bibitem[{\citenamefont{Abdel-Rehim et~al.}(2013)\citenamefont{Abdel-Rehim,
  Boucaud, Carrasco, Deuzeman, Dimopoulos et~al.}}]{Abdel-Rehim:2013yaa}
\bibinfo{author}{\bibfnamefont{A.}~\bibnamefont{Abdel-Rehim}},
  \bibinfo{author}{\bibfnamefont{P.}~\bibnamefont{Boucaud}},
  \bibinfo{author}{\bibfnamefont{N.}~\bibnamefont{Carrasco}},
  \bibinfo{author}{\bibfnamefont{A.}~\bibnamefont{Deuzeman}},
  \bibinfo{author}{\bibfnamefont{P.}~\bibnamefont{Dimopoulos}},
  \bibnamefont{et~al.}, \bibinfo{journal}{PoS}
  \textbf{\bibinfo{volume}{LATTICE2013}}, \bibinfo{pages}{264}
  (\bibinfo{year}{2013}), \eprint{1311.4522}.

\bibitem[{\citenamefont{Abdel-Rehim
  et~al.}(2014{\natexlab{b}})\citenamefont{Abdel-Rehim, Alexandrou, Dimopoulos,
  Frezzotti, Jansen et~al.}}]{Abdel-Rehim:2014nka}
\bibinfo{author}{\bibfnamefont{A.}~\bibnamefont{Abdel-Rehim}},
  \bibinfo{author}{\bibfnamefont{C.}~\bibnamefont{Alexandrou}},
  \bibinfo{author}{\bibfnamefont{P.}~\bibnamefont{Dimopoulos}},
  \bibinfo{author}{\bibfnamefont{R.}~\bibnamefont{Frezzotti}},
  \bibinfo{author}{\bibfnamefont{K.}~\bibnamefont{Jansen}},
  \bibnamefont{et~al.}, \bibinfo{journal}{PoS}
  \textbf{\bibinfo{volume}{LATTICE2014}}, \bibinfo{pages}{119}
  (\bibinfo{year}{2014}{\natexlab{b}}), \eprint{1411.6842}.

\bibitem[{\citenamefont{Sommer}(1994)}]{Sommer:1993ce}
\bibinfo{author}{\bibfnamefont{R.}~\bibnamefont{Sommer}},
  \bibinfo{journal}{Nucl. Phys.} \textbf{\bibinfo{volume}{B411}},
  \bibinfo{pages}{839} (\bibinfo{year}{1994}), \eprint{hep-lat/9310022}.

\bibitem[{\citenamefont{Alexandrou et~al.}(2012)\citenamefont{Alexandrou,
  Carbonell, Christaras, Drach, Gravina et~al.}}]{Alexandrou:2012xk}
\bibinfo{author}{\bibfnamefont{C.}~\bibnamefont{Alexandrou}},
  \bibinfo{author}{\bibfnamefont{J.}~\bibnamefont{Carbonell}},
  \bibinfo{author}{\bibfnamefont{D.}~\bibnamefont{Christaras}},
  \bibinfo{author}{\bibfnamefont{V.}~\bibnamefont{Drach}},
  \bibinfo{author}{\bibfnamefont{M.}~\bibnamefont{Gravina}},
  \bibnamefont{et~al.}, \bibinfo{journal}{Phys.Rev.}
  \textbf{\bibinfo{volume}{D86}}, \bibinfo{pages}{114501}
  (\bibinfo{year}{2012}), \eprint{1205.6856}.

\bibitem[{\citenamefont{Alexandrou
  et~al.}(2014{\natexlab{c}})\citenamefont{Alexandrou, Drach, Jansen,
  Kallidonis, and Koutsou}}]{Alexandrou:2014sha}
\bibinfo{author}{\bibfnamefont{C.}~\bibnamefont{Alexandrou}},
  \bibinfo{author}{\bibfnamefont{V.}~\bibnamefont{Drach}},
  \bibinfo{author}{\bibfnamefont{K.}~\bibnamefont{Jansen}},
  \bibinfo{author}{\bibfnamefont{C.}~\bibnamefont{Kallidonis}},
  \bibnamefont{and} \bibinfo{author}{\bibfnamefont{G.}~\bibnamefont{Koutsou}},
  \bibinfo{journal}{Phys.Rev.} \textbf{\bibinfo{volume}{D90}},
  \bibinfo{pages}{074501} (\bibinfo{year}{2014}{\natexlab{c}}),
  \eprint{1406.4310}.

\bibitem[{\citenamefont{Gasser et~al.}(1988)\citenamefont{Gasser, Sainio, and
  Svarc}}]{Gasser:1987rb}
\bibinfo{author}{\bibfnamefont{J.}~\bibnamefont{Gasser}},
  \bibinfo{author}{\bibfnamefont{M.}~\bibnamefont{Sainio}}, \bibnamefont{and}
  \bibinfo{author}{\bibfnamefont{A.}~\bibnamefont{Svarc}},
  \bibinfo{journal}{Nucl.Phys.} \textbf{\bibinfo{volume}{B307}},
  \bibinfo{pages}{779} (\bibinfo{year}{1988}).

\bibitem[{\citenamefont{Carrasco et~al.}(2014)}]{Carrasco:2014cwa}
\bibinfo{author}{\bibfnamefont{N.}~\bibnamefont{Carrasco}} \bibnamefont{et~al.}
  (\bibinfo{collaboration}{European Twisted Mass}),
  \bibinfo{journal}{Nucl.Phys.} \textbf{\bibinfo{volume}{B887}},
  \bibinfo{pages}{19} (\bibinfo{year}{2014}), \eprint{1403.4504}.

\bibitem[{\citenamefont{Alexandrou
  et~al.}(2014{\natexlab{d}})\citenamefont{Alexandrou, Constantinou,
  Hadjiyiannakou, Jansen, Kallidonis et~al.}}]{Alexandrou:2014wca}
\bibinfo{author}{\bibfnamefont{C.}~\bibnamefont{Alexandrou}},
  \bibinfo{author}{\bibfnamefont{M.}~\bibnamefont{Constantinou}},
  \bibinfo{author}{\bibfnamefont{K.}~\bibnamefont{Hadjiyiannakou}},
  \bibinfo{author}{\bibfnamefont{K.}~\bibnamefont{Jansen}},
  \bibinfo{author}{\bibfnamefont{C.}~\bibnamefont{Kallidonis}},
  \bibnamefont{et~al.} (\bibinfo{year}{2014}{\natexlab{d}}),
  \eprint{1411.3494}.

\bibitem[{\citenamefont{Alexandrou
  et~al.}(2014{\natexlab{e}})\citenamefont{Alexandrou, Constantinou, Jansen,
  Koutsou, and Panagopoulos}}]{Alexandrou:2013wka}
\bibinfo{author}{\bibfnamefont{C.}~\bibnamefont{Alexandrou}},
  \bibinfo{author}{\bibfnamefont{M.}~\bibnamefont{Constantinou}},
  \bibinfo{author}{\bibfnamefont{K.}~\bibnamefont{Jansen}},
  \bibinfo{author}{\bibfnamefont{G.}~\bibnamefont{Koutsou}}, \bibnamefont{and}
  \bibinfo{author}{\bibfnamefont{H.}~\bibnamefont{Panagopoulos}},
  \bibinfo{journal}{PoS} \textbf{\bibinfo{volume}{LATTICE2013}},
  \bibinfo{pages}{294} (\bibinfo{year}{2014}{\natexlab{e}}),
  \eprint{1311.4670}.

\bibitem[{\citenamefont{G$\ddot{\rm o}$ckeler
  et~al.}(1999)\citenamefont{G$\ddot{\rm o}$ckeler, Horsley, Oelrich, Perlt,
  Petters et~al.}}]{Gockeler:1998ye}
\bibinfo{author}{\bibfnamefont{M.}~\bibnamefont{G$\ddot{\rm o}$ckeler}},
  \bibinfo{author}{\bibfnamefont{R.}~\bibnamefont{Horsley}},
  \bibinfo{author}{\bibfnamefont{H.}~\bibnamefont{Oelrich}},
  \bibinfo{author}{\bibfnamefont{H.}~\bibnamefont{Perlt}},
  \bibinfo{author}{\bibfnamefont{D.}~\bibnamefont{Petters}},
  \bibnamefont{et~al.}, \bibinfo{journal}{Nucl.Phys.}
  \textbf{\bibinfo{volume}{B544}}, \bibinfo{pages}{699} (\bibinfo{year}{1999}),
  \eprint{hep-lat/9807044}.

\bibitem[{\citenamefont{Constantinou et~al.}(2009)\citenamefont{Constantinou,
  Lubicz, Panagopoulos, and Stylianou}}]{Constantinou:2009tr}
\bibinfo{author}{\bibfnamefont{M.}~\bibnamefont{Constantinou}},
  \bibinfo{author}{\bibfnamefont{V.}~\bibnamefont{Lubicz}},
  \bibinfo{author}{\bibfnamefont{H.}~\bibnamefont{Panagopoulos}},
  \bibnamefont{and}
  \bibinfo{author}{\bibfnamefont{F.}~\bibnamefont{Stylianou}},
  \bibinfo{journal}{JHEP} \textbf{\bibinfo{volume}{0910}}, \bibinfo{pages}{064}
  (\bibinfo{year}{2009}), \eprint{0907.0381}.

\bibitem[{\citenamefont{Alexandrou
  et~al.}(2011{\natexlab{d}})\citenamefont{Alexandrou, Constantinou, Korzec,
  Panagopoulos, and Stylianou}}]{Alexandrou:2010me}
\bibinfo{author}{\bibfnamefont{C.}~\bibnamefont{Alexandrou}},
  \bibinfo{author}{\bibfnamefont{M.}~\bibnamefont{Constantinou}},
  \bibinfo{author}{\bibfnamefont{T.}~\bibnamefont{Korzec}},
  \bibinfo{author}{\bibfnamefont{H.}~\bibnamefont{Panagopoulos}},
  \bibnamefont{and}
  \bibinfo{author}{\bibfnamefont{F.}~\bibnamefont{Stylianou}},
  \bibinfo{journal}{Phys.Rev.} \textbf{\bibinfo{volume}{D83}},
  \bibinfo{pages}{014503} (\bibinfo{year}{2011}{\natexlab{d}}),
  \eprint{1006.1920}.

\bibitem[{\citenamefont{Alexandrou et~al.}(2015)\citenamefont{Alexandrou,
  Constantinou, and Panagopoulos}}]{Alexandrou:2015sea}
\bibinfo{author}{\bibfnamefont{C.}~\bibnamefont{Alexandrou}},
  \bibinfo{author}{\bibfnamefont{M.}~\bibnamefont{Constantinou}},
  \bibnamefont{and}
  \bibinfo{author}{\bibfnamefont{H.}~\bibnamefont{Panagopoulos}}
  (\bibinfo{collaboration}{ETM}) (\bibinfo{year}{2015}), \eprint{1509.00213}.

\bibitem[{\citenamefont{Constantinou et~al.}(2010)}]{Constantinou:2010gr}
\bibinfo{author}{\bibfnamefont{M.}~\bibnamefont{Constantinou}}
  \bibnamefont{et~al.} (\bibinfo{collaboration}{ETM}), \bibinfo{journal}{JHEP}
  \textbf{\bibinfo{volume}{1008}}, \bibinfo{pages}{068} (\bibinfo{year}{2010}),
  \eprint{1004.1115}.

\bibitem[{\citenamefont{Wijesooriya et~al.}(2005)\citenamefont{Wijesooriya,
  Reimer, and Holt}}]{Wijesooriya:2005ir}
\bibinfo{author}{\bibfnamefont{K.}~\bibnamefont{Wijesooriya}},
  \bibinfo{author}{\bibfnamefont{P.}~\bibnamefont{Reimer}}, \bibnamefont{and}
  \bibinfo{author}{\bibfnamefont{R.}~\bibnamefont{Holt}},
  \bibinfo{journal}{Phys.Rev.} \textbf{\bibinfo{volume}{C72}},
  \bibinfo{pages}{065203} (\bibinfo{year}{2005}), \eprint{nucl-ex/0509012}.

\bibitem[{\citenamefont{Bali et~al.}(2015)\citenamefont{Bali, Collins, Glässle,
  Göckeler, Najjar et~al.}}]{Bali:2014nma}
\bibinfo{author}{\bibfnamefont{G.~S.} \bibnamefont{Bali}},
  \bibinfo{author}{\bibfnamefont{S.}~\bibnamefont{Collins}},
  \bibinfo{author}{\bibfnamefont{B.}~\bibnamefont{Glässle}},
  \bibinfo{author}{\bibfnamefont{M.}~\bibnamefont{Göckeler}},
  \bibinfo{author}{\bibfnamefont{J.}~\bibnamefont{Najjar}},
  \bibnamefont{et~al.}, \bibinfo{journal}{Phys.Rev.}
  \textbf{\bibinfo{volume}{D91}}, \bibinfo{pages}{054501}
  (\bibinfo{year}{2015}), \eprint{1412.7336}.

\bibitem[{\citenamefont{Bhattacharya et~al.}(2014)\citenamefont{Bhattacharya,
  Cohen, Gupta, Joseph, Lin et~al.}}]{Bhattacharya:2013ehc}
\bibinfo{author}{\bibfnamefont{T.}~\bibnamefont{Bhattacharya}},
  \bibinfo{author}{\bibfnamefont{S.~D.} \bibnamefont{Cohen}},
  \bibinfo{author}{\bibfnamefont{R.}~\bibnamefont{Gupta}},
  \bibinfo{author}{\bibfnamefont{A.}~\bibnamefont{Joseph}},
  \bibinfo{author}{\bibfnamefont{H.-W.} \bibnamefont{Lin}},
  \bibnamefont{et~al.}, \bibinfo{journal}{Phys.Rev.}
  \textbf{\bibinfo{volume}{D89}}, \bibinfo{pages}{094502}
  (\bibinfo{year}{2014}), \eprint{1306.5435}.

\bibitem[{\citenamefont{Bratt et~al.}(2010)}]{Bratt:2010jn}
\bibinfo{author}{\bibfnamefont{J.~D.} \bibnamefont{Bratt}} \bibnamefont{et~al.}
  (\bibinfo{collaboration}{LHPC}), \bibinfo{journal}{Phys. Rev.}
  \textbf{\bibinfo{volume}{D82}}, \bibinfo{pages}{094502}
  (\bibinfo{year}{2010}), \eprint{1001.3620}.

\bibitem[{\citenamefont{Green et~al.}(2014)\citenamefont{Green, Engelhardt,
  Krieg, Negele, Pochinsky et~al.}}]{Green:2012ud}
\bibinfo{author}{\bibfnamefont{J.}~\bibnamefont{Green}},
  \bibinfo{author}{\bibfnamefont{M.}~\bibnamefont{Engelhardt}},
  \bibinfo{author}{\bibfnamefont{S.}~\bibnamefont{Krieg}},
  \bibinfo{author}{\bibfnamefont{J.}~\bibnamefont{Negele}},
  \bibinfo{author}{\bibfnamefont{A.}~\bibnamefont{Pochinsky}},
  \bibnamefont{et~al.}, \bibinfo{journal}{Phys.Lett.}
  \textbf{\bibinfo{volume}{B734}}, \bibinfo{pages}{290} (\bibinfo{year}{2014}),
  \eprint{1209.1687}.

\bibitem[{\citenamefont{Gupta et~al.}(2014)\citenamefont{Gupta, Bhattacharya,
  Joseph, Cohen, and Lin}}]{Gupta:2014dla}
\bibinfo{author}{\bibfnamefont{R.}~\bibnamefont{Gupta}},
  \bibinfo{author}{\bibfnamefont{T.}~\bibnamefont{Bhattacharya}},
  \bibinfo{author}{\bibfnamefont{A.}~\bibnamefont{Joseph}},
  \bibinfo{author}{\bibfnamefont{S.~D.} \bibnamefont{Cohen}}, \bibnamefont{and}
  \bibinfo{author}{\bibfnamefont{H.-W.} \bibnamefont{Lin}},
  \bibinfo{journal}{PoS} \textbf{\bibinfo{volume}{LATTICE2013}},
  \bibinfo{pages}{409} (\bibinfo{year}{2014}), \eprint{1403.2447}.

\bibitem[{\citenamefont{Bhattacharya et~al.}(2015)\citenamefont{Bhattacharya,
  Cirigliano, Cohen, Gupta, Joseph et~al.}}]{Bhattacharya:2015wna}
\bibinfo{author}{\bibfnamefont{T.}~\bibnamefont{Bhattacharya}},
  \bibinfo{author}{\bibfnamefont{V.}~\bibnamefont{Cirigliano}},
  \bibinfo{author}{\bibfnamefont{S.}~\bibnamefont{Cohen}},
  \bibinfo{author}{\bibfnamefont{R.}~\bibnamefont{Gupta}},
  \bibinfo{author}{\bibfnamefont{A.}~\bibnamefont{Joseph}},
  \bibnamefont{et~al.} (\bibinfo{year}{2015}), \eprint{1506.06411}.

\bibitem[{\citenamefont{Green et~al.}(2012)\citenamefont{Green, Negele,
  Pochinsky, Syritsyn, Engelhardt et~al.}}]{Green:2012ej}
\bibinfo{author}{\bibfnamefont{J.}~\bibnamefont{Green}},
  \bibinfo{author}{\bibfnamefont{J.}~\bibnamefont{Negele}},
  \bibinfo{author}{\bibfnamefont{A.}~\bibnamefont{Pochinsky}},
  \bibinfo{author}{\bibfnamefont{S.}~\bibnamefont{Syritsyn}},
  \bibinfo{author}{\bibfnamefont{M.}~\bibnamefont{Engelhardt}},
  \bibnamefont{et~al.}, \bibinfo{journal}{Phys.Rev.}
  \textbf{\bibinfo{volume}{D86}}, \bibinfo{pages}{114509}
  (\bibinfo{year}{2012}), \eprint{1206.4527}.

\bibitem[{\citenamefont{Aoki et~al.}(2010)\citenamefont{Aoki, Blum, Lin, Ohta,
  Sasaki et~al.}}]{Aoki:2010xg}
\bibinfo{author}{\bibfnamefont{Y.}~\bibnamefont{Aoki}},
  \bibinfo{author}{\bibfnamefont{T.}~\bibnamefont{Blum}},
  \bibinfo{author}{\bibfnamefont{H.-W.} \bibnamefont{Lin}},
  \bibinfo{author}{\bibfnamefont{S.}~\bibnamefont{Ohta}},
  \bibinfo{author}{\bibfnamefont{S.}~\bibnamefont{Sasaki}},
  \bibnamefont{et~al.}, \bibinfo{journal}{Phys.Rev.}
  \textbf{\bibinfo{volume}{D82}}, \bibinfo{pages}{014501}
  (\bibinfo{year}{2010}), \eprint{1003.3387}.

\bibitem[{\citenamefont{Pleiter et~al.}(2010)}]{Pleiter:2011gw}
\bibinfo{author}{\bibfnamefont{D.}~\bibnamefont{Pleiter}} \bibnamefont{et~al.}
  (\bibinfo{collaboration}{QCDSF/UKQCD Collaboration}), \bibinfo{journal}{PoS}
  \textbf{\bibinfo{volume}{LATTICE2010}}, \bibinfo{pages}{153}
  (\bibinfo{year}{2010}), \eprint{1101.2326}.

\bibitem[{\citenamefont{Bali et~al.}(2014{\natexlab{b}})\citenamefont{Bali,
  Collins, Gläßle, Göckeler, Najjar et~al.}}]{Bali:2014gha}
\bibinfo{author}{\bibfnamefont{G.~S.} \bibnamefont{Bali}},
  \bibinfo{author}{\bibfnamefont{S.}~\bibnamefont{Collins}},
  \bibinfo{author}{\bibfnamefont{B.}~\bibnamefont{Gläßle}},
  \bibinfo{author}{\bibfnamefont{M.}~\bibnamefont{Göckeler}},
  \bibinfo{author}{\bibfnamefont{J.}~\bibnamefont{Najjar}},
  \bibnamefont{et~al.}, \bibinfo{journal}{Phys.Rev.}
  \textbf{\bibinfo{volume}{D90}}, \bibinfo{pages}{074510}
  (\bibinfo{year}{2014}{\natexlab{b}}), \eprint{1408.6850}.

\bibitem[{\citenamefont{Alekhin et~al.}(2012)\citenamefont{Alekhin, Blumlein,
  and Moch}}]{Alekhin:2012ig}
\bibinfo{author}{\bibfnamefont{S.}~\bibnamefont{Alekhin}},
  \bibinfo{author}{\bibfnamefont{J.}~\bibnamefont{Blumlein}}, \bibnamefont{and}
  \bibinfo{author}{\bibfnamefont{S.}~\bibnamefont{Moch}},
  \bibinfo{journal}{Phys. Rev.} \textbf{\bibinfo{volume}{D86}},
  \bibinfo{pages}{054009} (\bibinfo{year}{2012}), \eprint{1202.2281}.

\bibitem[{\citenamefont{Blumlein and Bottcher}(2010)}]{Blumlein:2010rn}
\bibinfo{author}{\bibfnamefont{J.}~\bibnamefont{Blumlein}} \bibnamefont{and}
  \bibinfo{author}{\bibfnamefont{H.}~\bibnamefont{Bottcher}},
  \bibinfo{journal}{Nucl. Phys.} \textbf{\bibinfo{volume}{B841}},
  \bibinfo{pages}{205} (\bibinfo{year}{2010}), \eprint{1005.3113}.

\bibitem[{\citenamefont{Bali et~al.}(2012)\citenamefont{Bali, Collins, Deka,
  Glassle, G$\ddot{\rm o}$ckeler et~al.}}]{Bali:2012av}
\bibinfo{author}{\bibfnamefont{G.~S.} \bibnamefont{Bali}},
  \bibinfo{author}{\bibfnamefont{S.}~\bibnamefont{Collins}},
  \bibinfo{author}{\bibfnamefont{M.}~\bibnamefont{Deka}},
  \bibinfo{author}{\bibfnamefont{B.}~\bibnamefont{Glassle}},
  \bibinfo{author}{\bibfnamefont{M.}~\bibnamefont{G$\ddot{\rm o}$ckeler}},
  \bibnamefont{et~al.}, \bibinfo{journal}{Phys.Rev.}
  \textbf{\bibinfo{volume}{D86}}, \bibinfo{pages}{054504}
  (\bibinfo{year}{2012}), \eprint{1207.1110}.

\end{thebibliography}

\end{document}